\documentclass[10pt]{article}
\usepackage{amssymb,amsmath}
\usepackage{amsthm,amsfonts}
\usepackage{mathrsfs}
\usepackage{dcolumn}
\usepackage{bm}
\usepackage{fullpage}
\usepackage{color}
\usepackage[all]{xy}
\usepackage[dvipdfmx]{graphicx,hyperref}
\usepackage{ulem}
\usepackage{ascmac}
\usepackage{caption}
\usepackage{here}

\DeclareMathOperator{\e}{e}

\DeclareMathOperator{\dr}{d}
\DeclareMathOperator{\bdr}{\textbf{d}}

\DeclareMathOperator{\Dr}{D}
\DeclareMathOperator{\Lr}{L}

\begin{document}
\theoremstyle{definition}
\newtheorem{Definition}{Definition}[section]
\newtheorem{Theorem}[Definition]{Theorem}
\newtheorem{Proposition}[Definition]{Proposition}
\newtheorem{Question}[Definition]{Question}
\newtheorem{Lemma}[Definition]{Lemma}
\newtheorem*{Proof}{Proof}
\newtheorem{Example}[Definition]{Example}
\newtheorem{Postulate}[Definition]{Postulate}
\newtheorem{Corollary}[Definition]{Corollary}
\newtheorem{Remark}[Definition]{Remark}
\newtheorem{Claim}[Definition]{Claim}
\newtheorem{Assumption}[Definition]{Assumption}

\theoremstyle{remark}
\newcommand{\beq}{\begin{equation}}
\newcommand{\beqa}{\begin{eqnarray}}
\newcommand{\eeq}{\end{equation}}
\newcommand{\eeqa}{\end{eqnarray}}
\newcommand{\non}{\nonumber}
\newcommand{\lb}{\label}
\newcommand{\fr}[1]{(\ref{#1})}
\newcommand{\bb}{\mbox{\boldmath {$b$}}}
\newcommand{\bbe}{\mbox{\boldmath {$e$}}}
\newcommand{\bt}{\mbox{\boldmath {$t$}}}
\newcommand{\bn}{\mbox{\boldmath {$n$}}}
\newcommand{\br}{\mbox{\boldmath {$r$}}}
\newcommand{\bC}{\mbox{\boldmath {$C$}}}
\newcommand{\bp}{\mbox{\boldmath {$p$}}}
\newcommand{\bx}{\mbox{\boldmath {$x$}}}
\newcommand{\bF}{\mbox{\boldmath {$F$}}}
\newcommand{\bT}{\mbox{\boldmath {$T$}}}
\newcommand{\bQ}{\mbox{\boldmath {$Q$}}}
\newcommand{\bS}{\mbox{\boldmath {$S$}}}
\newcommand{\balpha}{\mbox{\boldmath {$\alpha$}}}
\newcommand{\bomega}{\mbox{\boldmath {$\omega$}}}
\newcommand{\ve}{{\varepsilon}}
\newcommand{\f}{\mathrm{f}}
\newcommand{\s}{\mathrm{s}}
\newcommand{\B}{\mathrm{B}}
\newcommand{\E}{\mathrm{E}}
\newcommand{\G}{\tiny\mathrm{G}}
\newcommand{\R}{\mathrm{R}}
\newcommand{\W}{\mathrm{W}}
\newcommand{\Z}{\mathrm{Z}}
\newcommand{\HH}{\mathrm{H}}
\newcommand{\I}{\mathrm{I}}
\newcommand{\II}{\mathrm{II}}
\newcommand{\Id}{\mathrm{Id}}
\newcommand{\ddiv}{\mathrm{div}}
\newcommand{\Ising}{\footnotesize\mathrm{Ising}}
\newcommand{\hF}{\widehat F}
\newcommand{\hL}{\widehat L}
\newcommand{\tA}{\widetilde A}
\newcommand{\tB}{\widetilde B}
\newcommand{\tC}{\widetilde C}
\newcommand{\tL}{\widetilde L}
\newcommand{\tK}{\widetilde K}
\newcommand{\tX}{\widetilde X}
\newcommand{\tY}{\widetilde Y}
\newcommand{\tU}{\widetilde U}
\newcommand{\tZ}{\widetilde Z}
\newcommand{\talpha}{\widetilde \alpha}
\newcommand{\te}{\widetilde e}
\newcommand{\tv}{\widetilde v}
\newcommand{\ts}{\widetilde s}
\newcommand{\tx}{\widetilde x}
\newcommand{\ty}{\widetilde y}
\newcommand{\ud}{\underline{\delta}}
\newcommand{\uD}{\underline{\Delta}}
\newcommand{\chN}{\check{N}}
\newcommand{\cA}{{\cal A}}
\newcommand{\cB}{{\cal B}}
\newcommand{\cC}{{\cal C}}
\newcommand{\cD}{{\cal D}}
\newcommand{\cE}{{\cal E}}
\newcommand{\cF}{{\cal F}}
\newcommand{\cG}{{\cal G}}
\newcommand{\cH}{{\cal H}}
\newcommand{\cI}{{\cal I}}
\newcommand{\cJ}{{\cal J}}
\newcommand{\cK}{{\cal K}}
\newcommand{\cL}{{\cal L}}
\newcommand{\cM}{{\cal M}}
\newcommand{\cN}{{\cal N}}
\newcommand{\cO}{{\cal O}}
\newcommand{\cP}{{\cal P}}
\newcommand{\cQ}{{\cal Q}}
\newcommand{\cS}{{\cal S}}
\newcommand{\cT}{{\cal T}}
\newcommand{\cY}{{\cal Y}}
\newcommand{\cZ}{{\cal Z}}
\newcommand{\cU}{{\cal U}}
\newcommand{\cV}{{\cal V}}
\newcommand{\cW}{{\cal W}}
\newcommand{\vecX}{\mathfrak{X}}
\newcommand{\tcA}{\widetilde{\cal A}}
\newcommand{\DD}{{\cal D}}
\newcommand\TYPE[3]{ \underset {(#1)}{\overset{{#3}}{#2}}  }
\newcommand{\Qc}{\overset{\footnotesize\circ}{Q}}
\newcommand{\bfe}{\boldsymbol e} 
\newcommand{\bfb}{{\boldsymbol b}}
\newcommand{\bfd}{{\boldsymbol d}}
\newcommand{\bfh}{{\boldsymbol h}}
\newcommand{\bfj}{{\boldsymbol j}}
\newcommand{\bfn}{{\boldsymbol n}}
\newcommand{\bfA}{{\boldsymbol A}}
\newcommand{\bfB}{{\boldsymbol B}}
\newcommand{\bfJ}{{\boldsymbol J}}
\newcommand{\bfS}{{\boldsymbol S}}
\newcommand{\saddle}{\mathrm{saddle}}
\newcommand{\can}{\mathrm{can}}
\newcommand{\const}{\mathrm{const.}}
\newcommand{\eq}{\,\mathrm{eq}}
\newcommand{\wt}[1]{\widetilde{#1}}
\newcommand{\wh}[1]{\widehat{#1}}
\newcommand{\ch}[1]{\check{#1}}
\newcommand{\ol}[1]{\overline{#1}}
\newcommand{\ii}{\imath}
\newcommand{\ic}{\iota}
\newcommand{\mbbE}{\mathbb{E}}
\newcommand{\mbbJ}{\mathbb{J}}
\newcommand{\mbbP}{\mathbb{P}}
\newcommand{\mbbR}{\mathbb{R}}
\newcommand{\mbbN}{\mathbb{N}}
\newcommand{\mbbZ}{\mathbb{Z}}
\newcommand{\Leftrightup}[1]{\overset{\mathrm{#1}}{\Longleftrightarrow}}
\newcommand{\leftrightup}[1]{\overset{\mathrm{#1}}{\longleftrightarrow}}
\newcommand{\rightup}[1]{\overset{\mathrm{#1}}{\longrightarrow}}
\newcommand{\avg}[1]{\left\langle\,{#1}\, \right\rangle}
\newcommand{\step}{\lrcorner\hspace*{-0.55mm}\ulcorner}
\newcommand{\equp}[1]{\overset{\mathrm{#1}}{=}}
\newcommand{\nequp}[1]{\overset{\mathrm{#1}}{\tiny\neq}}
\newcommand{\equpg}[2]{\overset{\mathrm{#2}}{#1}}
\newcommand{\nin}{\in\hspace{-.78em}\setminus}
\newcommand{\fraX}{\mathfrak{X}}
\newcommand{\Gam}[1]{\Gamma{#1}}
\newcommand{\GamLamM}[1]{{\Gamma\Lambda^{{#1}}\cal{M}}}
\newcommand{\GamLam}[2]{{\Gamma\Lambda^{{#2}}{#1}}}
\newcommand{\GTM}{\Gamma T{\cal M}}
\newcommand{\GT}[1]{{\Gamma T{#1}}}
\newcommand{\inp}[2]{\left\langle\,  #1\, , \, #2\, \right\rangle}
\newcommand{\inpr}[2]{\left(\,  #1\, , \, #2\, \right)}
\newcommand{\rmt}[1]{{\mathrm{t}}({#1})}
\newcommand{\rmo}[1]{{\mathrm{o}}({#1})}

\makeatletter
\newcommand*{\da@rightarrow}{\mathchar"0\hexnumber@\symAMSa 4B }
\newcommand*{\da@leftarrow}{\mathchar"0\hexnumber@\symAMSa 4C }
\newcommand*{\xdashrightarrow}[2][]{%
  \mathrel{%
    \mathpalette{\da@xarrow{#1}{#2}{}\da@rightarrow{\,}{}}{}%
  }%
}
\newcommand{\xdashleftarrow}[2][]{%
  \mathrel{%
    \mathpalette{\da@xarrow{#1}{#2}\da@leftarrow{}{}{\,}}{}%
  }%
}
\newcommand*{\da@xarrow}[7]{%
  \sbox0{$\ifx#7\scriptstyle\scriptscriptstyle\else\scriptstyle\fi#5#1#6\m@th$}%
  \sbox2{$\ifx#7\scriptstyle\scriptscriptstyle\else\scriptstyle\fi#5#2#6\m@th$}%
  \sbox4{$#7\dabar@\m@th$}%
  \dimen@=\wd0 %
  \ifdim\wd2 >\dimen@
    \dimen@=\wd2 %
  \fi
  \count@=2 %
  \def\da@bars{\dabar@\dabar@}%
  \@whiledim\count@\wd4<\dimen@\do{%
    \advance\count@\@ne
    \expandafter\def\expandafter\da@bars\expandafter{%
      \da@bars
      \dabar@ 
    }%
  }%
  \mathrel{#3}%
  \mathrel{%
    \mathop{\da@bars}\limits
    \ifx\\#1\\%
    \else
      _{\copy0}%
    \fi
    \ifx\\#2\\%
    \else
      ^{\copy2}%
    \fi
  }%
  \mathrel{#4}%
}
\makeatother

\title{ From the Fokker-Planck equation to a contact Hamiltonian system}
\author{\large Shin-itiro GOTO 
 \\
Center for Mathematical Science and Artificial Intelligence,\\
Chubu University,\quad 
1200 Matsumoto-cho, Kasugai, Aichi 487-8501, Japan
}
%
%
\date{}
\maketitle
\noindent
ORCID:\ \verb|https://orcid.org/0000-0002-5249-1054|   \\


\begin{abstract}%
  The Fokker-Planck equation is one of the fundamental equations
  in nonequilibrium statistical mechanics,
  and this equation is known to be derived from  
  the Wasserstein gradient flow equation with a free energy.
  This gradient flow equation describes relaxation processes and is 
  formulated on a Riemannian manifold.   
  Meanwhile contact Hamiltonian systems are also known to 
  describe relaxation processes. Hence a relation 
  between these two equations is expected to be clarified, which gives
  a solid foundation in geometric statistical mechanics.   
  In this paper a class of contact Hamiltonian systems is derived from
  a class of the Fokker-Planck equations on Riemannian manifolds. 
  In the course of the derivation, the Fokker-Planck equation   
  is shown to be written as a diffusion equation
  with a weighted Laplacian without any approximation, which enables to  
  employ a theory of eigenvalue problems. 
\end{abstract}%
\tableofcontents

\section{Introduction}
Nonequilibrium statistical mechanics is a developing branch of physics,
studies   
relations between 
time-dependent processes of probability distribution
  functions and 
nonequilibrium thermodynamic phenomena\,\cite{Zwanzig2001},
and is consistent with equilibrium statistical mechanics\,\cite{Kubo1991}. 
In theoretical physics,  
considerable activity is being devoted to establish a reliable theory, 
where 
such a theory is expected to provide mathematical methodologies  
for analyzing nonequilibrium phenomena.  
To this end various fundamental equations have been proposed and analyzed,
where such fundamental equations 
are the Fokker-Planck equation, master equations, and so on\,\cite{Zwanzig2001}.
Their applications include not only to physical sciences, but also to 
mathematical engineering\,\cite{Risken1989,Nishimori2001}. 
In addition,
by applying these equations to toy models,  
explicit mathematical expressions of 
thermodynamic quantities have been obtained, 
where such toy models include the Glauber 
equation\,\cite{Glauber1963,Entov2023}, 
Brownian motion\,\cite{Hsu2002},
and gas systems\,\cite{Zwanzig2001}.  
Recalling that some of these developments 
lie in the applications of well-established mathematics, such as
stochastic partial differential equations\,\cite{Pavliotis2014},
dynamical systems theory\,\cite{Ruelle1999}, and
geometry\,\cite{Ezra2002,Gay2018}, 
one expects that further developments of this research field will lie in 
the application of advanced mathematics. 

Among various mathematical formulations that 
have been proposed so far, one class relies on geometric methods. 
Examples include Wasserstein geometry\,\cite{Vu2023PRX} and 
contact geometry\,\cite{Mrugala1990}, 
where Wasserstein geometry studies Wasserstein spaces 
for probability density functions\,\cite{Villani2008},
and 
contact geometry is an odd-dimensional   
cousin of symplectic geometry\,\cite{Silva2008}.
Contact geometry has been applied to thermodynamics and statistical mechanics
\,\cite{Mrugala2000chapter,Bravetti2019,Schaft2018,Haslach1997,Grmela2014,Gromov2015}, since it is a differential geometry endowed with 
an appropriate $1$-form.
This $1$-form describes the laws of thermodynamics.   
As mentioned above, 
since there are several theories, any comparison among them is needed
to establish a solid foundation of geometric nonequilibrium
statistical mechanics. 
As a first step, a Wasserstein gradient flow
was compared with a contact Hamiltonian one for a particular class of
models\,\cite{Entov2023arxiv}, where relaxation was investigated.  
Since relaxation process is a simple nonequilibrium phenomenon and
this links  nonequilibrium states and equilibrium states, 
the clarification of how to express relaxation processes should  
be a meaningful target to advance our understanding of 
nonequilibrium systems. 
Beyond this pioneering study, 
more general approaches should be explored. 
One of the advantages of involving Wasserstein geometry is that 
techniques developed in Riemannian geometry can be employed,
where Riemannian geometry has been applied to various physical
problems\,\cite{Abraham1991}. This
advantage is expected to be used efficiently
in future explorations.

In this paper a class of contact Hamiltonian equation
as ODEs (Ordinary Differential Equations)  
is derived from the
Fokker-Planck equation by means of Riemannian geometry,
where this Fokker-Planck equation is 
a PDE (Partial Differential Equation). This PDE is 
the Wasserstein gradient flow of a free energy.  
Here a set of 
thermodynamic  variables
 is represented as a point of a contact manifold, 
where these variables are
defined by integrating some functions with the weight of a solution of the
Fokker-Planck equation. 
The most principal step for deriving a contact Hamiltonian system 
is to put fast time-scales appeared in a solution of 
the Wasserstein gradient equation zero.  
Then main theorems in this paper are 
informally summarized as follows:
\begin{Claim}
\label{claim:diffusion}
  (Informal version of Theorem\,\ref{fact:diffusion-from-Fokker-Planck}).
  A diffusion equation with a weighted Laplacian is derived 
  from the Fokker-Planck equation without any approximation,
  where the weighted Laplacian is of the form $A^{\dagger}A$ with
  $A$ and $A^{\dagger}$ being an operator and its adjoint. 
\end{Claim}
\begin{Claim}
  \label{claim:contact}
  (Informal version of Theorem\,\ref{fact:Fokker-Planck-contact-Hamiltonian}).
  The Fokker-Planck equation is equivalent to
  a contact Hamiltonian system when fast
  time-scales in the diffusion equation are neglected. 
\end{Claim}
In addressing Claim\,\ref{claim:diffusion},
a weighted Laplacian is introduced in this paper.
It should be emphasized that 
there are several weighted Laplacians in the literature, 
and they are sometimes called the Witten Laplacians. 
On the one hand, 
as summarized in \,\cite{Helffer2005}, 
although one existing Witten Laplacian is 
closely related to 
the Fokker-Planck equation, there is no exact one-to-one correspondence 
between a heat equation 
and the Fokker-Planck equation, 
where heat equation is a synonym of diffusion equation.  
On the other hand, the weighted Laplacian in this paper 
is shown to be equivalent to 
another existing self-adjoint operator acting on a
function\,\cite{Villani2000}
(see Section\,\ref{section:appendix-Fokker-Planck-d-and-d-dagger} of 
this paper).  
This existing operator is shown to be a Witten Laplacian, and
yields another form of the Fokker-Planck equation. 
  
The rest of this paper is organized as follows.
In Section\,\ref{section:Fokker-Planck-to-Diffusion},
how the Fokker-Planck equation is derived from the
Wasserstein gradient flow is summarized. Then a diffusion equation
with a weighted Laplacian is derived without any approximation.
Since this diffusion equation exactly corresponds to the Fokker-Planck
equation, several properties of the introduced Laplacian are clarified here  
and employed for later sections.  
In Section\,\ref{section:diffusion-to-contact},
after contact geometry is briefly summarized, 
the contact Hamiltonian flow is derived
with the use of the information about a spectrum of the
weighted Laplacian. 
In addition, an explicit example of how the theory is applied is shown. 
Finally, Section\,\ref{section:conclusion} summarizes
the present study, and possible future
directions are presented.

\section{From the Fokker-Planck equation to a diffusion equation}
\label{section:Fokker-Planck-to-Diffusion}
As shown in \cite{Otto1997}, the Fokker-Planck equation
can be derived from a variational principle with a free-energy functional 
in the context of Wasserstein geometry.

In this section after recalling how to derive the Fokker-Planck equation, 
it is 
shown that this equation induces a diffusion equation with
a weighted Laplacian.

\subsection{The Fokker-Planck equation as a Wasserstein gradient flow}
In this subsection necessary notations in Riemannian geometry are fixed,
and then how the Fokker-Planck equation is derived is summarized. 

Let $(\cM,g)$ be an $m$-dimensional Riemannian manifold with
$g$ being a Riemannian metric tensor field, 
and $\star 1$ a canonical volume-form on $\cM$. The boundary of $\cM$
is denoted by $\partial\cM$. 
An interpretation of $\cM$ in this paper 
is phase space of a 
dynamical  system. 
An example of $\cM$ is a symplectic manifold,
and $g$ is a compatible 
metric with respect to an endowed symplectic form on $\cM$,  
in the sense that $g(-,-)=\omega(-,J-)$ holds 
with $J$ being an almost complex structure.  
In the discussion below the manifold need not be symplectic. 
The spaces of vector fields and $k$-forms on $\cM$ are denoted by $\GTM$ and
$\GamLamM{k}$, respectively.
Since $0$-forms are identified with functions, 
$\GamLamM{0}$ denotes the space of functions on $\cM$.  
The Hodge map is introduced and denoted by 
$\star:\GamLamM{k}\to\GamLamM{m-k}$, ($k=0,1,\ldots,m$).
Given $\star$, its inverse is denoted by $\star^{-1}$. 
Let $\Id$ be the identical operator:$\GamLamM{k}\to\GamLamM{k}$, 
($k=0,1,\ldots,m$).   
Vector fields in a patch on $\cM$ can explicitly be 
written in coordinates locally,
and coordinate expressions give systems
of ordinary differential equations (ODEs). 
Since ODEs describe continuous-time dynamical systems, 
vector fields are often identified with dynamical systems in this paper. 
Then $\dr:\GamLamM{k}\to\GamLamM{k+1}$, ($k=0,1,\ldots,m$) 
is the exterior 
derivative and $\ii_{X}$ is the interior product with respect to $X\in\GTM$.   
After defining a global inner product for forms, 
the so-called co-derivative $\dr^{\dagger}$, is introduced as in 
the standard way. This $\dr^{\dagger}$ is often expressed in the literature 
as $\delta$\,\cite{Choquet-Bruhat1997}, but $\delta$ is not used in this paper. 
To clearly distinguish nationally 
the adjoint of $\dr$ and the Hodge map  
$\star$, $\dr^{*}$ is not used in this paper, where $\dr^{*}$ is often 
employed to denote the adjoint of $\dr$. 
Some of the details of $\dr^{\dagger}$ are summarized in this section 
so that a cousin $\dr_{\G}^{\dagger}$ is analogously introduced in this paper.  
Throughout the paper   
integrals over $\cM$ are assumed to commute with 
$\partial/\partial t$, where $t\in\mbbR$ denotes time. 

To describe the Fokker-Planck equation on $\cM$, 
let  $h$ be a function on $\cM$ called a 
{\it Hamiltonian},   
$\beta$ denote the inverse temperature with the Boltzmann constant 
$k_{\B}$ being unity. 
Throughout this paper $\beta>0$ is constant in time and 
over $\cM$.   
The space of other externally applied fields, $\mbbR^{n}$,   
will be introduced in later sections.  
It is known that the Fokker-Planck equation can be derived as a Wasserstein 
gradient flow associated with a free energy 
on the Euclidean spaces\,\cite{Otto1997}. 
Some extension to Riemannian manifolds can be established\,\cite{Lotto2008}.  
To recall this derivation and to fix notations, in the following,  
the Fokker-Planck equation on $\cM$ for $\rho_{t}\in\GamLamM{0}$,
($\forall t\in\mbbR$) 
\beq
\frac{\partial}{\partial t}\rho_{t}
=-\frac{1}{\beta}\dr^{\dagger}\dr\rho_{t}-\dr^{\dagger}\rho_{t}\dr h,
\label{Fokker-Planck-0}
\eeq
is derived as a Wasserstein gradient flow. 
The $\rho_{t}$ is interpreted as a time-dependent distribution function. 
Hence $\rho_{t}$ is demanded to be normalized: 
\beq
\int_{\cM}\rho_{t}\star 1
=1,\qquad \forall t\in \mbbR.
\label{rho-normalization}
\eeq
For a simple case where $\cM=\mbbR$ and $g=\dr x\otimes\dr x$ with $x\in\mbbR$, 
a coordinate expression of \fr{Fokker-Planck-0} 
will be given in \fr{Fokker-Planck-R}.  

To derive \fr{Fokker-Planck-0}, preliminary formulae are shown below. 
Let $\dr^{\dagger}$ be the co-derivative 
that is  
\beqa
\dr^{\dagger}
:\GamLamM{k}
&\to&\GamLamM{m-1},\qquad k=0,1,\ldots,m
\non\\
\alpha
&\mapsto&\dr^{\dagger}\alpha,
\non
\eeqa
with \cite{Choquet-Bruhat1997}  
$$
\dr^{\dagger}\alpha 
:=  (-1)^{k}\star^{-1}\dr\star\alpha, 
\qquad k=0,1,\ldots,m. 
$$
From this definition it follows that $\dr^{\dagger}\Phi=0$ 
for any function $\Phi$.  
To see that $\dr^{\dagger}$ is an adjoint operator of $\dr$,  
introduce the global inner-product 
\beqa
\inp{}{}:\GamLamM{k}\times\GamLamM{k}&\to&\mbbR,\qquad k=0,1,\ldots,m
\non\\
(\alpha^{\I},\alpha^{\II})&\mapsto&\inp{\alpha^{\I}}{\alpha^{\II}},
\qquad
\non
\eeqa
with
\beq
\inp{\alpha^{\I}}{\alpha^{\II}}
:=\int_{\cM}\alpha^{\I}\wedge\star\alpha^{\II}, 
\label{inner-product-standard-k}
\eeq
if the support of either $\alpha^{\I}$ or $\alpha^{\II}$ is compact 
  such that \fr{inner-product-standard-k} is finite. This finiteness condition
is assumed throughout implicitly. 
For instance, for any $\alpha\in\GamLamM{1}$ and $\Phi\in\GamLamM{0}$ 
with boundary values on $\partial\cM$, it follows that 
\beqa
\inp{\dr^{\dagger}\alpha}{\Phi}
&=&\int_{\cM}(\dr^{\dagger}\alpha)\Phi\star1
=-\int_{\cM}(\star^{-1}\dr\star\alpha)\Phi\star1
=-\int_{\cM}\Phi(\dr\star\alpha)
\non\\
&=&-\int_{\cM}\dr(\Phi\star\alpha)+\int_{\cM}\dr\Phi\wedge\star\alpha
=-[\Phi\star\alpha]_{\partial\cM}
+\inp{\dr\Phi}{\alpha}
\non\\
&=&\inp{\alpha}{\dr\Phi},
\label{d-dagger-for-1-forms}
\eeqa
where the boundary term on $\partial\cM$ has been assumed to vanish. 
The operator $\ddiv$ is defined as 
$$
\ddiv 
:=-\dr^{\dagger}. 
$$
In addition the corresponding potential 
$\phi_{\dot{\rho}_{t}}\in\GamLamM{0}$ 
is introduced such that the continuity equation holds\,\cite{Abraham1991}: 
\beq
\dot{\rho}_{t}
=-\ddiv(\rho_{t}\dr\phi_{\dot{\rho}_{t}})
=\dr^{\dagger}\rho_{t}\dr\phi_{\dot{\rho}_{t}},\quad
\forall\ t\in\mbbR, 
\label{continuity-equation}
\eeq
where the abbreviation has been introduced: 
$$
\dot{\rho}_{t}
:=\frac{\partial}{\partial t}\rho_{t}.
$$

Using these introduced operators and integrals, one can derive
\fr{Fokker-Planck-0} as follows. Recall that $\cM$ is phase space of a
dynamical system, 
and that $g$ is given on $\cM$ as a Riemannian metric.  
In addition, $h\in\GamLamM{0}$ has been given as a Hamiltonian. 
Let the free energy $F$ be written as  
\beqa
F&=&\beta^{-1}S-H,
\label{F=S-H}\\
S&=&-\inp{\rho_{t}}{\ln\rho_{t}}
=-\int_{\cM}\rho_{t}\ln\rho_{t}\,\star1,
\non\\
H&=&\inp{h}{\rho_{t}}
=\int_{\cM}h\rho_{t}\star1.
\non
\eeqa
Then from
\fr{continuity-equation}, \fr{d-dagger-for-1-forms}, and     
$$
\int_{\cM}\dot{\rho}_{t}\star1
=0,
$$
it follows that 
\beqa
\dot{S}
&=&-\int_{\cM}\dot{\rho}_{t}(\ln\rho_{t}+1)\star1
=-\int_{\cM}\dot{\rho}_{t}\ln\rho_{t}\star1
=-\inp{\dr^{\dagger}\rho_{t}\dr\phi_{\dot{\rho}_{t}}}{\ln\rho_{t}}
\non\\
&=&
-\inp{\rho_{t}\dr\phi_{\dot{\rho_{t}}}}{\dr \ln\rho_{t}}
=-\inp{\dr\phi_{\dot{\rho}_{t}}}{\rho_{t}\dr\ln\rho_{t}}
=-\inp{\phi_{\dot{\rho}_{t}}}{\dr^{\dagger}\rho_{t}\dr\ln\rho_{t}}.
\non
\eeqa
To reduce this equation further, by employing
$$
\rho_{t}\dr\ln\rho_{t}
=\rho_{t}\frac{1}{\rho_{t}}\dr\rho_{t}
=\dr\rho_{t},
$$
one has that 
$$
\dot{S}
=-\inp{\phi_{\dot{\rho}_{t}}}{\dr^{\dagger}\dr\rho_{t}}
=-\inp{\dr^{\dagger}\dr\rho_{t}}{\phi_{\dot{\rho}_{t}}}.
$$
In addition, simple calculations yield 
\beqa
\dot{H}
&=&\int_{\cM}h\dot{\rho}_{t}\star1
=\int_{\cM}h(\dr^{\dagger}\rho_{t}\dr\phi_{\dot{\rho}_{t}})\star1
=\inp{h}{\dr^{\dagger}\rho_{t}\dr\phi_{\dot{\rho}_{t}}}
\non\\
&=&\inp{\rho_{t}\dr h}{\dr\phi_{\dot{\rho}_{t}}}
=\inp{\dr^{\dagger}\rho_{t}\dr h}{\phi_{\dot{\rho}_{t}}}.
\non
\eeqa
By defining $\text{grad}_{W}F\in\GamLamM{0}$ such that 
$$
\dot{F}
=\inp{\text{grad}_{W}F}{\phi_{\dot{\rho}_{t}}},
$$
one identifies
$$
\text{grad}_{W}F
=-\beta^{-1}\dr^{\dagger}\dr \rho_{t}
-\dr^{\dagger}\rho_{t}\dr h.
$$
Thus the equation called the {\it gradient flow equation}, 
$$
\frac{\partial}{\partial t}\rho_{t}
=\text{grad}_{W}F, 
$$
is the Fokker-Planck equation,\fr{Fokker-Planck-0}. 
See \cite{Entov2023arxiv} for the case of discrete phase space.  

For a stationary solution $\rho_{\G}$ that satisfies $\dot{\rho}_{\G}=0$ 
for \fr{Fokker-Planck-0}, 
introduce the partition function $Z_{\G}$ such that 
\beq
\int_{\cM}\rho_{\G}\star 1
=1.
\label{rho-G-normalization}
\eeq
The stationary solution satisfies 
$$
\dr^{\dagger}\left(\beta^{-1}\dr\rho_{\G}+\rho_{\G}\dr h\right)
=0,
$$
and it is the Gibbs distribution function:
\beq
\rho_{\G}
=Z_{\G}^{-1} \exp(-\beta h),
\label{Gibbs-distribution}
\eeq
where
$Z_{\G}$ is expressed as
$$
Z_{\G}=\int_{\cM}\e^{-\beta h}\star 1,
$$
and the stationary property can be verified from
\fr{Gibbs-distribution} with   
\beq
\beta^{-1}\dr\rho_{\G}
=-Z_{\G}^{-1}\exp(-\beta h)\dr h
=-\rho_{\G}\dr h.
\label{d-Gibbs-distribution}
\eeq

\subsection{Diffusion equation with a weighted Laplacian}
In this subsection a diffusion equation with a weighted Laplacian
is derived from the Fokker-Planck equation, where no approximation is made. 
The main steps for this derivation 
are to employ a change of variables and to define
an appropriate weighted Laplacian.
After deriving the diffusion equation, its basic properties are shown.

To derive a version of the diffusion equation in an appropriate variable
from the Fokker-Planck equation,  introduce $\Phi_{t}$ such that 
\beq
\rho_{t}
=\rho_{\G}\Phi_{t}
\label{rho-rho-G-Phi},
\eeq
where $\rho_{\G}$ has been defined in \fr{Gibbs-distribution}.
To write the equation for $\Phi_{t}$ in terms of a symmetric
operator 
with respect to a global inner product, several operators and
a global inner product are defined below.
In this paper any symmetric operator is assumed to become a self-adjoint one. 
First, 
let $\star_{\G}$ and its inverse be given by 
$$
\star_{\G}:\GamLamM{k}\to\GamLamM{m-k}\quad\text{and}\quad
\star_{\G}^{-1}:\GamLamM{m-k}\to\GamLamM{k},\qquad
k=0,1,\ldots,m
$$
with 
$$
\star_{\G}
:=\rho_{\G}\star,\quad\text{and}\quad
\star_{\G}^{-1}
=\rho_{\G}^{-1}\star^{-1}, 
$$
for $\rho_{\G}\neq 0$. Notice that
$$
\star_{\G}^{-1}\ \star_{\G}
=\star_{\G}\ \star_{\G}^{-1}
=\Id.
$$
In addition, let $\dr_{\G}^{\dagger}$ be the map such that 
\beqa
\dr_{\G}^{\dagger}:
\GamLamM{k}
&\to&\GamLamM{m-1},\qquad k=0,1,\ldots,m
\non\\
\alpha&\mapsto&\dr_{\G}^{\dagger}\alpha,\qquad
\non
\eeqa
with
$$
\dr_{\G}^{\dagger}\alpha
:=(-1)^{k}\star_{\G}^{-1}\dr \star_{\G}\alpha,\qquad k=0,1,\ldots,m.
$$
From this definition it follows that $\dr_{\G}^{\dagger}\Phi=0$ for any  
function $\Phi$. 
Thus,
\beq
\dr_{\G}^{\dagger}
=\rho_{\G}^{-1}\dr^{\dagger}\rho_{\G},\qquad\text{or equivalently}\qquad
\dr^{\dagger}
=\rho_{\G}\dr_{\G}^{\dagger}\rho_{\G}^{-1},
\label{d-dagger-conversion}
\eeq
from which $\dr_{\G}^{\dagger}\dr_{\G}^{\dagger}=0$. 
This $\dr_{\G}^{\dagger}$ is called the {\it co-derivative associated with $\rho_{\G}$}, and $\rho_{\G}$ is called a {\it weight} 
in this paper.
    
Introduce  
the {\it weighted Laplacian} $\triangle_{\G}$ acting on functions on $\cM$ 
\beq
\triangle_{\G}
:=\dr_{\G}^{\dagger}\dr,
\label{weighted-Laplacian}
\eeq
which reduces to the standard Laplacian on $\cM$ when $\rho_{\G}=1$.
To discuss properties of the weighted Laplacian $\triangle_{\G}$, introduce
the global inner product with $\rho_{\G}$, 
\beqa
\inp{}{}_{\G}:\GamLamM{k}\times \GamLamM{k}&\to&\mbbR
\non\\
(\alpha^{\I},\alpha^{\II})&\mapsto&\inp{\alpha^{\I}}{\alpha^{\II}}_{\G}
\non
\eeqa
where 
\beq
\inp{\alpha^{\I}}{\alpha^{\II}}_{\G}
=\int_{\cM}\left(\alpha^{\I}\wedge\star \alpha^{\II}\right)\,\rho_{\G},\quad
\forall\, \alpha^{\I},\alpha^{\II}\in\GamLamM{k},\quad
k=0,1,\ldots,m. 
\label{inner-product-G}
\eeq
  The integral in \fr{inner-product-G} is assumed to be finite.
  This finiteness condition on $\alpha^{\I}$, $\alpha^{\II}$ and $\rho_{\G}$
  is assumed throughout this paper implicitly, which is similar to that for
  $\inp{}{}$. 
For the case of $k=0$, this definition is written as 
$$
\inp{\Phi^{\I}}{\Phi^{\II}}_{\G}
=\int_{\cM}\Phi^{\I}\Phi^{\II}\rho_{\G}\star 1,
\quad\forall\ \Phi^{\I},\Phi^{\II}\in\GamLamM{0}.
$$

Then $\inp{}{}_{\G}$, $\dr_{\G}^{\dagger}$, $\triangle_{\G}$ have
the following properties.
These properties are employed 
below without mentioning explicitly. 
\begin{Lemma}
\label{Fact:weighted-operators-formulae}
\begin{enumerate}
\item 
$$
\inp{\alpha^{\I}}{\alpha^{\II}}_{\G}
=\inp{\alpha^{\II}}{\alpha^{\I}}_{\G},
\qquad \forall\ \alpha^{\I},\alpha^{\II}\in\GamLamM{k},\quad
k=0,1,\ldots,m.
$$

\item
(Adjoint of $\dr_{\G}^{\dagger}$ is $\dr$). 
When the following boundary condition is satisfied: 
$$
\Phi\dr\star\rho_{\G}\alpha|_{\partial\cM}
=0,
$$
it follows that 
$$
\inp{\dr_{\G}^{\dagger}\alpha}{\Phi}_{\G}
=\inp{\alpha}{\dr\Phi}_{\G},\qquad   
\forall\ \alpha\in\GamLamM{1},\ \Phi\in\GamLamM{0}.
$$

\item
(Adjoint of $\dr$ is $\dr_{\G}^{\dagger}$).
When the following boundary condition is satisfied: 
$$
\alpha^{\I}\wedge\star\rho_{\G}\alpha^{\II}|_{\partial\cM}
=0,
$$
it follows that 
$$
\inp{\dr\alpha^{\I}}{\alpha^{\II}}_{\G}
=\inp{\alpha^{\I}}{\dr_{\G}^{\dagger}\alpha^{\II}}_{\G},\qquad
\forall \alpha^{\I}\in\GamLamM{k},\quad
\alpha^{\II}\in\GamLamM{k+1},\quad k=0,1,\ldots,m.
$$

\item
(Decomposition of $\triangle_{\G}$ into $\dr$). 
When the boundary condition is satisfied:
$$
\Phi^{\II}\dr\star\rho_{\G}\dr\Phi^{\I}|_{\partial\cM}
=0, 
$$
it follows that 
$$
\inp{\triangle_{\G}\Phi^{\I}}{\Phi^{\II}}_{\G}
=\inp{\dr\Phi^{\I}}{\dr\Phi^{\II}}_{\G},\qquad   
\forall\ \Phi^{\I},\Phi^{\II}\in\GamLamM{0}.
$$

\item
(Symmetric property of $\triangle_{\G}$). 
When the boundary conditions are satisfied:
$$
\Phi^{\II}\dr\star\rho_{\G}\dr\Phi^{\I}|_{\partial\cM}
=0,\quad\text{and}\quad
\Phi^{\I}\dr\star\rho_{\G}\dr\Phi^{\II}|_{\partial\cM}
=0,   
$$ 
it follows that 
$$
\inp{\triangle_{\G}\Phi^{\I}}{\Phi^{\II}}_{\G}
=\inp{\Phi^{\I}}{\triangle_{\G}\Phi^{\II}}_{\G},\qquad
\forall\ \Phi^{\I},\Phi^{\II}\in\GamLamM{0}.
$$
\end{enumerate}
\end{Lemma}
\begin{Proof}
\begin{enumerate}
\item
  The statement follows by
  $\alpha^{\I}\wedge\star\alpha^{\II}=\alpha^{\II}\wedge\star\alpha^{\I}$
  for all $\alpha^{\I},\alpha^{\II}\in\GamLamM{k}$, $k=0,1,\ldots,m$.
\item
Since the co-derivative associated with $\rho_{\G}$ acting on 
$\alpha\in\GamLamM{1}$ is
$$
\dr_{\G}^{\dagger}\alpha
=-\rho_{\G}^{-1}\star^{-1}\dr\star\rho_{\G}\alpha,
$$
it follows that 
\beqa
\inp{\dr_{\G}^{\dagger}\alpha}{\Phi}_{\G}
&=&
-\int_{\cM}(\rho_{\G}^{-1}\star^{-1}\dr\star\rho_{\G}\alpha)
\wedge\star \Phi\rho_{\G}
=-\int_{\cM}\Phi(\dr\star\rho_{\G}\alpha)
\non\\
&=&-\int_{\cM}\dr(\Phi\star\rho_{\G}\alpha)
+\int_{\cM}\dr\Phi\wedge\star\rho_{\G}\alpha
=-\Phi\dr\star\rho_{\G}\alpha|_{\partial\cM} 
+\int_{\cM}\dr\Phi\wedge\star\rho_{\G}\alpha
\non\\
&=&0+\inp{\dr\Phi}{\alpha}_{\G}
=\inp{\alpha}{\dr\Phi}_{\G}.
\non
\eeqa

\item
  It follows that 
\beqa
\inp{\dr\alpha^{\I}}{\alpha^{\II}}_{\G}
&=&\int_{\cM}\dr\alpha^{\I}\wedge\star\rho_{\G}\alpha^{\II}
=\int_{\cM}\dr(\alpha^{\I}\wedge\star\rho_{\G}\alpha^{\II})
-(-1)^{k}\int_{\cM}\alpha^{\I}\wedge\dr\star\rho_{\G}\alpha^{\II}
\non\\
&=&0-(-1)^{k}\int_{\cM}\alpha^{\I}\wedge\dr\star\rho_{\G}\alpha^{\II}
=(-1)^{k+1}\int_{\cM}\alpha^{\I}\wedge\star(\rho_{\G}\rho_{\G}^{-1}) 
\star^{-1}\dr\star\rho_{\G}\alpha^{\II}
\non\\
&=&
(-1)^{k+1}\int_{\cM}\alpha^{\I}\wedge\star\rho_{\G}
(\rho_{\G}^{-1}\star^{-1}\dr\star\rho_{\G})\alpha^{\II}
\non\\
&=&(-1)^{k+1}\int_{\cM}\alpha^{\I}\wedge\star\rho_{\G}(-1)^{k+1}\dr_{\G}^{\dagger}\alpha^{\II}
\non\\
&=&\int_{\cM}\alpha^{\I}\wedge\star\rho_{\G}\dr_{\G}^{\dagger}\alpha^{\II}
=\inp{\alpha^{\I}}{\dr_{\G}^{\dagger}\alpha^{\II}}_{\G}.
\non
\eeqa

\item
It follows from item 1 that 
$$
\inp{\triangle_{\G}\Phi^{\I}}{\Phi^{\II}}_{\G}
=\inp{\dr_{\G}^{\dagger}\dr\Phi^{\I}}{\Phi^{\II}}_{\G}
=\inp{\dr\Phi^{\I}}{\dr\Phi^{\II}}_{\G}.
$$

\item
Combining the proved equations, one has
$$
\inp{\triangle_{\G}\Phi^{\I}}{\Phi^{\II}}_{\G}
=\inp{\dr_{\G}^{\dagger}\dr\Phi^{\I}}{\Phi^{\II}}_{\G}
=\inp{\dr\Phi^{\I}}{\dr\Phi^{\II}}_{\G}
=\inp{\Phi^{\I}}{\dr_{\G}^{\dagger}\dr\Phi^{\II}}_{\G}
=\inp{\Phi^{\I}}{\triangle_{\G}\Phi^{\II}}_{\G}.
$$
\end{enumerate}
\qed
\end{Proof}
In Lemma\,\ref{Fact:weighted-operators-formulae},
several boundary conditions have been imposed.
These conditions are satisfied on closed manifolds due to
$\partial\cM=\emptyset$. For non-closed manifolds, such as $\mbbR^{m}$, 
these conditions are not strong, because for the case that
$h$ is divergent sufficiently fast at the boundary, 
$\rho_{\G}=Z_{\G}^{-1}\exp(-\beta h)$ in \fr{Gibbs-distribution}
makes the boundary terms zero.  
In what follows all of these boundary conditions are assumed to be satisfied. 

In terms of the weighted Laplacian $\triangle_{\G}$
and the change of variables \fr{rho-rho-G-Phi}, 
it is shown below, as the main theorem in this subsection, that 
the Fokker-Planck equation is equivalent to 
the diffusion equation with the weighted Laplacian.

\begin{Theorem}
\label{fact:diffusion-from-Fokker-Planck}
(Diffusion equation from the Fokker-Planck equation). 
The function $\Phi_{t}$ in \fr{rho-rho-G-Phi} satisfies
the modified diffusion equation:
\beq
\frac{\partial}{\partial t}\Phi_{t}
=-\,\beta^{-1}\triangle_{\G}\Phi_{t},  
\label{diffusion-from-Fokker-Planck} 
\eeq
where $\rho_{t}=\rho_{\G}\Phi_{t}$ solves the Fokker-Planck equation
\fr{Fokker-Planck-0} 
with $\rho_{\G}\neq 0$ on $\cM\setminus\partial\cM$,   and 
$\triangle_{G}=\dr_{\G}^{\dagger}\dr$ defined in \fr{weighted-Laplacian}  
is the weighted Laplacian.
\end{Theorem}
\begin{Proof}
  By integrating $\dot{\rho}_{t}\zeta$ over $\cM$ with $\zeta$ being
  an arbitrary function,  and employing
\fr{Fokker-Planck-0}, one has
\beq
\frac{\partial}{\partial t}\inp{\rho_{t}}{\zeta}
=-\beta^{-1}\inp{\dr^{\dagger}\dr\rho_{t}}{\zeta}
-\inp{\dr^{\dagger}\rho_{t}\dr h}{\zeta}.
\label{diffusion-from-Fokker-Planck-proof1}
\eeq
After expressing \fr{diffusion-from-Fokker-Planck-proof1} in terms of
$\inp{}{}_{\G}$ in \fr{inner-product-G}, and recalling
that $\zeta$ is arbitrary, one completes the proof.
The details of calculations are as follows.  

The left hand side of \fr{diffusion-from-Fokker-Planck-proof1} is written as
$$
\frac{\partial}{\partial t}\inp{\rho_{t}}{\zeta}
=\frac{\partial}{\partial t}\inp{\Phi_{t}}{\zeta}_{\G}.
$$
The right hand side of \fr{diffusion-from-Fokker-Planck-proof1}
is written as follows.
First, using \fr{d-dagger-conversion} and \fr{d-Gibbs-distribution}, one has  
\beqa
-\beta^{-1}\inp{\dr^{\dagger}\dr\rho_{t}}{\zeta}
&=&-\beta^{-1}\inp{\rho_{\G}\dr_{\G}^{\dagger}\rho_{\G}^{-1}\dr\rho_{t}}{\zeta}
=-\beta^{-1}\inp{\dr_{\G}^{\dagger}\rho_{\G}^{-1}\left(\dr\rho_{t}\right)}{\zeta}_{\G}
\non\\
&=&-\beta^{-1}\inp{\dr_{\G}^{\dagger}\rho_{\G}^{-1}
  \left(\rho_{\G}\dr\Phi_{t}+\Phi_{t}\dr\rho_{\G}\right)}{\zeta}_{\G}
\non\\
&=&-\beta^{-1}\inp{\dr_{\G}^{\dagger}
  \left(\dr\Phi_{t}+\rho_{\G}^{-1}\Phi_{t}\dr\rho_{\G}\right)}{\zeta}_{\G}
\non\\
&=&-\beta^{-1}\inp{\triangle_{\G}\Phi_{t}+\dr_{\G}^{\dagger}\rho_{\G}^{-1}\Phi_{t}\dr\rho_{\G}}{\zeta}_{\G},
\non
\eeqa
and
\beqa
-\inp{\dr^{\dagger}\rho_{t}\dr h}{\zeta}
&=&-\inp{\rho_{\G}^{-1}\dr^{\dagger}(\rho_{\G}\rho_{\G}^{-1})\rho_{t}\dr h}{\zeta}_{\G}
=-\inp{\dr_{\G}^{\dagger}\rho_{\G}^{-1}(\rho_{\G}\Phi_{t})\dr h}{\zeta}_{\G}
\non\\
&=&-\inp{\dr_{\G}^{\dagger}\Phi_{t}\dr h}{\zeta}_{\G}
=\beta^{-1}\inp{\dr_{\G}^{\dagger}\rho_{\G}^{-1}\Phi_{t}\dr\rho_{\G}}{\zeta}_{\G}.
\non
\eeqa
Hence, by summing the two obtained equations, 
the right hand side of \fr{diffusion-from-Fokker-Planck-proof1}
is expressed as 
$$
-\beta^{-1}\inp{\triangle_{\G}\Phi_{t}}{\zeta}_{\G}.
$$
Since the right hand side equals the left hand side, one has
$$
\frac{\partial}{\partial t}\inp{\Phi_{t}}{\zeta}_{\G}
=-\beta^{-1}\inp{\triangle_{\G}\Phi_{t}}{\zeta}_{\G}.
$$
Since $\zeta$ is arbitrary, one has \fr{diffusion-from-Fokker-Planck}.
\qed
\end{Proof}
Another derivation of \fr{diffusion-from-Fokker-Planck}
is shown in Section\,\ref{section:appendix-simplified-proof-1}, in which, 
it is proved without the explicit use of the global inner product. 
The operator $\triangle_{\G}$ is 
a Witten Laplacian, where  
Witten Laplacians are  
widely discussed in several contexts. 
Here, in this paper, a weighted 
Laplacian associated with a global inner product is
also called the 
{\it Witten Laplacian}, when a weight is not identical.  
Since several weights can be considered,    
several Witten Laplacians can be defined.  
\begin{Remark}
There are several studies on how to rewrite the Fokker-Planck equation
and related ones in the literature. In this Remark some of them are briefly discussed. 
In Theorem\,\ref{fact:diffusion-from-Fokker-Planck}
the change of variables, \fr{rho-rho-G-Phi}, is employed. This sort of
change of variables was also employed in \cite{Goto2020JMP} for
deriving a diffusion equation from the
continuous-time master equation without
approximation.
In addition, some changes of variables are discussed in
Sections 5 and 6 of \cite{Risken1989}. In particular in Section 6.3 
of \cite{Risken1989}, a transformation is presented in which 
a method of eigenfunction expansions can be applied to     
the Fokker-Planck equation, and one operator in \cite{Risken1989}
is not self-adjoint (or equivalently, that operator is non-hermitian).
Meanwhile the present operator $\Delta_{\G}$ is
symmetric 
as shown in Lemma\,\ref{Fact:weighted-operators-formulae}, and
hence a general theory of self-adjoint operators on manifolds
can fully be applied to 
the study of the Fokker-Planck equation 
if the symmetric operator is self-adjoint.  
In the literature another self-adjoint operator acting on 
$\Phi_{t}=\rho_{\G}^{-1}\rho_{t}$ has been known \cite{Villani2000,Li2011}.
To briefly discuss such an operator, let $\Lr_{\beta,h}$
be the operator such that
$-\Lr_{\beta,h}\Phi_{t}$ is the right hand side of
\fr{diffusion-from-Fokker-Planck}  
(see Propositions\,\ref{fact:F-self-adjoint}
and \ref{fact:F-weighted-Laplacian} in  
Section\,\ref{section:appendix-Fokker-Planck-d-and-d-dagger}
of this paper for properties of $\Lr_{\beta,h}$).
It is unclear how to generalize $\Lr_{\beta,h}$ 
so that it can be applied to a $k$-form with $k\geq 1$. 
An advantageous point of $\triangle_{\G}$ is that  
it can be generalized so that its action to $k$-forms is defined,  
$\GamLamM{k}\to\GamLamM{k}$, $k=0,\ldots,m$.
This generalization should be $\dr_{\G}^{\dagger}\dr+\dr\dr_{\G}^{\dagger}$,
by following the definition of the standard Laplacian for $k$-forms 
\cite{Choquet-Bruhat1997}. Another advantageous point is
as follows.
As shown in Proposition\,\ref{fact:F-weighted-Laplacian}, 
$\triangle_{\G}$ is equal to  
a version of the Witten Laplacian studied in \cite{Futaki2013} up to some
constant.   
Hence existing theorems about    
the Witten Laplacian can be applied to the Fokker-Planck equation.    
\end{Remark}
In the rest of this subsection properties of $\Phi_{t}$ 
in   
Theorem\,\ref{fact:diffusion-from-Fokker-Planck} are shown. 

In theorem\,\ref{fact:diffusion-from-Fokker-Planck}, the diffusion equation
is derived. Conversely, as shown below,
the diffusion equation with some conditions yield
the Fokker-Planck equation. 
\begin{Proposition}
Let $\rho_{\infty}$ be an equilibrium distribution function 
associated with a function $h$ such that
$$
\dot{\rho}_{\infty}=0,\quad\text{and}\quad
\dr{\rho}_{\infty}=-\beta \rho_{\infty}\dr h.
$$
In addition, let 
$\triangle_{\infty}=\dr_{\infty}^{\dagger}\dr$
be the weighted Laplacian
with $\dr_{\infty}^{\dagger}=\rho_{\infty}^{-1}\dr\rho_{\infty}$. 
Then the diffusion equation associated with $\triangle_{\infty}$,   
\beq
\frac{\partial}{\partial t}\Phi_{t}
=-\beta^{-1}\triangle_{\infty}\Phi_{t}
\label{diffusion-eq-0}
\eeq
induces the Fokker-Planck equation.
\end{Proposition}
\begin{Proof}
Let $\rho_{t}=\Phi_{t}\rho_{\infty}$.  
Multiplying both sides of \fr{diffusion-eq-0}  by $\rho_{\infty}$,  
one has
$$
\frac{\partial}{\partial t}\rho_{t}
=-\beta^{-1}\rho_{\infty}\dr_{\infty}^{\dagger}\dr \Phi_{t}
=-\beta^{-1}\rho_{\infty}(\rho_{\infty}^{-1}\dr^{\dagger}\rho_{\infty})\dr \Phi_{t}
=-\beta^{-1}\dr^{\dagger}\rho_{\infty}\dr \Phi_{t}.
$$
Since $\dr\rho_{t}=\rho_{\infty}\dr\Phi_{t}+\Phi_{t}\dr\rho_{\infty}$ 
and $\dr\rho_{\infty}=-\beta \rho_{\infty}\dr h$, one has
$$
\frac{\partial}{\partial t}\rho_{t}
=-\beta^{-1}\dr^{\dagger}(\dr\rho_{t}-\Phi_{t}\dr\rho_{\infty})
=-\beta^{-1}\dr^{\dagger}\dr\rho_{t}-\dr^{\dagger}\rho_{t}\dr h,  
$$
which is the Fokker-Planck equation, \fr{Fokker-Planck-0}.
\qed
\end{Proof}

There are several properties of $\Phi_{t}$ as shown below.
\begin{Lemma}
$$
\inp{\Phi_{t}}{1}_{\G}
=\inp{1}{1}_{\G}=1,\quad\text{and}\quad
\inp{\dot{\Phi}_{t}}{1}_{\G}
=0.
$$
\end{Lemma}
\begin{Proof}
Straightforward calculations show that 
$$
\inp{\Phi_{t}}{1}_{\G}
=\int_{\cM}\Phi_{t}\rho_{\G}\star1
=\int_{\cM}\rho_{t}\star1
=1,\quad\text{and}\quad
\inp{1}{1}_{\G}
=\int_{\cM}\rho_{\G}\star1
=1,
$$
due to \fr{rho-normalization} and \fr{rho-G-normalization}. 
In addition, it follows that 
$$
\inp{\dot{\Phi}_{t}}{1}_{\G}
=\inp{-\beta^{-1}\triangle_{\G}\Phi}{1}_{\G}
=-\beta^{-1}\inp{\Phi}{\triangle_{\G}1}_{\G}
=-\beta^{-1}\inp{\Phi}{0}_{\G}
=0,
$$
or when $\partial/\partial t$ and the  
integral over $\cM$ commutes,  
$$
\inp{\dot{\Phi}_{t}}{1}_{\G}
=\frac{\dr}{\dr t}
\inp{\Phi_{t}}{1}_{\G}
=\frac{\dr}{\dr t}1
=0.
$$
\qed
\end{Proof}

To show that $\Phi_{t}=1$ 
is stable under \fr{diffusion-from-Fokker-Planck},
one can construct a Lyapunov function,
where the state $\Phi_{t}=1$ is equivalent to $\rho_{t}=\rho_{\G}$ due to
$\rho_{t}=\rho_{\G}\Phi_{t}$. To show this property of $\Phi_{t}$,  
let $\Upsilon_{t}$ be the function of $t$ defined as  
$$
\Upsilon_{t}
:=\frac{\beta}{2}\int_{\cM}\dr\Phi_{t}\wedge\rho_{\G}\star\dr \Phi_{t}
=\frac{\beta}{2}\inp{\dr\Phi_{t}}{\dr\Phi_{t}}_{\G},\quad
\forall t\in\mbbR.
$$
Then the following statements show 
that $\rho_{\G}$ is attractive under the time-development.
\begin{Proposition}
\label{fact:Lyapunov-function-exists}
The function $\Upsilon_{t}$ is a Lyapunov function for 
\fr{diffusion-from-Fokker-Planck}, 
and $\Phi_{t}=\const$ is asymptotically stable. 
\end{Proposition}
\begin{Proof}
It immediately follows that $\Upsilon_{t}\geq 0$, $\forall t\in\mbbR$.  
The derivative of $\Upsilon_{t}$ with respect to $t$ is calculated 
with \fr{diffusion-from-Fokker-Planck}  to be 
\beqa
\frac{\dr}{\dr t}\Upsilon_{t}
&=&
\beta\inp{\dr\dot{\Phi}_{t}}{\dr\Phi_{t}}_{\G}
=\beta\inp{-\beta^{-1}\dr\triangle_{\G}\Phi_{t}}{\dr\Phi_{t}}_{\G}
=-\inp{\triangle_{\G}\Phi_{t}}{\dr_{\G}^{\dagger}\dr\Phi_{t}}_{\G}
\non\\
&=&-\inp{\triangle_{\G}\Phi_{t}}{\triangle_{\G}\Phi_{t}}_{\G}
\leq 0,
\non
\eeqa
for all $t\in\mbbR$. Combining 
$\Upsilon_{t}\geq 0$ and $\dot{\Upsilon}_{t}\leq 0$, one concludes that 
$\Upsilon_{t}$ is a Lyapunov function.
The equality holds when $\triangle_{\G}\Phi_{t}=0$.
Notice for $c$ being a constant that 
$\triangle_{\G}c=\dr_{\G}^{\dagger}\dr c=\dr_{\G}^{\dagger}0=0$. 
Applying the
Lyapunov theorem to this inequality, one completes the proof. 
\qed
\end{Proof}
From Proposition\,\ref{fact:Lyapunov-function-exists},
the following holds:
\begin{Corollary}
\label{fact:limit-Phi}
$$
\lim_{t\to\infty}\Phi_{t}
=1,\qquad\text{and}\qquad
\lim_{t\to\infty}\rho_{t}
=\rho_{\G}.
$$
\end{Corollary}
\begin{Proof}
By Proposition\,\ref{fact:Lyapunov-function-exists},
$\Phi_{t}\to c$ as $t\to\infty$, where $c$ is constant,  one has that 
$$
\lim_{t\to\infty}\rho_{t}
=\rho_{\G}\lim_{t\to\infty}\Phi_{t}
=\rho_{\G} c.
$$
This and the normalization condition \fr{rho-normalization} 
yield
$$
\int_{\cM}\lim_{t\to\infty}\rho_{t}\star 1
=c\int_{\cM}\rho_{\G}\star 1
=1,
$$
Applying \fr{rho-G-normalization} to this obtained equation,  
one has that $c=1$.
This completes the proof. 
\qed
\end{Proof}

\subsection{Spectrum associated with the Fokker-Planck equation}
In Section\,\ref{section:diffusion-to-contact}, 
ODEs will be derived by reducing  
the diffusion equation
\fr{diffusion-from-Fokker-Planck}, 
where \fr{diffusion-from-Fokker-Planck} has been derived from  the Fokker-Planck equation. To this end several preliminary arguments are necessary, and
they are discussed in this subsection.  
Recall how the standard diffusion equation associated
with the unweighted Laplacian 
reduces to a set of ODEs. The strategy for reducing the 
modified diffusion equation is to modify this reduction for the standard 
diffusion equation. Then, 
one recognizes that the spectrum of  
the self-adjoint operator $\triangle_{\G}$ plays a central role in 
this reduction associated with the Fokker-Planck equation.  
Hence in this subsection attention is focused on the eigenvalue problem
of $\triangle_{\G}$ 
first. Then its basic properties and applications are shown. 

Let $\phi_{s}$ be an eigenfunction of $\triangle_{\G}$
labeled by $s$, 
\beq
\triangle_{\G}\phi_{s}
=\lambda_{s}\phi_{s},
\qquad s=0,1,2,\ldots,  
\label{eigenvalue-equation}
\eeq
where $\lambda_{s}$ is an eigenvalue, and 
all the elements of $\{\phi_{s}\}$ are assumed to be orthonormal:      
\beq
\inp{\phi_{s}}{\phi_{s^{\prime}}}_{\G}
=\delta_{ss^{\prime}},
\qquad s,s^{\prime}=0,1,2,\ldots,  
\label{orthonormal-phi}
\eeq
with $\delta_{ss^{\prime}}$ being the Kronecker delta,
giving unity if $s=s^{\prime}$ and $0$ otherwise. 
In addition all the elements of $\{\phi_{s}\}$ are assumed 
to be non-identically vanishing.   
  To employ standard techniques in functional analysis,
  several properties on $\triangle_{\G}$ are needed to be 
  verified. Since the focus of this study is not such mathematical analysis,
  needed properties are assumed to be satisfied in this paper.
  In the context of geometric analysis 
  rigorous statements are known for manifolds with $\rho_{\G}=1$ \cite{Jost2017,Chavel1984}.  
  Meanwhile rigorous statements for manifolds with non-constant $\rho_{\G}$ 
  should be clarified in future.

The following simple statement 
plays one of the central roles
in the later sections. 
\begin{Lemma}
\label{fact:eigenvalue-0-positive}
In\,\fr{eigenvalue-equation}, every eigenvalue $\lambda_{s}$ 
is equal or greater than zero. 
\end{Lemma}
\begin{Proof}
From\,\fr{eigenvalue-equation}, it follows for each $s$ that
$$
\lambda_{s}\inp{\phi_{s}}{\phi_{s}}_{\G}
=\inp{\lambda_{s}\phi_{s}}{\phi_{s}}_{\G}
=\inp{\triangle_{\G}\phi_{s}}{\phi_{s}}_{\G}
=\inp{\dr^{\dagger}_{\G}\dr\phi_{s}}{\phi_{s}}_{\G}
=\inp{\dr\phi_{s}}{\dr\phi_{s}}_{\G}.
$$
Combining this, 
$$
\inp{\phi_{s}}{\phi_{s}}_{\G}
=1, 
\qquad\text{and}\qquad
\inp{\dr\phi_{s}}{\dr\phi_{s}}_{\G}\geq 0, 
$$
one has that $\lambda_{s}\geq 0$.
\qed
\end{Proof}
In this paper $\phi_{s}$ is called the {\it $s$th mode}, and
the following ordering is assumed:
$$
\lambda_{0}=0 \leq 
\lambda_{1} \leq \lambda_{2}\leq \cdots.
$$
The following is a basic statement for the eigenvalue problem.
\begin{Proposition}
\label{fact:eigen-0-function}
In\,\fr{eigenvalue-equation},
the $0$th mode always exists, and it is given by
$$
\phi_{0}=1.
$$
\end{Proposition}
\begin{Proof}
The $0$th mode $\phi_{0}$ is obtained
from $\triangle_{\G}\phi_{0}=0$ as $\phi_{0}=1$, 
where $\inp{\phi_{0}}{\phi_{0}}_{\G}=\inp{1}{1}_{\G}=1$ 
is satisfied as required in \fr{orthonormal-phi}.  
\qed
\end{Proof}

To obtain ODEs that are reduced from the Fokker-Planck equation,   
assume that $\Phi_{t}$ can be expanded  
in terms of $\{\phi_{s}\}$ as
\beq
\Phi_{t}
=\sum_{s=0}^{\infty}a_{t}^{s}\phi_{s},\qquad \phi_{s}\in\GamLamM{0}
\label{decomposition-Phi-a-phi}
\eeq
where 
\beqa
a^{s}:\mbbR&\to&\mbbR \qquad s=0,1,\ldots,
\non\\
t&\mapsto&a_{t}^{s}.\non
\eeqa
In this paper \fr{decomposition-Phi-a-phi} is called the
{\it eigenfunction expansion} of the Fokker-Planck equation or that of 
$\Phi_{t}$. 
As shown below, this expansion 
yields the dynamical system for $a_{t}^{s}$, and 
its time-scale is $\beta^{-1}\lambda_{s}$ for each $s$. 
\begin{Proposition}
\label{fact:dynamics-a}
For each $s$, $a_{t}^{s}$ satisfies the ODE 
$$
\frac{\dr }{\dr t}a_{t}^{s}
=- \beta^{-1}\lambda_{s}a_{t}^{s},
$$
whose solution is given by
$$
a_{t}^{s}
=a_{0}^{s}\exp(- \beta^{-1}\lambda_{s}\,t).
$$
\end{Proposition}
\begin{Proof}
  To derive the equation of $a_{t}^{s}$, take the global 
  inner product of 
\fr{diffusion-from-Fokker-Planck} and $\phi_{s}$ with a fixed $s$:    
$$
\inp{\phi_{s}}{\dot{\Phi}_{t}}_{\G}
=\inp{\phi_{s}}{-\beta^{-1}\triangle_{\G}\Phi_{t}}_{\G}.
$$
The left and right hand sides of this are   
\beqa
\inp{\phi_{s}}{\frac{\partial\Phi_{t}}{\partial t}}_{\G}
&=&\sum_{s^{\prime}}\frac{\dr a_{t}^{s^{\prime}}}{\dr t}\inp{\phi_{s}}{\phi_{s^{\prime}}}_{\G}
=\frac{\dr a_{t}^{s}}{\dr t},
\non\\
\inp{\phi_{s}}{-\beta^{-1}\triangle_{\G}\Phi_{t}}_{\G}
&=&-\beta^{-1}
\sum_{s^{\prime}}\lambda_{s^{\prime}}a_{t}^{s^{\prime}}\inp{\phi_{s}}{\phi_{s^{\prime}}}_{\G}
=-\beta^{-1}\lambda_{s}a_{t}^{s},
\non
\eeqa
respectively. Since they are equal, one has that 
$$
\frac{\dr }{\dr t}a_{t}^{s}
=- \beta^{-1}\lambda_{s}a_{t}^{s},
\qquad s=0,1,\ldots  
$$
whose solution is immediately obtained. This yields the desired expression
of the solution.
\qed
\end{Proof}
\begin{Remark}
The $0$th mode is interpreted as the equilibrium state. To see this,  
consider the special case where there is only the $0$th mode,
that is, $\Phi_{t}=a_{t}^{0}\phi_{0}$ with 
$\phi_{0}=1$.
Then, it follows from Proposition\,\ref{fact:dynamics-a} that
$\Phi_{t}=a_{0}^{0}$, where $a_{0}^{0}$ is constant.
From \fr{rho-rho-G-Phi}, it follows that 
$\rho_{t}=\rho_{\G}\Phi_{t}=a_{0}^{0}\rho_{\G}$. This and   
normalization conditions, \fr{rho-normalization} and  
\fr{rho-G-normalization}, 
yield $a_{0}^{0}=1$.  
Hence this special case consisting only of the $0$th mode yields 
$\rho_{t}=\rho_{\G}$, and the $0$th mode 
expresses the equilibrium distribution
function $\rho_{\G}$.
\end{Remark}
\begin{Remark}
\label{remark:Phi-all-time-scales}
It follows from Proposition\,\ref{fact:dynamics-a} with  
\fr{decomposition-Phi-a-phi} that $\Phi_{t}$ has the
   countably infinite   
time-scales $\{\beta^{-1}\lambda_{s}\}$:
$$
\Phi_{t}
=\sum_{s=0}^{\infty}a_{0}^{s}
\exp\left(-\beta^{-1}\lambda_{s}t\right)\phi_{s}.
$$
In addition,
since $\rho_{t}=\rho_{\G}\Phi_{t}$ in \fr{rho-rho-G-Phi},
$\rho_{t}$ also has the countably infinite time-scales. 
\end{Remark}

There are several applications of the 
eigenfunction expansion  
of the Fokker-Planck equation.
In what follows one of them is considered. 
Recall that the Fokker-Planck equation describes the time-evolution of the
probability distribution function, and hence one can define
expectation value (or expected value)  
with respect to the solution to the Fokker-Planck equation by
integrating a function over possible states with a weight. 
In this paper these expectation 
values are assumed to be identified with 
nonequilibrium thermodynamic variables. 

To write the expectation 
values in terms of a geometric language,  
let $B_{1}$ be a function on $\cM$.
The expectation value of $B_{1}$ 
at $t$ is expressed in terms of $\mbbE_{t}$, where 
\beqa
\mbbE_{t}:\GamLamM{0}&\to&\mbbR
\non\\
B_{1}&\mapsto&\mbbE_{t}[B_{1}]
\non
\eeqa
with    
\beq
\mbbE_{t}[B_{1}]
:=\int_{\cM}B_{1}\rho_{t}\star 1.
\label{expected-value-B}
\eeq
This time-dependent quantity can be written in terms of the
global inner product:
$$
\mbbE_{t}[B_{1}]
=\int_{\cM}B_{1}\rho_{t}\star 1
=\int_{\cM}B_{1}\Phi_{t}\rho_{\G}\star1
=\inp{B_{1}}{\Phi_{t}}_{\G}.
$$
If $B_{1}$ does not depend on $t$, then the following holds:
$$
\frac{\dr}{\dr t}\mbbE_{t}[B_{1}]
=\int_{\cM}B_{1}\rho_{\G}\dot{\Phi}_{t}\star 1
=-\beta^{-1}\inp{B_{1}}{\triangle_{\G}\Phi_{t}}_{\G}
=-\beta^{-1}\inp{\triangle_{\G}B_{1}}{\Phi_{t}}_{\G}.
$$
\begin{Remark}
  When $B_{1}=1$, the
  conservation property   
  for $\rho_{t}$ 
is reconstructed in the sense that 
$$
\frac{\dr}{\dr t}\int_{\cM}\rho_{t}\star1
=\frac{\dr}{\dr t}\mbbE_{t}[1]
=-\beta^{-1}\inp{\triangle_{\G}1}{\Phi_{t}}_{\G}
=0,
$$
due to $\triangle_{\G}1=\dr_{\G}^{\dagger}\dr 1=\dr_{\G}^{\dagger}0=0$.
\end{Remark}

There are several formulae regarding the expansion 
of $B_{1}$ in terms of eigenfunctions. 
To show these, expand $B_{1}\in\GamLamM{0}$ in terms of $\{\phi_{s}\}$:
$$
B_{1}=\sum_{s^{\prime}=0}^{\infty}
b_{s^{\prime}}\phi_{s^{\prime}},
\quad\text{where}\quad
b_{s}=\inp{\phi_{s}}{B_{1}}_{\G}\, \in\mbbR,\qquad s=0,1,\ldots 
$$
This and 
$$
\rho_{t}
=\rho_{\G}\Phi_{t}
=\rho_{\G}\sum_{s=0}^{\infty}a_{t}^{s}\phi_{s},
$$
yield 
\beqa
\mbbE_{t}[B_{1}]
&=&\int_{\cM}B_{1}\rho_{t}\star 1
=\sum_{s^{\prime}}b_{s^{\prime}}\int_{\cM}\phi_{s^{\prime}}\sum_{s}a_{t}^{s}\phi_{s}
\rho_{\G}\star1
\non\\
&=&
\sum_{s}\sum_{s^{\prime}}b_{s^{\prime}}a_{t}^{s}
\inp{\phi_{s^{\prime}}}{\phi_{s}}_{\G}
=\sum_{s}b_{s}a_{t}^{s}.
\non
\eeqa
 If $\dot{b}_{s}=0$ for all $s$, then it follows that 
$$
\frac{\dr}{\dr t}\mbbE_{t}[B_{1}]
=-\beta^{-1}\inp{\triangle_{\G}B_{1}}{\Phi_{t}}_{\G}
=-\beta^{-1}\sum_{s}\lambda_{s}b_{s}a_{t}^{s}.
$$
In addition, one has
$$
\mbbE_{t}[\phi_{s}]
=\int_{\cM}\phi_{s}\rho_{t}\star1
=\sum_{s^{\prime}}\inp{\phi_{s^{\prime}}}{\phi_{s}}_{\G}a_{t}^{s^{\prime}}
=a_{t}^{s},\quad\text{and}\quad
\frac{\dr}{\dr t}\mbbE_{t}[\phi_{s}]
=-\beta^{-1}\lambda_{s}a_{t}^{s}
=-\beta^{-1}\lambda_{s}\,\mbbE_{t}[\phi_{s}].
$$
The first equation above and \fr{decomposition-Phi-a-phi} yield 
$$
\Phi_{t}
=\sum_{s}(\mbbE_{t}[\phi_{s}])\phi_{s}.
$$

Before closing this section,
it should be mentioned that, in the literature, 
there are several operators that are analogous to
$\dr_{\G}^{\dagger}\dr$\,\cite{Helffer2005}.   
To compare the approach with $\dr_{\G}^{\dagger}\dr$  in this paper
with the existing approaches, several calculations are shown in
appendix of this paper
(see Section\,\ref{section:appendix-various-Laplacian}). 
In addition in the case where
the standard exterior derivative and co-derivative $\dr$ and $\dr^{\dagger}$
are only employed, one can also have a PDE for
$\Phi_{t}=\rho_{\G}^{-1}\rho_{t}$, as shown in
Section\,\ref{section:appendix-Fokker-Planck-d-and-d-dagger}.

\section{From the diffusion equation to a contact Hamiltonian system}
\label{section:diffusion-to-contact}
In Section\,\ref{section:Fokker-Planck-to-Diffusion}, a diffusion equation
associated with a weighted Laplacian has been derived. In this section,
this diffusion equation is shown to induce a contact Hamiltonian system.
In contact geometry, contact Hamiltonian systems are defined on contact
manifolds and employed
to describe thermodynamics, where contact geometry is an odd-dimensional
analogue of symplectic geometry\,\cite{Silva2008,Mcduff2016}. 
After recalling contact geometry briefly
and fix our notation, it is shown how the diffusion equation induces 
a contact Hamiltonian system.

\subsection{Contact geometry}
In this subsection necessary background of contact geometry is summarized.

A contact manifold is a pair $(\cN,\ker\alpha)$, where
$\cN$ is an odd-dimensional manifold and
$\ker\alpha=\{X\in \GT{\cN}|\alpha(X)=0\}$ with $\alpha$ being a contact form. 
The interpretation of $\cN$ in this section is thermodynamic phase space,
whose coordinates represent 
thermodynamic variables. 
Let $\dim\cN=2n+1$, and denote by $\bdr$ the exterior derivative,
$\bdr:\GamLam{\cN}{k}\to\GamLam{\cN}{k+1}$, ($k=0,1,\ldots,2n+1$).  
A contact form $\alpha$ on $\cN$ is a $1$-form that the form 
$\alpha\wedge\bdr \alpha\wedge\cdots\wedge\bdr\alpha\in\GamLam{\cN}{2n+1}$
does not vanish anywhere.  
There are canonical coordinates $(p,q,z)$
with $p=(p_{1},\ldots,p_{n})$ and
$q=(q^{1},\ldots,q^{n})$ such that the contact form is written as 
$$
\alpha=\bdr z-\sum_{j=1}^{n} p_{j}\bdr q^{j}.
$$
As an example, the pair consisting of $T^{*}\mbbR^{n}\times \mbbR$ and 
$\ker\alpha$ is a contact manifold, where $T^{*}\mbbR^{n}$ is the
cotangent bundle of $\mbbR^{n}$, $q$ denotes a point of $\mbbR^{n}$,
$p$ coordinates of $T_{q}^{*}\mbbR^{n}$, $z$ coordinate of $\mbbR$, and
$\alpha$ is given above. 
Given a function $\cH$ on $\cN$, a vector field $X_{\cH}$ satisfying 
$$
\ii_{X_{\cH}}\bdr\alpha
=-\bdr \cH 
+(R\cH)\alpha,\quad\text{and}\quad
\ii_{X_{\cH}}\alpha
=\cH,  
$$
is called a {\it contact Hamiltonian vector field}, where $\cH$ is called
a {\it contact Hamiltonian}, 
$R$ is called the {\it Reeb vector field} that is the unique vector field
satisfying
$\ii_{R}\alpha=1$ and $\ii_{R}\bdr \alpha=0$.  
In the canonical coordinates $R$ and $X_{\cH}$ are written as
$$
R=\frac{\partial}{\partial z},\qquad
X_{\cH}=\dot{z}\frac{\partial}{\partial z}
+\sum_{j=1}^{n}\left(\dot{p}_{j}\frac{\partial}{\partial p_{j}}
+\dot{q}^{j}\frac{\partial}{\partial q^{j}}\right),
$$
where
\beq
\dot{q}^{j}
=-\frac{\partial \cH}{\partial p_{j}},\quad
\dot{p}_{j}
=\frac{\partial \cH}{\partial q^{j}}+p_{j}\frac{\partial \cH}{\partial z},\quad
\dot{z}
=\cH-\sum_{j=1}^{n}p_{j}\frac{\partial \cH}{\partial p_{j}},\qquad
j=1,\ldots,n.
\label{contact-Hamiltonian-coordinates}
\eeq
\begin{Remark}
\label{remark:time-scale-general-Hamiltonian}
To relate \fr{contact-Hamiltonian-coordinates} with a dynamical system, 
$\dot{\ }$ in the left hand side  
is identified with $\dr/\dr t$. 
If $\cH$ is written as $\cH=\gamma_{1}\cH^{(0)}$ 
with some function $\cH^{(0)}$ and constant $\gamma_{1}>0$, 
then $\gamma_{1}$ gives a natural 
time-scale. To see this, let $t_{1}:=\gamma_{1}t$. Then 
$t_{1}$ can be recognized as 
a scaled time with a time scale $\gamma_{1}$.  
If introducing $t_{1}$, then 
$\gamma_{1}$ does not explicitly appear in the dynamical system   
\fr{contact-Hamiltonian-coordinates}. 
\end{Remark}
In the applications of contact geometry to thermodynamics,
$\alpha$ is employed to express 
the laws of thermodynamics, and then 
$X_{\cH}$ preserves $\ker\alpha$.  
Equilibrium thermodynamic phase space is modeled by
a class of Legendrian submanifolds\,\cite{Mrugala2000chapter,Bravetti2019}.
The Legendrian submanifold $\cL$ is a submanifold, where 
$\dim\cL=n$ and the pullback of $\alpha$ to $\cL$ vanishes.
One example of $\cL$ is expressed with a function $\psi$ of $q$ in
the canonical coordinates as 
$$
\cL_{\psi}
:=\bigcup_{q\in\mbbR^{n}}\cL_{\psi}(q)\ 
\subset T^{*}\mbbR^{n}\times\mbbR,\quad\text{with}\quad
\cL_{\psi}(q)
:=\left\{\ (p,q,z)\ \bigg|\ p_{j}=\frac{\partial\psi}{\partial q^{j}}(q),
\quad
z=\psi(q),\quad j=1,\ldots,n \ \right\}.
$$
There are several approaches to express
nonequilibrium thermodynamic processes by means of contact
geometry,\,\cite{Goto2015,Entov2023,GLP}.

  In Section\,\ref{section:derivation-contact-Hamiltonian} 
  the class of contact Hamiltonians of the form 
  \beq
  \cH(q,z)=c(\psi(q)-z),
  \label{simple-contact-Hamiltonian}
  \eeq
  will be derived, where $c>0$ is constant and $\psi$ a function.
  It follows from \fr{contact-Hamiltonian-coordinates} with
  \fr{simple-contact-Hamiltonian} that 
  $\dot{q}^{j}=0$, ($j=1,\ldots,n$). The static property $\dot{q}^{j}=0$,
  ($j=1,\ldots,n$) 
  is appropriate when $q$ is treated as a set of static thermodynamic
  variables (see Section\,\ref{section:derivation-contact-Hamiltonian}).  
  The contact Hamiltonian \fr{simple-contact-Hamiltonian} has been 
  focused in \cite{Goto2015} and its related one in \cite{Entov2023}.
  In \cite{Goto2015},  
  integral curves of $X_{\cH}$ 
  induced from \fr{simple-contact-Hamiltonian}  
  are interpreted as relaxation processes, because
  the long-time limit of a point of an integral curve of $X_{\cH}$
  is on $\cL_{\psi}$,
  where $\cL_{\psi}$ is interpreted as equilibrium thermodynamic phase space.
  This interpretation still holds in this paper.
  In \cite{Entov2023}, several generalizations of
  \fr{simple-contact-Hamiltonian}  and related contact topology
  have been studied.

\subsection{Derivation of a contact Hamiltonian system}
\label{section:derivation-contact-Hamiltonian}
To derive a contact Hamiltonian system from the Wasserstein gradient 
flow or equivalently the Fokker-Planck equation, 
thermodynamic variables are needed to be defined,   
since contact Hamiltonian equations are equations for thermodynamic variables. 
In general, thermodynamic variables form pairs, and they consist of 
intensive and extensive parameters\,\cite{Callen}. 
In this paper they are called {\it primal and conjugate (or dual) variables}, 
where primal variables 
often express externally applied fields, and the conjugate or dual variables 
express expectation values. 
An example of this pair is an applied magnetic field 
and the magnetization for an Ising spin system, where the magnetization is 
obtained by integrating a function over a state space 
with the weight of a distribution function. 
As is known in equilibrium thermodynamics, given primal variables, 
thermodynamic conjugate variables 
are obtained by differentiation of a moment generating function 
with respect to the primal variables  
(see \cite{Kubo1962,Salazar2023} 
for applications of moment generating functions in statistical mechanics).  
In this paper, this known approach for equilibrium systems is 
extended to that for nonequilibrium systems. 

In this subsection, after introducing time-dependent thermodynamic variables 
and a time-dependent moment generating function,  
it is shown how a contact Hamiltonian equation is derived 
from the Fokker-Planck equation. 
The principal step in this derivation, 
from a PDE to a system of ODEs, 
is to integrate functions over $\cM$ with a weight. 
In calculating various time-dependent quantities,   
the Fokker-Planck equation \fr{Fokker-Planck-0}
is identified with the diffusion equation
\fr{diffusion-from-Fokker-Planck} 
via the change of variables \fr{rho-rho-G-Phi}.

To describe a set of primal variables $q$ and its conjugate ones, let 
$$
q=(q^{1},q^{2},\ldots,q^{n})\ \in\mbbR^{n},\qquad
B=(B_{1},B_{2},\ldots,B_{n})\quad \text{with}\quad B_{j}\in\GamLamM{0},\quad
j=1,\ldots,n
$$
and
$$
q\cdot B 
:=\sum_{j=1}^{n}q^{j}B_{j}\ \in\GamLamM{0}.
$$
The {\it time-dependent moment generating function},  
\beqa
M:\mbbR\times\mbbR^{n}&\to&\mbbR,
\non\\
(t,q)&\mapsto& M_{t}(q), 
\non
\eeqa
is defined as 
\beq
M_{t}(q)
:=\mbbE_{t}[\exp(q\cdot B)]
=\int_{\cM}\rho_{\G}\Phi_{t}\exp(q\cdot B)\star 1
=\inp{\Phi_{t}}{\exp(q\cdot B)}_{\G}.
\label{M-t-q}
\eeq
This generating function is employed to obtain 
the expectation values of $B_{1},\ldots,B_{n}$.  
By assuming the commutability between $\partial/\partial q^{j}$ and
the integral over $\cM$, one has from \fr{M-t-q} that 
\beq
\frac{\partial}{\partial q^{j}}M_{t}(q)
=\mbbE_{t}[B_{j}\exp(q\cdot B)]
=\int_{\cM}\rho_{\G}\Phi_{t}B_{j}\exp(q\cdot B)\star 1
=\inp{\Phi_{t}}{B_{j}\exp(q\cdot B)}_{\G},
\label{M-t-q-diff}
\eeq
where $\Phi_{t}$ has been given in \fr{rho-rho-G-Phi} and this $\Phi_{t}$
is independent of $q\cdot B$ but depends on $h$. 
An interpretation of $q^{j}$ and
$$
\mbbE_{t}[B_{j}\exp(q\cdot B)]\quad
\text{or}\quad
\mbbE_{t}[B_{j}]
$$
for each $j$ 
is a pair of thermodynamic primal and time-dependent conjugate variables.

Since time-evolution is governed by the Fokker-Planck equation,
asymptotic limits of \fr{M-t-q} and \fr{M-t-q-diff} are
obtained with Corollary\,\ref{fact:limit-Phi} as 
\beqa
\lim_{t\to\infty}M_{t}(q)
&=&\inp{\exp(q\cdot B)}{1}_{\G}
=\psi_{\G}(q),
\non\\
\lim_{t\to\infty}\frac{\partial}{\partial q^{j}}M_{t}(q)
&=&\inp{B_{j}\exp(q\cdot B)}{1}_{\G}
=\frac{\partial \psi_{\G}}{\partial q^{j}}(q),\qquad j=1,\ldots,n
\non
\eeqa
where $\psi_{\G}:\mbbR^{n}\to\mbbR$ has been introduced as   
\beq
\psi_{\G}(q)
:=\inp{\exp(q\cdot B)}{1}_{\G}
=\int_{\cM}\exp(q\cdot B)\rho_{\G}\star 1.
\label{psi-G}
\eeq
In addition, the asymptotic limit of $M_{t}(q)$ is expressed as 
\beqa
\lim_{t\to\infty}M_{t}(q)
&=&\frac{1}{Z_{\G}}\int_{\cM}\e^{-\beta\wt{h}}\star 1,
\qquad
\wt{h}:=h-\beta^{-1}q\cdot B, 
\non\\
&=&\frac{Z_{\G}(q)}{Z_{\G}(0)},\qquad
Z_{\G}(q)
:=\int_{\cM}\e^{-\beta h+q\cdot B}\star 1
\non
\eeqa
where an interpretation of $\wt{h}$ 
is the sum of a 
Hamiltonian and an energy due to an
externally applied field $-\beta^{-1}q\cdot B$, where 
$\wt{h}=-\beta^{-1}\ln(\rho_{\G}\exp(q\cdot B))+\const$.  
To see why this interpretation  is valid with $n=1$,
consider a system consisting of interacting Ising spins as an example.
Then $h$ describes spin-spin interactions,   
$\beta^{-1}q$ a constant expressing an externally applied magnetic field,  
and $B$ denotes magnetization, the mean of Ising spins. Analogously,
$-\beta^{-1}\ln(\rho_{t}\exp(q\cdot B))$, the log 
of the integrand of \fr{M-t-q} at $t$, is possibly   
interpreted as the sum of a 
Hamiltonian and an energy due to an 
externally applied field up to an additional constant. 

\begin{Remark}
The function $\psi_{\G}$ is convex, since for any $q\in\mbbR^{n}$
the Hessian matrix
$(\partial^{2}\psi_{\G}/\partial q^{j}\partial q^{k})$
is positive semi-definite: 
$$
\sum_{j=1}^{n}\sum_{k=1}^{n}q^{j}
\frac{\partial^{2}\psi_{\G}}{\partial q^{j}\partial q^{k}}q^{k}
=\sum_{j=1}^{n}\sum_{k=1}^{n}q^{j}q^{k}\int_{\cM}B_{j}B_{k}\e^{q\cdot B}\rho_{\G}\star 1
=\int_{\cM}(q\cdot B)^{2}\e^{q\cdot B}\rho_{\G}\star 1
\geq 0.
$$
\end{Remark}

Under these preliminary definitions, attention can now be focused on  
a contact geometric description of a system, where this system  
is derived from the Fokker-Planck equation.
In particular, 
  a class of contact Hamiltonian systems is derived below,
  where this class is such that the set of primal variables $q$ is constant. 
  This static property of $q$ is consistent when considering the case that
  the primal variables  are static.    
  In the case that $q\in\mbbR^{n}$ is constant in time, 
  such a Hamiltonian does not depend on $p$, that is, $\cH=\cH(q,z)$. 
  One known contact Hamiltonian is given by 
  \fr{simple-contact-Hamiltonian}. Below, this form of $\cH$ is 
  derived. A strategy for deriving this is given as follows.
\begin{itemize}
\item
  First, 
  identify a contact manifold and a Legendrian submanifold. 
\item
  Second, 
  derive a set of equations $\dot{q}^{j}=0$, $\dot{p}_{j}=\cdots$, and
  $\dot{z}=\cdots$, $j=1,\ldots,n$.
  This set of ODEs is derived by integrating 
  some functions with 
  a solution to a given Fokker-Planck equation. 
\item
  Third, find a way in which
  the right hand sides of the derived equations above 
  are written in terms of $(p,q,z)$. 
  In other words,
  the system should be closed in $(p,q,z)$, otherwise the system is not
  on a contact manifold. 
\item
  Finally, find a contact Hamiltonian $\cH$
  by comparing the obtained closed equations with 
  \fr{contact-Hamiltonian-coordinates}. 
\end{itemize}

Discussions below are divided into two cases. One is
full-time scale dynamics, and the other is
slowest-time scale dynamics. In the latter,
a contact Hamiltonian system is derived. 

\subsubsection*{Full times-scale dynamics: a non-closed system}

In the following several variables are introduced so that
a contact geometric description of the Fokker-Planck equation is established.
Meanwhile the derivation of a contact Hamiltonian is not given here.

Introduce $z(t;q),p_{j}(t;q)$ with $q=(q^{1},q^{2},\ldots,q^{n})$ as 
\beqa
z(t;q)
&:=&M_{t}(q)
=\inp{\Phi_{t}}{\exp(q\cdot B)}_{\G},
\non\\
p_{j}(t;q)
&:=&\frac{\partial}{\partial q^{j}}M_{t}(q)
=\inp{\Phi_{t}}{B_{j}\exp(q\cdot B)}_{\G},\qquad j=1,\ldots,n.
\non
\eeqa
These variables $z$ and $p$ contain all the time-scales
$\{\beta^{-1}\lambda_{s}\}$ as
discussed in Remark\,\ref{remark:Phi-all-time-scales},
therefore  
its corresponding dynamics is called {\it full time-scale dynamics}
in this paper. As is shown below, this dynamics on a contact manifold 
is not written as a simple
contact Hamiltonian system. 
Meanwhile making an approximation discussed later 
allows to write it as a simple contact Hamiltonian system,  
where this approximation scheme is viewed as a projection of a function onto 
a subspace spanned by a set of lower eigenfunctions. 

Let $(p,q,z)$ be coordinates of $T^{*}\mbbR^{n}\times\mbbR$, and 
let $\alpha$ be the $1$-form on $T^{*}\mbbR^{n}\times\mbbR$ as 
$$
\alpha
=\bdr z-\sum_{j=1}^{n}p_{j}\bdr q^{j}.
$$
Then $(T^{*}\mbbR^{n}\times\mbbR,\ker\alpha)$ is a contact manifold.
Points on the Legendrian submanifold expressed in coordinates,  
\beq
\cL_{\psi_{\G}}
:=\bigcup_{q\in\mbbR^{n}}\cL_{\psi_{\G}}(q),\quad      
\cL_{\psi_{\G}}(q)
:=\left\{\ (p,q,z)\ \bigg|\ z=\psi_{\G}(q),\quad
p_{j}=\frac{\partial\psi_{\G}}{\partial q^{j}}(q),\ j=1,\ldots,n
\ \right\}, 
\label{Legendrian-psi-G}
\eeq
are interpreted as points of  
the equilibrium state. 
Consider dynamical behavior of $z(t;q)$ on 
the $z$-axis, and that of $p(t;q)$ on the $p$-axis.   
A natural question is whether or not the long-time limit of the point 
$(p(t;q),q,z(t;q))$ is on the Legendrian submanifold. The answer is as follows. 
\begin{Proposition}
\label{fact:exact-Fokker-Planck-on-Legendrian}
The long-time limit of the point $(p(t;q),q,z(t;q))$ is
on the Legendrian submanifold $\cL_{\psi_{\G}}(q)$,  
where dynamics of $z(t;q)$ and that of $p(t;q)$ obey 
the diffusion equation
\fr{diffusion-from-Fokker-Planck} associated with 
the Fokker-Planck  equation.
\end{Proposition}
\begin{Proof}
  The long-time limits of $z$ and $p$ are obtained by using $\Phi_{t}\to1$,
($t\to\infty$) shown in Corollary\,\ref{fact:limit-Phi} as 
\beqa
\lim_{t\to\infty}z(t;q)
&=&\inp{\exp(q\cdot B)}{1}_{\G}
=\psi_{\G}(q),
\non\\
\lim_{t\to\infty}p_{j}(t;q)
&=&\inp{B_{j}\exp(q\cdot B)}{1}_{\G}
=\frac{\partial }{\partial q^{j}}\inp{\exp(q\cdot B)}{1}_{\G}
=\frac{\partial\psi_{\G}}{\partial q^{j}}(q),\qquad j=1,\ldots,n.
\non
\eeqa
Hence the point in the limit $t\to\infty$ is on the Legendrian submanifold 
$\cL_{\psi_{\G}}(q)$.
\qed
\end{Proof}
\begin{Remark}
  In many cases, $q=0$ is substituted after evaluating $p_{j}$, and
  it is useful to expand $z$ and $p$ in terms of $q$. They are calculated to be
\beqa
\lim_{t\to\infty}z(t;q)
&=&\psi_{\G}(q)
=1+\sum_{j=1}^{n}q^{j}\inp{B_{j}}{1}_{\G}
+\cO(q^{2}),
\non\\
\lim_{t\to\infty}p_{j}(t;q)
&=&
\frac{\partial\psi_{\G}}{\partial q^{j}}
=\frac{\partial }{\partial q^{j}}\inp{1+(q\cdot B)+\frac{(q\cdot B)^{2}}{2!}+\cO(q^{3})}{1}_{\G}
\non\\
&=&\inp{B_{j}}{1}_{\G}+\sum_{k=1}^{n}q^{k}\inp{B_{j}}{B_{k}}_{\G}
+\cO(q^{2}).
\non
\eeqa
\end{Remark}

After clarifying the long-time limit
of $z(t)$ and that of $p(t)$,   
the next point of discussion is dynamical behavior of $z$ and $p$
governed by the 
Fokker-Planck equation. 
First, observe that 
\beqa
\frac{\dr}{\dr t}z
&=&\inp{\dot{\Phi}_{t}}{\exp(q\cdot B)}_{\G}
=-\beta^{-1}\inp{\triangle_{\G}\Phi_{t}}{\exp(q\cdot B)}_{\G}
=-\beta^{-1}\inp{\Phi_{t}}{\triangle_{\G}\exp(q\cdot B)}_{\G},
\non\\
\frac{\dr}{\dr t}p_{j}
&=&\inp{\dot{\Phi}_{t}}{B^{j}\exp(q\cdot B)}_{\G}
=-\beta^{-1}\inp{\Phi_{t}}{\triangle_{\G}(B^{j}\exp(q\cdot B))}_{\G},
\quad j=1,\ldots,n.
\non
\eeqa
In general, the right hand sides of the equations above are not
written in terms of $z$ and $p$. In this sense, they are not closed. 

\subsubsection*{Slowest times-scale dynamics: a contact Hamiltonian system}
To obtain a closed dynamical system on $T_{q}^{*}\mbbR^{n}\times\mbbR$, 
some approximation is introduced below.   
This approximation is to neglect the fast-time scales, and leads to  
a contact Hamiltonian system. 
More precisely, 
this approximation scheme is based on a projection of $\Phi_{t}$ 
onto  the subspace spanned by $\phi_{0}$ and $\phi_{1}$. This 
projection of $\Phi_{t}$ then induces  
the corresponding projections of $z$ and $p$.  

Before
addressing this approximation, recall that the $0$th mode exists as shown in 
Proposition\,\ref{fact:eigen-0-function}. 
Then the next point of discussion is the $1$st mode. 
The following two cases for the $1$st mode are considered separately.
However the conclusion on one side is equal to a conclusion on the other.   
\begin{itemize}
\item
Case I:
There is no degeneracy: 
\beq
0=\lambda_{0}<\lambda_{1}<\lambda_{2}<\cdots.
\label{spectrum-assumptions-no-degeneracy}
\eeq

In the case of nodegeneracy, consider the following formal expansion:
\beqa
\Phi_{t}
&=&\sum_{s=0}^{\infty}a_{t}^{s}\phi_{s}
=\sum_{s=0}^{\infty}a_{0}^{s}\exp(-\beta^{-1}\lambda_{s}t) \phi_{s}
\non\\
&=&a_{t}^{0}\phi_{0}+a_{t}^{1}\phi_{1}+\cdots
=a_{0}^{0}\phi_{0}\e^{-\beta^{-1}\lambda_{0}t}
+a_{0}^{1}\phi_{1}\e^{-\beta^{-1}\lambda_{1}t}
+\cdots
\non\\
&=&\Phi_{\infty}+a_{0}^{1}\phi_{1}\e^{-\beta^{-1}\lambda_{1}t}
+\cdots, 
\label{Phi-expanded}
\eeqa 
where $\Phi_{\infty}:=\lim_{t\to\infty}\Phi_{t}$ has been defined. 
Note that
the terms 
$a_{0}^{s}\phi_{s}$, $(s=1,2,\ldots)$ are written from \fr{Phi-expanded} as  
$$
a_{0}^{1}\phi_{1}
=\lim_{t\to\infty}\e^{\beta^{-1}\lambda_{1}t}
\left(\Phi_{t}-\Phi_{\infty}\right),\quad
a_{0}^{2}\phi_{2}
=\lim_{t\to\infty}\e^{\beta^{-1}\lambda_{2}t}
\left(\Phi_{t}-\Phi_{\infty}-a_{0}^{1}\phi_{1}\e^{-\beta^{-1}\lambda_{1}t}\right),\quad\cdots.
$$
Then, differentiating $\Phi_{t}$ in \fr{Phi-expanded} with respect to $t$ and 
recalling $\Phi_{\infty}=1$ from Corollary\,\ref{fact:limit-Phi}, 
one has
\beqa
\dot{\Phi}_{t}
&=&-\beta^{-1}\lambda_{1}a_{0}^{1}\phi_{1}\e^{-\beta^{-1}\lambda_{1}t}+\cdots
\non\\
&=&-\beta^{-1}\lambda_{1}(\Phi_{t}-1)+\cdots.
\non
\eeqa
Let us define the equation associated with the approximation by this neglect: 
\begin{Definition}
Dynamics generated by
\beq
\ol{\Phi}_{t}(\xi)
=1+a_{t}^{1}\phi_{1}(\xi),\quad\forall \xi\in\cM,\ \forall t\in\mbbR
\label{slowest-mode}
\eeq
that satisfies the PDE  
\beq
\frac{\partial}{\partial t}\ol{\Phi}_{t}
=-\beta^{-1} \lambda_{1}(\ol{\Phi}_{t}-1)
\label{slowest-mode-ODE}
\eeq
is called the {\it slowest time-scale dynamics}, its solution   
$\ol{\Phi}_{t}$ is called the {\it slowest time-scale mode}, and 
the equation the {\it slowest time-scale mode equation},  
in this paper. 
\end{Definition}
Notice that \fr{slowest-mode} can be viewed as a projection of $\Phi_{t}$ 
onto the subspace spanned by $\phi_{0}$ and $\phi_{1}$, and can also be viewed 
as a truncation of the eigenvalue function expansion. 
\item
Case II:  There is degeneracy of the form, 
$$
0=\lambda_{0}
<\lambda_{1}
=\lambda_{2}
=\cdots
=\lambda_{\nu}
<\lambda_{\nu+1}
<\cdots.
$$
Analogous to \fr{slowest-mode},  let 
$$
\ol{\Phi}_{t}^{\text{dgt}}
=1+(a_{0}^{1}\phi_{1}+\cdots +a_{0}^{\nu}\phi_{\nu})
\exp(-\beta^{-1}\lambda_{1}t).
$$
Then this $\ol{\Phi}_{t}^{\text{dgt}}$ satisfies
\beq
\frac{\partial}{\partial t}\ol{\Phi}_{t}^{\text{dgt}}
=-\beta^{-1}\lambda_{1}\left(\ol{\Phi}_{t}^{\text{dgt}}-1\right).
\label{slowest-mode-ODE-degenerate}
\eeq
\end{itemize}

Since \fr{slowest-mode-ODE-degenerate} in Case II can be 
identical to \fr{slowest-mode-ODE} in Case I,  
these two cases are not distinguished below. 

The following is a remarkable property of \fr{slowest-mode}.
\begin{Proposition}
\label{fact:slowest-mode-diffusion-equation}
The slowest mode \fr{slowest-mode} satisfies the
modified diffusion equation \fr{diffusion-from-Fokker-Planck}:
  \beq
  \frac{\partial}{\partial t}\overline{\Phi}_{t}
  =-\beta^{-1}\triangle_{\G}\overline{\Phi}_{t}.
\label{slowest-mode-diffusion-equation}  
  \eeq
\end{Proposition}
\begin{Proof}
Differentiation of \fr{slowest-mode} yields \fr{slowest-mode-ODE} 
and 
$$
\triangle_{\G}\ol{\Phi}_{t}
=a_{t}^{1}\triangle_{\G}\phi_{1}
=\lambda_{1}a_{t}^{1}\phi_{1}
=\lambda_{1}\left(\ol{\Phi}_{t}-1\right).
$$
Combining these equations, one has that
$$
\frac{\partial}{\partial t}\ol{\Phi}_{t}
=-\beta^{-1}\triangle_{\G}\ol{\Phi}_{t},
$$
which is \fr{slowest-mode-diffusion-equation}. 
\qed
\end{Proof}
Proposition\,\ref{fact:slowest-mode-diffusion-equation} shows that 
the approximation, or the projection, 
\fr{slowest-mode} is adequate in the sense that
full time-scale mode $\Phi_{t}$ and slowest time-scale mode $\ol{\Phi}_{t}$ 
obey the same PDE. Hence,
this adequate approximation preserves,
for example, Lie symmetries of
differential equations\,\cite{Olver1986}. 

It is straightforward to verify by 
\fr{slowest-mode} and Proposition\,\ref{fact:dynamics-a}  
 that
$$
\ol{\Phi}_{t}
=1+a_{0}^{1}\exp(-\beta^{-1}\lambda_{1}t)\phi_{1}.
$$
It follows from this and $\beta^{-1}\lambda_{1}>0$ that 
\beq
\lim_{t\to\infty}\ol{\Phi}_{t}
=1.
\label{limit-Phi-bar}
\eeq
To compare the asymptotic behavior of $\Phi_{t}$ and $\ol{\Phi}_{t}$,
the following can be employed:
$$
\left|\ \Phi_{t}(\xi)-\ol{\Phi}_{t}(\xi)\ \right|
=\left| \ a_{0}^{2}\phi_{2}(\xi)\e^{-\beta^{-1}\lambda_{2}t}+\cdots \ \right|,
\quad\forall\ \xi\in\cM.
$$
This yields
$$
\lim_{t\to\infty}\left|\ \Phi_{t}(\xi)-\ol{\Phi}_{t}(\xi)\ \right|
=0,\quad \forall\ \xi\in\cM.
$$
To see the approximation in the original variable, introduce 
\beq
\ol{\rho}_{t}
=\rho_{\G}\ol{\Phi}_{t}.
\label{rho-bar}
\eeq
Then this $\ol{\rho}_{t}$ satisfies the PDE
\beq
\frac{\partial}{\partial t}\ol{\rho}_{t}
=-\beta^{-1}\lambda_{1}\left(\ol{\rho}_{t}-\rho_{\G}\right),
\label{rho-bar-PDE}
\eeq
which is derived by 
differentiating both sides of \fr{rho-bar}
with respect to $t$. In addition, $\ol{\rho}_{t}$ is a relevant approximation
in the sense that the asymptotic limits coincide:  
$$
\lim_{t\to\infty}\left|\ \rho_{t}(\xi)-\ol{\rho}_{t}(\xi)\ \right|
=\rho_{\G}(\xi)
\lim_{t\to\infty}\left|\ \Phi_{t}(\xi)-\ol{\Phi}_{t}(\xi)\ \right|
=0,\quad\forall \xi\in\cM. 
$$
  \begin{Remark}
    The discussions so far in this subsection are briefly summarized here. 
    To express the projection of
    $\Phi_{t}$,
    introduce the projection 
    operator 
    \beqa
    \cP:\GamLamM{0}
    &\to&\GamLamM{0}
    \non\\
    \Phi_{t}^{\I}
    &\mapsto&\cP\Phi_{t}^{\I}
    \non
    \eeqa
    acting on an eigenfunction expanded function 
    $$
    \Phi_{t}^{\I}
    =1+\sum_{s=1}^{\infty}a_{t}^{s}\phi_{s}\ \in\GamLamM{0}, \qquad
    \forall t\in\mbbR
    $$
    with 
    $$
    \cP\Phi_{t}^{\I}
    :=1+a_{t}^{1}\phi_{1} \ \in\GamLamM{0}, \qquad
    \forall t\in\mbbR.
    $$
    Then it follows that the  diagrams for $t>0$ 
\beq
\xymatrix@R=25pt@C=43pt{
\rho_{\G}\ar@{|->}[r]^-{\rho_{\G}^{-1}\cdot}
&\Phi_{\infty}=1
\ar@{|->}[r]^-{\cP}
&\ol{\Phi}_{\infty}=1
\ar@{|->}[r]^{\rho_{\G}\cdot}
&\rho_{\G}
\\
\rho_{t}\ar[u]^-{t\to\infty}
\ar@{|->}[r]^-{\rho_{\G}^{-1}\cdot}
&\Phi_{t}\ar[u]^-{t\to\infty}
\ar@{|->}[r]^-{\cP}
&\ol{\Phi}_{t}
\ar@{->}[u]^-{t\to\infty}
\ar@{|->}[r]^{\rho_{\G}\cdot}
&\overline{\rho}_{t}\ar[u]^-{t\to\infty}
\\
\rho_{0}\ar@{~>}[u]^-{\fr{Fokker-Planck-0}}
\ar@{|->}[r]^-{\rho_{\G}^{-1}\cdot}
&\Phi_{0}\ar@{~>}[u]^-{\fr{diffusion-from-Fokker-Planck}}
\ar@{|->}[r]^-{\cP}
&\ol{\Phi}_{0}\ar@{~>}[u]^-{\fr{slowest-mode-diffusion-equation}}
\ar@{|->}[r]^{\rho_{\G}\cdot}
&\overline{\rho}_{0}
\ar@{~>}[u]^-{\fr{rho-bar-PDE}}
}
\label{diagrams-relations-variables}
\eeq
commute, where $A\equpg{\leadsto}{C}B$ denotes that $B$ is obtained  
from $A$ with C. 
\end{Remark}

When neglecting higher order (or fast time-scale) modes,
i.e., $\phi_{s}=0$ is substituted for 
$s\geq 2$, as shown below, a closed system
for analogues of $z$ and $p$ 
is obtained.
To obtain such a closed system of ODEs, introduce
the slowest time-scale moment generating function 
\beqa
\ol{M}:\mbbR\times\mbbR^{n}&\to&\mbbR
\non\\
(t,q)&\mapsto&\ol{M}_{t}(q),
\non
\eeqa
with
$$
\ol{M}_{t}(q)
:=\inp{\ol{\Phi}_{t}}{\exp(q\cdot B)}_{\G},
$$
and introduce the dynamical variables 
$\ol{z}$ and $\ol{p}$ with $\ol{p}=(\ol{p}_{1},\ldots,\ol{p}_{n})$ such that
\beqa
\ol{z}(t;q)
&:=&\ol{M}_{t}(q)
=\inp{\ol{\Phi}_{t}}{\exp(q\cdot B)}_{\G},
\label{z-bar}\\
\ol{p}_{j}(t;q)
&:=&\frac{\partial}{\partial q^{j}}\ol{M}_{t}(q)
=\inp{\ol{\Phi}_{t}}{B_{j}\exp(q\cdot B)}_{\G},\qquad
j=1,\ldots,n.
\label{p-bar}
\eeqa
Consider dynamical behavior of $\ol{z}(t;q)$ on the $z$-axis,
and that of  $\ol{p}(t;q)$ 
on the $p$-axis. One then finds
that there are a similarity between $z(t;q)$ and $\ol{z}(t;q)$, and 
a similarity between $p(t;q)$ and $\ol{p}(t;q)$. These are described as follows.
\begin{Proposition}
\label{fact:approximate-Fokker-Planck-on-Legendrian}
The long-time limit of the point $(\ol{p}(t;q),q,\ol{z}(t;q))$  
is on the Legendrian submanifold $\cL_{\psi_{\G}}(q)$ of the contact manifold, 
where 
dynamics of $\ol{z}(t;q)$ and that of $\ol{p}(t;q)$ obey 
the slowest time-scale mode equation\,\fr{slowest-mode-ODE}. 
\end{Proposition}
\begin{Proof}
A proof is similar to that of 
Proposition\,\ref{fact:exact-Fokker-Planck-on-Legendrian}. 
\qed
\end{Proof}

The main theorem in this paper is as follows. 
With the variables $\ol{z}$ and $\ol{p}$,
one has a contact Hamiltonian system.
\begin{Theorem}
\label{fact:Fokker-Planck-contact-Hamiltonian}
(Approximate Fokker-Planck equation and contact Hamiltonian system). 
The time-development of  $\ol{z}$, $\ol{p}$ and $q$ 
is identified with 
a contact Hamiltonian system on $T^{*}\mbbR^{n}\times\mbbR$, where 
dynamics of $\ol{z}$ and that of $\ol{p}$ are governed by
the slowest time-scale dynamics.
\end{Theorem}
\begin{Proof}
Substituting  the slowest time-scale mode \fr{slowest-mode-ODE}
into the time-derivative of
$\ol{z}$ and $\ol{p}$, one has the closed dynamical system: 
\beqa
\frac{\dr}{\dr t}\ol{z}
&=&-\beta^{-1}\lambda_{1}\inp{\ol{\Phi}_{t}-1}{\exp(q\cdot B)}_{\G}
=\beta^{-1}\lambda_{1}(\psi_{\G}(q)-\ol{z})
\non\\
\frac{\dr}{\dr t}\ol{p}_{j}
&=&-\beta^{-1}\lambda_{1}\inp{\ol{\Phi}_{t}-1}{B^{j}\exp(q\cdot B)}_{\G}
=\beta^{-1}\lambda_{1}\left(
\frac{\partial\psi_{\G}}{\partial q^{j}}-\ol{p}_{j}
\right),\quad j=1,2,\ldots,n
\non
\eeqa
To imitate this dynamics of $\ol{z}$ and that of $\ol{p}$   
by means of contact geometry, one  
introduces a dynamical system on a contact manifold. 
More specifically, a contact Hamiltonian system is introduced as follows.  
Choose $\cH$ as a contact Hamiltonian to be\,\cite{Goto2015,Entov2023}  
\beq
\cH(\ol{p},q,\ol{z})
=\gamma_{1}(\psi_{\G}(q)-\ol{z}), 
\label{contact-Hamiltonian-derived}
\eeq
with $\gamma_{1}\neq 0$ being some constant.
From the coordinate expression of the contact Hamiltonian vector
field\,\fr{contact-Hamiltonian-coordinates},   
the corresponding contact Hamiltonian system is expressed as 
$$
\dot{q}^{j}
=0,\quad
\dot{\ol{p}}_{j}
=\gamma_{1} 
\left(\frac{\partial \psi_{\G}}{\partial q^{j}}-\ol{p}_{j}\right),\quad
\dot{\ol{z}}
=\gamma_{1}(\psi_{\G}(q)-\ol{z}),
\qquad
j=1,\ldots,n.
$$
By identifying
$$
\gamma_{1} 
=\beta^{-1}\lambda_{1},
$$
one recognizes that the Fokker-Planck equation with the slowest time-scale
is exactly the same as the contact Hamiltonian system.
\qed
\end{Proof}

Integral curves of the contact Hamiltonian flow
in Theorem\,\ref{fact:Fokker-Planck-contact-Hamiltonian} 
are explicitly written in coordinates as 
$$
q_{j}(t)
=q_{j}(0),\qquad
\ol{p}_{j}(t;q)
=\frac{\partial \psi_{\G}}{\partial q^{j}}(q)
+\left(\ol{p}_{j}(0;q)-\frac{\partial \psi_{\G}}{\partial q^{j}}(q)
\right)\e^{-\gamma_{1} t},
\quad j=1,\ldots,n, 
$$
and
$$
\ol{z}(t;q)
=\psi_{\G}(q)+
\left(\ol{z}(0;q)-\psi_{\G}(q)\right) 
\e^{-\gamma_{1} t}.
$$
As can be seen from this explicit expression, 
$\gamma_{1}$ is the natural time-scale 
(see also Remark\,\ref{remark:time-scale-general-Hamiltonian}). 

From the expression of $\ol{p}_{j}(t;q)$, 
one can express an approximated
expectation value of $B_{1}$, as well as
the exact expectation value that
has been defined in \fr{expected-value-B}. 
To express the approximated one for a given $B_{1}\in\GamLamM{1}$ at $t$,
introduce 
\beqa
\ol{\mbbE}_{t}:\GamLamM{1}&\to&\mbbR
\non\\
B_{1}&\mapsto&\ol{\mbbE}_{t}[B_{1}]
\non
\eeqa
as
\beq
\ol{\mbbE}_{t}[B_{1}]
:=\int_{\cM}B_{1}\ol{\rho}_{t}\star 1 
=\int_{\cM}B_{1}\rho_{\G}\ol{\Phi}_{t}\star 1.
\label{expected-value-B-bar}
\eeq
From \fr{expected-value-B-bar} with $\ol{p}_{j}(t;q)$ 
in \fr{p-bar}, one has
$$
\ol{\mbbE}_{t}[B_j]
=\ol{p}_{j}(t;0)
=\frac{\partial \psi_{\G}}{\partial q^{j}}(0)
+\left(
\ol{p}_{j}(0;0)-\frac{\partial \psi_{\G}}{\partial q^{j}}(0)
\right)\e^{-\gamma_{1} t},
\quad j=1,\ldots,n. 
$$
In addition, from \fr{z-bar} and \fr{expected-value-B-bar}, 
$\ol{z}(t;q)$ is interpreted
as the approximated moment generating function:
$$
\ol{\mbbE}_{t}\left[\exp(q\cdot B)\right]
=\ol{z}(t;q).
$$
\begin{Remark}
\label{remark:Fokker-Planck-contact-Hamiltonian}
By Theorem\,\ref{fact:Fokker-Planck-contact-Hamiltonian},
a procedure is obtained, in which,  
for a given pair $(h,q)\in\GamLamM{0}\times\mbbR^{n}$, 
a contact Hamiltonian $\cH\in\GamLam{(T^{*}\mbbR^{n}\times\mbbR)}{0}$  
is obtained systematically. 
This $\cH$ is given by  
$$
\cH(\ol{p},q,\ol{z}) 
=\beta^{-1}\lambda_{1}(\psi_{\G}(q)-\ol{z}), 
$$ 
where $\psi_{\G}(q)$ has been given by \fr{psi-G} as the integral 
involving $h$ over $\cM$,
$\ol{z}$ and $\ol{p}$ have been given by \fr{z-bar} and \fr{p-bar}
respectively,  
and $\lambda_{1}$ has been
the $1$st eigenvalue of $\triangle_{\G}$. 
\end{Remark}
\begin{Remark}
  How the contact Hamiltonian system is obtained from the Fokker-Planck equation
  and the time-evolution of these systems are summarized as the following diagrams:
  $$
  \xymatrix@R=25pt@C=43pt{
    (1,q,B)\ar@{~>}[r]^-{\fr{z-bar},\fr{p-bar}}
    &\cL_{\psi_{\G}}
    \\
    (\overline{\Phi}_{t},q,B)\ar@{~>}[r]^-{\fr{z-bar},\fr{p-bar}}
    \ar[u]^-{t\to\infty}
    &(\overline{p}(t),q,\overline{z}(t))\ar[u]^-{t\to\infty}
    \\
    (\overline{\Phi}_{0},q,B)\ar@{~>}[r]^-{\fr{z-bar},\fr{p-bar}}
    \ar@{~>}[u]^-{\fr{slowest-mode-diffusion-equation}}
    &(\overline{p}(0),q,\overline{z}(0))\ .
    \ar@{~>}[u]^-{X_{\cH}}
  }
  $$
\end{Remark}
\begin{Remark}
\label{remark:averaged-values-asymptotic-same}
Given a function $B_{1}\in\GamLamM{0}$ that does not depend on $t$,
consider $\mbbE_{t}[B_{1}]$ in \fr{expected-value-B}
and $\ol{\mbbE}_{t}[B_{1}]$ in \fr{expected-value-B-bar}, where 
they are unrelated to a contact Hamiltonian system in general.
Then from Corollary\,\ref{fact:limit-Phi} and \fr{limit-Phi-bar}, 
it follows that
the asymptotic
limits of these variables coincide:
$$
\lim_{t\to\infty}\mbbE_{t}[B_{1}]
=\lim_{t\to\infty}\ol{\mbbE}_{t}[B_{1}]
=\inp{1}{B_{1}}_{\G}. 
$$  
\end{Remark}
  
In this subsection the contact geometric description of
the Fokker-Planck equation has been discussed, where the variable
$\ol{z}$ is chosen to be the moment generating function, and
the equilibrium distribution function does not depend on $q$ and $B$. 
Meanwhile in the literature dynamics of the minus of a free-energy
$F$ in \fr{F=S-H}   
is often considered\,\cite{Goto2015,GLP,Entov2023arxiv}. 
To compare the present choice of $z$ with the existing one,
the case with $z=-F$ is considered in Section\,\ref{section:appendix-z=-F}. 
Note that in \cite{Goto2020}, another choice of $z$ was studied.  

\subsection{Dynamics depending on externally applied fields}
\label{section:dynamics-depends-on-q}
In Section\,\ref{section:derivation-contact-Hamiltonian},  
$\Phi_{t}=\rho_{\G}^{-1}\rho_{t}$ given by
\fr{rho-rho-G-Phi} is employed to show the relation between
the Fokker-Planck equation and a contact Hamiltonian system, where 
$\rho_{t}$ obeys \fr{Fokker-Planck-0}. 
The asymptotic limit of $\rho_{t}$ is the stationary distribution function
$\rho_{\G}$ given by 
\fr{Gibbs-distribution}, and $\rho_{\G}$ is independent of $q\cdot B$.
Here, recall that $q$ is a set of primal variables  
and the integrals of $B$ yield
the conjugate thermodynamic variables, as mentioned in 
Section\,\ref{section:derivation-contact-Hamiltonian}.  
In addition, $q$ is interpreted as a set of externally applied fields.   
Meanwhile, one can consider the case that the stationary distribution depends
on $(q\cdot B)$, and it is discussed in this subsection. 

Consider the case that the stationary distribution depends
on externally applied fields. 
For this case, one introduces 
  a set of constants $q\in\mbbR^{n}$ and
\beqa
\wt{h}
&:=&
h-\beta^{-1}(q\cdot B)\quad\in\GamLam{(\cM\times\mbbR^{n})}{0},
\non\\
\wt{\rho}_{\G}
&:=&
\wt{Z}_{\G}^{-1}\exp(-\beta \wt{h})
=\wt{Z}_{\G}^{-1}\exp(-\beta h+q\cdot B) 
\quad \in\GamLam{(\cM\times\mbbR^{n})}{0},
\non
\eeqa
with $h,B_{j}\in\GamLamM{0}$, $j=1,\ldots,n$, 
where $\wt{Z}_{\G}$ depends on $\beta$ and $q$.
The principal role of $\wt{Z}_{\G}$ is 
to normalize $\wt{\rho}_{\G}$ such that 
$$
\int_{\cM}\wt{\rho}_{\G}\star1
=1.
$$
Note that the integral is over $\cM$, not $\cM\times \mbbR^{n}$.
The reason for it is that $q\in\mbbR^{n}$ is treated as a set of
  given time-independent parameters, hence phase space of the dynamical system associated with $\wt{h}$ is
  $\cM$.
In addition, $\star 1$ is a volume form on $\cM$,  
not on $\cM\times\mbbR^{n}$. 
  To describe the Fokker-Planck equation for this case in the differential
  geometric manner,  
  the exterior derivative $\dr$ is employed here,
  where this $\dr$ is the same as that on $\cM$.
  To be precise, let $\wt{\dr}$ be the exterior derivative on the ambient manifold:  
  $$
  \wt{\dr}:\GamLam{(\cM\times\mbbR^{n})}{k}\to
  \GamLam{(\cM\times\mbbR^{n})}{k+1},\quad k=0,1,\ldots,m+n-1,  
  $$
  and $\bdr_{\mbbR^{n}}$ be the one on $\mbbR^{n}$:  
  $$
  \bdr_{\mbbR^{n}}:\GamLam{\mbbR^{n}}{k}\to\GamLam{\mbbR^{n}}{k+1},
  \quad k=0,\ldots,n-1.   
  $$
  One can write $\wt{\dr}$ with $\dr$ and $\bdr_{\mbbR^{n}}$ such that 
  $$
  \wt{\dr}=\dr +\bdr_{\mbbR^{n}}
  $$
  whose analogue can be found in Chapter 4.3b of \cite{Frankel2012}.
  For instance, given a function $f\in\GamLam{(\cM\times\mbbR^{n})}{0}$ with coordinates $x$
  for $\cM$, it follows that
  $$
  \wt{\dr}f
  =\dr f+\bdr_{\mbbR^{n}} f,\qquad
  \text{where}\quad
  \dr f
  =\sum_{k=1}^{m}\frac{\partial f}{\partial x^{k}}\dr x^{k},
  \quad\text{and}\quad
  \bdr_{\mbbR^{n}} f
  =\sum_{j=1}^{n}\frac{\partial f}{\partial q^{j}}\bdr_{\mbbR^{n}} q^{j}.
  $$
  Since phase space is $\cM$ and $q\in\mbbR^{n}$ is fixed,
  the Fokker-Planck equation involves $\dr$, not
  $\bdr_{\mbbR^{n}}$.  
The relation,
$$
\beta^{-1}\dr\wt{\rho}_{\G}
=-\wt{\rho}_{\G}\dr\wt{h}, 
$$
holds and is used to derive a diffusion equation from the Fokker-Planck
equation.   
The Fokker-Planck equation for 
$\wt{\rho}_{t}\in\GamLam{(\cM\times\mbbR^{n})}{0}$ is
\beqa
\frac{\partial}{\partial t}\wt{\rho}_{t}
&=&-\beta^{-1}\dr^{\dagger}\dr\wt{\rho}_{t}
-\dr^{\dagger}\wt{\rho}_{t}\dr \wt{h}
\non\\
&=&-\beta^{-1}\dr^{\dagger}\dr\wt{\rho}_{t}
-\dr^{\dagger}\wt{\rho}_{t}(\dr h-\beta^{-1}q\cdot \dr B).
\label{Fokker-Planck-0-q}
\eeqa
Notice that the time-evolution of $\wt{\rho}_{t}$ depends on $q$ and $B$, 
which is dissimilar to the time-evolution of $\rho_{t}$.
    
To derive the corresponding diffusion equation,
introduce $\wt{\Phi}_{t}$ such that
$$
\wt{\rho}_{t}
=\wt{\rho}_{\G}\wt{\Phi}_{t}, 
$$
and the global inner product
$$
\inp{\wt{\alpha}^{\I}}{\wt{\alpha}^{\II}}_{\wt{\G}}
=\int_{\cM}\left(\wt{\alpha}^{\I}\wedge\star\wt{\alpha}^{\II}\right)\wt{\rho}_{\G}
$$
for $\wt{\alpha}^{\I},\wt{\alpha}^{\II}\in\GamLam{(\cM\times\mbbR^{n})}{k}$, 
$k=0,1,\ldots,m$, such that  
$$
\ii_{\partial/\partial q^{j}}\wt{\alpha}^{\I}
=\ii_{\partial/\partial q^{j}}\wt{\alpha}^{\II}
=0,\qquad j=1,\ldots,n. 
$$
In addition, the weighted Laplacian
can be defined
as
$$
\triangle_{\wt{\G}}
:=
\dr_{\wt{\G}}^{\dagger}\dr,
$$   
with
$\dr_{\wt{\G}}^{\dagger}\alpha:=\wt{\rho}_{\G}^{\,-1}\dr^{\dagger}\wt{\rho}_{\G}\alpha$ 
for all $\alpha\in\GamLam{(\cM\times\mbbR^{n})}{k}$, $k=0,1,\ldots,m$.
It is then natural to consider the eigenvalue problem:
$$
\triangle_{\wt{\G}}\wt{\phi}_{s}
=\wt{\lambda}_{s}\wt{\phi}_{s},\qquad s=0,1,\ldots,
$$
and assume the formal eigenfunction expansion:
$$
\wt{\Phi}_{t}
=\sum_{s=0}^{\infty}\wt{a}_{t}^{s}\wt{\phi}_{s},\qquad
\wt{\phi}_{s}\in\GamLam{(\cM\times\mbbR^{n})}{0},
$$
with
$\wt{a}_{t}^{s}$ being a function $t\mapsto \wt{a}_{t}^{s}$,
$s=0,1,\ldots$.  
Note that, unlike $\lambda_{s}$, $\wt{\lambda}_{s}$ depends on
$q$ for each $s$ in general. 
Normalizations are imposed such that
$$
\inp{\wt{\phi}_{s}}{\wt{\phi}_{s^{\prime}}}_{\wt{\G}}
=\delta_{ss^{\prime}}.\qquad s,s^{\prime}=0,1,\ldots.
$$
In addition, one can show that $\wt{\lambda}_{s}\geq 0$ 
for all $s$ by arguing as in Lemma\,\ref{fact:eigenvalue-0-positive}.  

Repeating the derivation of the diffusion equation for $\Phi_{t}$
(see Theorem\,\ref{fact:diffusion-from-Fokker-Planck}),  
one derives
\beq
\frac{\partial}{\partial t}\wt{\Phi}_{t}
=-\beta^{-1}\triangle_{\wt{\G}}\wt{\Phi}_{t}.
\label{diffusion-from-Fokker-Planck-q}
\eeq
  An ODE for $\wt{a}_{t}^{s}$ for each $s$ is derived
  by arguing as in Proposition\,\ref{fact:dynamics-a}.
  First, notice the identity  
  $$
  \wt{a}_{t}^{s}
  =\inp{\wt{\Phi}_{t}}{\wt{\phi}_{s}}_{\wt{\G}},\qquad t\in\mbbR.
  $$
  Then, from this and \fr{diffusion-from-Fokker-Planck-q} it follows for each $s$ that  
  $$
  \frac{\dr}{\dr t}\wt{a}_{t}^{s}
  =-\beta^{-1}\inp{\triangle_{\wt{G}}\wt{\Phi}_{t}}{\wt{\phi}_{s}}_{\wt{\G}}
  =-\beta^{-1}\wt{\lambda}_{s}\wt{a}_{t}^{s},\qquad t\in\mbbR. 
  $$
To see the asymptotic limit of $\wt{\rho}_{t}$, let
$$
\wt{\Upsilon}_{t}
:=\frac{\beta}{2}\int_{\cM}
\dr\wt{\Phi}_{t}\wedge\wt{\rho}_{\G}\star\dr\wt{\Phi}_{t}.
$$
Then applying a similar argument
in Proposition\,\ref{fact:Lyapunov-function-exists}, 
one has that $\wt{\Upsilon}_{t}$ is a Lyapunov function.  
By following the same argument as in
Corollary\,\ref{fact:limit-Phi},   
the asymptotic limit of $\wt{\rho}_{t}$ is $\wt{\rho}_{\G}$: 
$$
\lim_{t\to\infty}\wt{\Phi}_{t}(\wt{\xi})
=1,\quad\text{and}\quad
\lim_{t\to\infty}\wt{\rho}_{t}(\wt{\xi})
=\wt{\rho}_{\G}(\wt{\xi}),
\quad\forall\wt{\xi}\in\cM\times\mbbR^{n}.
$$

To derive a contact Hamiltonian system, define
the slowest time-scale mode from $\wt{\Phi}_{t}$ as 
$$
\wt{\Phi}_{t}^{(1)}
:=1+\wt{a}_{t}^{1}\wt{\phi}_{1}.
$$
This $\wt{\Phi}_{t}^{(1)}$  satisfies 
$$
\frac{\partial}{\partial t}\wt{\Phi}_{t}^{(1)}
=-\beta^{-1}\wt{\lambda}_{1}\left(\wt{\Phi}_{t}^{(1)}-1\right).
$$
Dynamics generated by this PDE is called,
by abuse of terminology, the 
{\it slowest mode dynamics} in this paper.
In addition $\wt{\Phi}_{t}^{(1)}$ satisfies 
$$
\triangle_{\wt{\G}}\wt{\Phi}_{t}^{(1)}
=\wt{\lambda}_{1}\left(\wt{\Phi}_{t}^{(1)}-1\right),
$$
and the diffusion equation: 
\beq
\frac{\partial}{\partial t}\wt{\Phi}_{t}^{(1)}
=-\beta^{-1}\triangle_{\wt{\G}}\wt{\Phi}_{t}^{(1)}.
\label{slowest-mode-diffusion-equation-q}
\eeq
Since $\wt{\lambda}_{1}>0$, this $\wt{\Phi}_{t}^{(1)}$ satisfies
$$
\lim_{t\to\infty}\wt{\Phi}_{t}^{(1)}(\wt{\xi})
=1,\quad \forall\wt{\xi}\in\cM\times\mbbR^{n}.
$$
  To see the approximation in $\wt{\rho}_{t}^{(1)}$, introduce
  $$
  \wt{\rho}_{t}^{(1)}
  =\wt{\rho}_{\G}\wt{\Phi}_{t}^{(1)}.
  $$
  Then this $\wt{\rho}_{t}^{(1)}$ satisfies the PDE:
  \beq
  \frac{\partial}{\partial t}\wt{\rho}_{t}^{(1)}
  =-\beta^{-1}\wt{\lambda}_{1}\left(\wt{\rho}_{t}^{(1)}-\wt{\rho}_{\G}\right).
  \label{rho-bar-PDE-q}
  \eeq
  The discussions so far in this subsection are briefly summarized as 
  the diagrams for $t>0$:
$$
\xymatrix@R=25pt@C=43pt{
\wt{\rho}_{\G}\ar@{|->}[r]^-{\wt{\rho}_{\G}^{\,-1}\cdot}
&\wt{\Phi}_{\infty}=1
\ar@{|->}[r]^-{\cP}
&\wt{\Phi}_{\infty}^{(1)}=1
\ar@{|->}[r]^{\wt{\rho}_{\G}\cdot}
&\wt{\rho}_{\G}
\\
\wt{\rho}_{t}\ar[u]^-{t\to\infty}
\ar@{|->}[r]^-{\wt{\rho}_{\G}^{\,-1}\cdot}
&\wt{\Phi}_{t}\ar[u]^-{t\to\infty}
\ar@{|->}[r]^-{\cP}
&\wt{\Phi}_{t}^{(1)}
\ar@{->}[u]^-{t\to\infty}
\ar@{|->}[r]^{\wt{\rho}_{\G}\cdot}
&\wt{\rho}_{t}^{(1)}\ar[u]^-{t\to\infty}
\\
\wt{\rho}_{0}\ar@{~>}[u]^-{\fr{Fokker-Planck-0-q}}
\ar@{|->}[r]^-{\wt{\rho}_{\G}^{\,-1}\cdot}
&\wt{\Phi}_{0}\ar@{~>}[u]^-{\fr{diffusion-from-Fokker-Planck-q}}
\ar@{|->}[r]^-{\cP}
&\wt{\Phi}_{0}^{(1)}\ar@{~>}[u]^-{\fr{slowest-mode-diffusion-equation-q}}
\ar@{|->}[r]^{\wt{\rho}_{\G}\cdot}
&\wt{\rho}_{0}^{(1)}
\ar@{~>}[u]^-{\fr{rho-bar-PDE-q}}. 
}
$$
Note that these diagrams are similar to the ones in
\fr{diagrams-relations-variables}.

To construct a contact Hamiltonian system, introduce 
\beqa
\wt{z}^{(1)}(t;q)
&:=&\int_{\cM}\rho_{\G}\wt{\Phi}_{t}^{(1)}\exp(q\cdot  B)\star 1
=\inp{\wt{\Phi}_{t}^{(1)}}{\exp(q\cdot B)}_{\G},
\label{z-bar-tilde}\\
\wt{p}_{j}^{(1)}(t;q)
&:=&\int_{\cM}\rho_{\G}\wt{\Phi}_{t}^{(1)}B_{j}\exp(q\cdot  B)\star 1
=\inp{\wt{\Phi}_{t}^{(1)}}{B_{j}\exp(q\cdot B)}_{\G},\quad
j=1,\ldots,n.
\label{p-bar-tilde}
\eeqa
Note that $\rho_{\G}\exp(q\cdot B)=\wt{Z}_{G}\wt{\rho}_{\G}(\neq\wt{\rho}_{\G})$
and that the employed global inner product has been 
$\inp{}{}_{\G}$ defined in \fr{inner-product-G}, not $\inp{}{}_{\wt{\G}}$.   
The reason why $\inp{}{}_{\wt{\G}}$ is not employed is that 
the quantity $\inp{B_{i}}{B_{j}}_{\wt{\G}}$ depends on $q$ even 
for $q$-independent functions $B_{i}$ and $B_{j}$. 
To be precise, this $q$-dependence of $\inp{}{}_{\wt{\G}}$ yields  
$$
  \frac{\partial}{\partial q^{i}} \inp{\wt{\Phi}_{t}^{(1)}}{B_{j}}_{\wt{\G}}
  \neq \inp{\frac{\partial}{\partial q^{i}}{\wt{\Phi}}_{t}^{(1)}}{B_{j}}_{\wt{\G}}, \quad
  i,j=1,\ldots,n,
  $$
which may not lead to a closed equation. 
Thus we do {\it not} employ
  $\psi_{\wt{\G}}\in\GamLam{(\cM\times\mbbR^{n})}{0}$ that is defined by   
  $$
  \psi_{\wt{\G}}(q)
  :=\inp{\exp(q\cdot B)}{1}_{\wt{\G}}
  =\int_{\cM}\exp(q\cdot B)\wt{\rho}_{\G}\star 1.
  $$
  To show that $\psi_{\wt{\G}}$ is not appropriate for expressing
  the expectation values of $B$, one differentiates $\psi_{\wt{\G}}$:
  $$
  \frac{\partial\psi_{\wt{\G}}}{\partial q^{i}}
  =\int_{\cM}\left(B_{i}\wt{\rho}_{\G}+\frac{\partial \wt{\rho}_{\G}}{\partial q^{i}}
  \right)\exp(q\cdot B)\star 1
  ,\quad i=1,\ldots,n.
  $$
  The right hand side of the equation above does not express
  the expectation values of $B_{i}$, $i=1,\ldots,n$. 

Consider dynamical behavior of $\wt{z}^{(1)}$ on
the $z$-axis, and that of $\wt{p}^{(1)}$ 
on the $p$-axis.  
Regarding the time-evolution of \fr{z-bar-tilde} and \fr{p-bar-tilde}, 
one has the following. 
\begin{Proposition}
\label{fact:Fokker-Planck-contact-Hamiltonian-q-dynamics}  
(Approximate Fokker-Planck equation depending on externally applied field).  
The time-development of $\wt{z}^{(1)}$ and $\wt{p}_{j}^{(1)}$, $(j=1,\ldots,n)$
is described 
on $T^{*}\mbbR^{n}\times\mbbR$, 
where dynamics of $\wt{z}^{(1)}$ and that of $\wt{p}_{j}^{(1)}$ are governed by the  
slowest time-scale dynamics.  
Their asymptotic limits
are on the Legendrian submanifold generated by $\psi_{\G}$.
\end{Proposition}
\begin{Proof} 
From \fr{z-bar-tilde} and \fr{p-bar-tilde}, one has that
\beqa
\frac{\dr}{\dr t}\wt{z}^{(1)}
&=&\inp{\dot{\wt{\Phi}}_{t}^{(1)}}{\exp(q\cdot B)}_{\G}
=-\beta^{-1}\wt{\lambda}_{1}
\inp{\wt{\Phi}_{t}^{(1)}-1}{\exp(q\cdot B)}_{\G}
=\beta^{-1}\wt{\lambda}_{1}\left(\psi_{\G}(q)-\wt{z}^{(1)}\right),
\non\\
\frac{\dr}{\dr t}\wt{p}_{j}^{(1)}
&=&
\inp{\dot{\wt{\Phi}}_{t}^{(1)}}{B_{j}\exp(q\cdot B)}_{\G}
=-\beta^{-1}\wt{\lambda}_{1}\inp{\wt{\Phi}_{t}^{(1)}-1}{B_{j}\exp(q\cdot B)}_{\G}
\non\\
&=&\beta^{-1}\wt{\lambda}_{1}\left(
\frac{\partial\psi_{\G}}{\partial q^{j}}(q)-\wt{p}_{j}^{(1)}\right), 
\qquad j=1,\ldots,n, 
\non
\eeqa
where $\psi_{\G}$ has been defined in \fr{psi-G}.  
Then it follows from \fr{z-bar-tilde} and \fr{p-bar-tilde} that 
\beqa
\lim_{t\to\infty}\wt{z}^{(1)}(t;q)
&=&\inp{1}{\exp(q\cdot B)}_{\G}
=\psi_{\G}(q),
\non\\
\lim_{t\to\infty}
\wt{p}_{j}^{(1)}(t;q)
&=&\inp{1}{B_{j}\exp(q\cdot B)}_{\G}
=\frac{\partial\psi_{\G}}{\partial q^{j}}(q),\quad
j=1,\ldots,n. 
\non
\eeqa
This shows that, for a fixed $q$, 
the asymptotic limit of a point is on a point of 
the Legendrian submanifold $\cL_{\psi_{\G}}(q)$.  
\qed
\end{Proof}
Notice that the obtained system of ODEs 
in Proposition\,\ref{fact:Fokker-Planck-contact-Hamiltonian-q-dynamics} 
is different to 
that in Theorem\,\ref{fact:Fokker-Planck-contact-Hamiltonian}, since
the eigenvalue $\wt{\lambda}_{1}$ depends on $q$ in general. 
To see this difference, consider the contact Hamiltonian  
$$
\wt{\cH}^{(1)}(\wt{p}^{(1)},q,\wt{z}^{(1)})
=\beta^{-1}\wt{\lambda}_{1}(q)\left(\psi_{\G}(q)-\wt{z}^{(1)}\right),
$$
which is analogous to but different to $\cH$ 
in \fr{contact-Hamiltonian-derived}.  
From \fr{contact-Hamiltonian-coordinates}, one has the contact Hamiltonian system:
\beqa
\dot{\wt{z}}^{(1)}
&=&\wt{\cH}^{(1)}
=\beta^{-1}\wt{\lambda}_{1}(q)\left(\psi_{\G}(q)-\wt{z}^{(1)}\right),
\non\\
\dot{q}^{j}
&=&0,\quad
\dot{\wt{p}}^{(1)}_{j}
=\beta^{-1}\wt{\lambda}_{1}\left(
\frac{\partial\psi_{\G}}{\partial q^{j}}-\wt{p}_{j}^{(1)}\right)
+\beta^{-1}\frac{\partial\wt{\lambda}_{1}}{\partial q^{j}}
\left(\psi_{\G}(q)-\wt{z}^{(1)}\right),\quad
j=1,\ldots,n.
\non
\eeqa
From this discussion associated with
Theorem\,\ref{fact:Fokker-Planck-contact-Hamiltonian-q-dynamics}, 
one notices the following.
\begin{Remark}
  \label{remark:Fokker-Planck-contact-q-dynamics}
  If $\partial\wt{\lambda}_{1}/\partial q^{j}=0$ for all $j$, then
  the time-development of
  $\wt{z}^{(1)}$ in \fr{z-bar-tilde} and that of 
  $\wt{p}^{(1)}$  in \fr{p-bar-tilde}
  obey a contact Hamiltonian system. 
\end{Remark}

The differences 
between the system in Theorem\,\ref{fact:Fokker-Planck-contact-Hamiltonian}
and the system in Proposition\,\ref{fact:Fokker-Planck-contact-Hamiltonian-q-dynamics}
in the asymptotic limit can be discussed as follows. 
The differences between $\dot{\wt{p}}^{(1)}_{j}$ and $\dot{\ol{p}}_{j}$ 
on $T_{q}^{*}\mbbR^{n}\times\mbbR$ at each $q\in\mbbR^{n}$  
are 
$$
\dot{\wt{p}}^{(1)}_{j}-\dot{\ol{p}}_{j}
=\beta^{-1}\wt{\lambda}_{1}\left(
\frac{\partial\psi_{\G}}{\partial q^{j}}-\wt{p}_{j}^{(1)}\right)
-\lambda_{1}\left(
\frac{\partial\psi_{\G}}{\partial q^{j}}-\ol{p}_{j}^{(1)}\right)
+
\beta^{-1}\frac{\partial\wt{\lambda}_{1}}{\partial q^{j}}
\left(\psi_{\G}(q)-\wt{z}^{(1)}\right),\quad
j=1,\ldots,n,
$$
and the difference between $\dot{\wt{z}}^{(1)}$ and $\dot{\ol{z}}$ is
$$
\dot{\wt{z}}^{(1)}-\dot{\ol{z}}
=\beta^{-1}\wt{\lambda}_{1}(q)\left(\psi_{\G}(q)-\wt{z}^{(1)}\right)
-\lambda_{1}\left(\psi_{\G}(q)-\ol{z}\right).
$$
The both differences vanish in the asymptotic limit, $t\to\infty$.

\subsection{Example}
\label{section:Example}
In this subsection the case of $\cM=\mbbR$ is considered as an example. 
After discussing basic properties for this case, 
a Hamiltonian $h$ is chosen and focused, and  
its corresponding contact Hamiltonian $\cH$ and its
variables are explicitly obtained. 
This analysis demonstrates  
how the general theory of this paper is applied.

Let $x$ be the
coordinate of $\mbbR$, where $x$ also denotes a point of $\mbbR$.
In addition let 
the Riemannian metric and its canonical volume-form to be
$$
g=\dr x\otimes \dr x,\qquad
\star 1=\dr x.
$$
The equilibrium state is totally characterized by the 
Gibbs distribution function in \fr{Gibbs-distribution}. 
Then the next target is to analyze nonequilibrium states 
with the Fokker-Planck equation in the present framework.    
From straightforward calculations, 
the  Fokker-Planck equation \fr{Fokker-Planck-0} is expressed 
in coordinates as 
$$
\frac{\partial}{\partial t}\rho_{t}
=\beta^{-1}\frac{\partial^{2}}{\partial x^{2}}\rho_{t}
+\frac{\partial}{\partial x}\left(\rho_{t}\frac{\partial h}{\partial x}\right),
$$
that is
\beq
\frac{\partial}{\partial t}\rho_{t}
=\frac{\partial}{\partial x}
\left(\beta^{-1}\frac{\partial}{\partial x}+\frac{\partial h}{\partial x}\right)
\rho_{t}.
\label{Fokker-Planck-R}
\eeq
The point of departure is to discuss the eigenvalues of $\triangle_{\G}$. 
The eigenvalue problem $\triangle_{\G}\phi_{s}=\lambda_{s}\phi_{s}$ for 
an $s$th mode is equivalent to 
$$
-\rho_{\G}^{-1}\star^{-1}\dr\star\rho_{\G}\dr\phi_{s}
=\lambda_{s}\phi_{s},
$$
which is,
$$
\dr\star\rho_{\G}\dr \phi_{s}
=-\,\lambda_{s}\rho_{\G}\phi_{s}\star1.
$$
This reduces further. To ease of notation, ${\ }^{\prime}$ denotes 
$\partial/\partial x$, with  the use of 
$$
\rho_{\G}^{\prime}
=-\beta h^{\prime}\rho_{\G},\quad\text{and}\quad
\star\dr x=1,
$$
one calculates 
\beqa
\dr\star\rho_{\G}\dr \phi_{s}
&=&\dr(\star\rho_{\G}\phi_{s}^{\prime}\dr x)
=\dr(\rho_{\G}\phi_{s}^{\prime})
=(\rho_{\G}\phi_{s}^{\prime\prime}+\rho_{\G}^{\prime}\phi_{s}^{\prime})\dr x
\non\\
&=&(\rho_{\G}\phi_{s}^{\prime\prime}-\beta h^{\prime}\rho_{\G}\phi_{s}^{\prime})\dr x,
\non\\
\lambda_{s}\rho_{\G}\phi_{s}\star1
&=&\lambda_{s}\rho_{\G}\phi_{s}\dr x.
\non
\eeqa
Hence the eigenvalue problem is equivalent to
\beq
\phi_{s}^{\prime\prime}-\beta h^{\prime}\phi_{s}^{\prime}
+\lambda_{s}\phi_{s}
=0.
\label{eigenvalue-problem-on-R}
\eeq
This equation is a special case of the Strum-Liouville equation,
and various theorems are known\,\cite{Zetti2005}.  
For $s=0$ with $\lambda_{0}=0$, \fr{eigenvalue-problem-on-R} reduces to
$$
\phi_{0}^{\prime\prime}
=\beta h^{\prime}\phi_{0}^{\prime}.
$$
Its solution is found to be
$$
\phi_{0}(x)
=1,
$$
where $\inp{\phi_{0}}{\phi_{0}}_{\G}=1$ is satisfied so that 
\fr{orthonormal-phi} holds for $s=s^{\prime}=0$.  
The  modes $\phi_{s}$ with $s\geq 1$ should satisfy
$\inp{\phi_{s}}{\phi_{0}}_{\G}=0$. 

On this manifold, we consider the Hamiltonian on $\mbbR$, 
\beq
h(x)=\frac{x^{2}}{2\mu}.
\label{microscopic-hamiltonian-example}
\eeq
Physically this $h$ is the total energy of a free particle
with $\mu>0$ being mass.
In addition, $x$ is interpreted as momentum of the free particle, 
the metric $g$ measures the norm of the momentum
for a unit mass particle, and 
\fr{microscopic-hamiltonian-example}
can be written as
$$
h(x)=\frac{1}{2\mu}g\left(
\mathfrak{p}^{\sharp},\mathfrak{p}^{\sharp}\right),\quad
\mathfrak{p}
=x\dr x=g(\mathfrak{p}^{\sharp},-)\ \in T_{\mathfrak{q}}^{*}Q,\quad
\mathfrak{p}^{\sharp}
=x\frac{\partial}{\partial x}\ \in T_{\mathfrak{q}}Q,\quad
\mathfrak{q}\in Q,
$$
with $Q$ being a $1$-dimensional manifold.
For this Hamiltonian the Gibbs distribution function \fr{Gibbs-distribution}
is 
$$
\rho_{\G}(x)
=Z_{\G}^{-1}\exp\left(-\beta \frac{x^{2}}{2\mu}\right),\qquad \text{with}\qquad
Z_{\G}=\sqrt{\frac{2\pi\mu}{\beta}}, 
$$
the weighted Laplacian $\triangle_{\G}$ acting on a function $f$ on $\mbbR$ is
$$
\triangle_{\G}f
=-\rho_{\G}^{-1}\star^{-1}\dr\star\rho_{\G}\dr f
=-\e^{\beta x^{2}/(2\mu)}
\frac{\partial}{\partial x}
\left(\e^{-\beta x^{2}/(2\mu)}\frac{\partial f}{\partial x}\right)
=-\frac{\partial^{2}f }{\partial x^{2}}
+\beta\mu^{-1} x \frac{\partial f}{\partial x}, 
$$
and the eigenvalue equation is written as
$$
-\phi_{s}^{\prime\prime}+\beta\mu^{-1} x\phi_{s}^{\prime}
=\lambda_{s}\phi_{s},\quad s=0,1,\ldots.
$$
This equation is the Hermite differential equation up to some coefficients,
and the solutions for lower eigenvalue labels are 
\beqa
&&\lambda_{0}
=0,\qquad
\phi_{0}(x)
=1,
\non\\
&&\lambda_{1}
=\beta\mu^{-1},\qquad
\phi_{1}(x)
=\phi_{1}(1)\, x,\quad \cdots,
\non
\eeqa
where $\phi_{1}(1)$ is the normalization constant. 
Note that $\inp{\phi_{0}}{\phi_{0}}_{\G}=1$ and $\inp{\phi_{1}}{\phi_{0}}_{\G}=0$
are satisfied, and  
$\phi_{1}(1)$ 
is determined by $\inp{\phi_{1}}{\phi_{1}}_{\G}=1$. The explicit
form of $\phi_{1}(1)$ is obtained with  
$$
\int_{-\infty}^{\infty}x^{2}\e^{-\beta x^{2}/(2\mu)}\dr x
=\frac{1}{2}\sqrt{\pi \frac{(2\mu)^{3}}{\beta^{3}}},
$$
as
$$
\phi_{1}(1)
=\left(\frac{\beta^{3}}{2\pi\mu^{3}}\right)^{1/4}.
$$
The slowest time-scale mode equation \fr{slowest-mode-ODE}
is expressed for this model with $\lambda_{1}=\beta\mu^{-1}$ as 
$$
\frac{\partial}{\partial t}\ol{\Phi}_{t}
=-\mu^{-1}(\ol{\Phi}_{t}-1).
$$
The slowest time-scale mode is then
$$
\ol{\Phi}_{t}(x)
=1+c\,x\exp\left(-\mu^{-1}t\right),\quad\text{where}\quad
c:=\ol{\Phi}_{0}(1)-1
=a_{0}^{0}\phi_{1}(1).
$$
This $\ol{\Phi}_{t}$ satisfies the modified diffusion equation, as stated 
in Proposition\,\ref{fact:slowest-mode-diffusion-equation} and verified from 
$$
\frac{\partial}{\partial t}\ol{\Phi}_{t}
=-\mu^{-1}c\,x\e^{-\mu^{-1}t}
\quad\text{and}\quad
-\beta^{-1}\triangle_{\G}\ol{\Phi}_{t}
=-\beta^{-1}c\e^{-\mu^{-1}t}\left(-\frac{\partial^{2}x}{\partial x^{2}}+\beta\mu^{-1}x\frac{\partial x}{\partial x}
\right).
$$
In addition, from \fr{rho-bar} with this $\ol{\Phi}_{t}$, one has 
the approximate distribution function $\ol{\rho}_{t}=\rho_{\G}\ol{\Phi}_{t}$: 
$$
\ol{\rho}_{t}(x)
=Z_{\G}^{-1}\exp(-\beta x^{2}/(2\mu))
\left[1+c\,x\exp\left(-\mu^{-1}t\right)
\right],
  $$
which is a truncation of the exact distribution function 
$\rho_{t}(x)=\rho_{\G}\Phi_{t}(x)$:
$$
\rho_{t}(x)
=Z_{\G}^{-1}\exp(-\beta x^{2}/(2\mu))
\left[1+c\,x\exp\left(-\mu^{-1}t\right)+\phi_{2}(x)\exp(-\beta^{-1}\lambda_{2}t)+\cdots
\right].
$$
This $\ol{\rho}_{t}$ captures the asymmetry
$\ol{\rho}_{t}(-x)-\ol{\rho}_{t}(x)\neq 0$ at finite $t$ with
$x\neq 0$, and
shows relaxation  $\ol{\rho}_{t}(-x)-\ol{\rho}_{t}(x)\to0$, $t\to\infty$. 
These properties also hold for $\rho_{t}$.  
For a set of some functions $B=(B_{1},\ldots,B_{n})$ and
$q=(q^{1},\ldots,q^{n})\in\mbbR^{n}$, let  
\beqa
\psi_{\G}(q)
&=&\inp{\exp(q\cdot B)}{1}_{\G}
=\int_{\mbbR}\exp(q\cdot B(x))\rho_{\G}(x)\dr x
\non\\
&=&Z_{\G}^{-1}\int_{\mbbR}
\exp\left(-\beta \frac{x^{2}}{2\mu}+q\cdot B(x)\right) \dr x.
\non
\eeqa
and
\beqa
\ol{z}(t;q)
&=&\inp{\ol{\Phi}_{t}}{\exp(q\cdot B)}_{\G}
=Z_{\G}^{-1}\int_{\mbbR}\ol{\Phi}_{t}\exp(-\beta x^{2}/(2\mu)+q\cdot B)\dr x,
\non\\
\ol{p}_{j}(t;q)
&=&\inp{\ol{\Phi}_{t}}{B_{j}\exp(q\cdot B)}_{\G}
\non\\
&=&Z_{\G}^{-1}\int_{\mbbR}\ol{\Phi}_{t}B_{j}(x)
\exp(-\beta x^{2}/(2\mu)+q\cdot B)\dr x,
\quad j=1,\ldots,n.
\non
\eeqa
By Theorem\,\ref{fact:Fokker-Planck-contact-Hamiltonian}, the  
dynamical system for $(\ol{p},q,\ol{z})$ is written with  
the contact Hamiltonian  
$$
\cH(\ol{p},q,\ol{z})
=\mu^{-1}(\psi_{\G}(q)-\ol{z}).
$$
In particular choose $n=1$ and $B=x$ as a simple case.
In this case,  one has
\beqa
\psi_{\G}(q)
&=&Z_{\G}^{-1} 
\int_{-\infty}^{\infty}
\exp\left(-\beta \frac{x^{2}}{2\mu}+qx\right) \dr x
\non\\
&=&Z_{\G}^{-1}
\exp\left(\frac{\mu}{2\beta}q^{2}\right)
\int_{-\infty}^{\infty}
\exp\left(- \frac{\beta}{2\mu}(x-\beta^{-1}\mu q)^{2}\right) \dr x
=\exp\left(\frac{\mu}{2\beta}q^{2}\right).
\non
\eeqa
The asymptotic limit of a point $(\ol{p},q,\ol{z})$
on $T^{*}\mbbR\times\mbbR$ is a point on 
the Legendre submanifold generated by this $\psi_{\G}$: 
$$
\cL_{\psi_{\G}}
=\left\{
\left(\ol{p},q,\ol{z}\right)\ \bigg|\
\ol{p}=\frac{\mu}{\beta}q\exp\left(\frac{\mu}{2\beta}q^{2}\right),\
\ol{z}=\exp\left(\frac{\mu}{2\beta}q^{2}\right)
\right\}.
$$
Then the approximated expectation 
value of $B_{1}=x$ at $t$, defined
in \fr{expected-value-B-bar}, is calculated as 
\beqa
\ol{\mbbE}_{t}[x]
&=&\ol{p}(t;0)
=Z_{\G}^{-1}\int_{-\infty}^{\infty}\ol{\Phi}_{t}x\exp(-\beta x^{2}/(2\mu))\,\dr x
\non\\
&=&\sqrt{\frac{\beta}{2\pi\mu}}\int_{-\infty}^{\infty}
\left(1+cx\e^{-\mu^{-1}t}\right)x\e^{-\beta x^{2}/(2\mu)}\dr x
=c\sqrt{\frac{\beta}{2\pi\mu}}\e^{-\mu^{-1}t}
\int_{-\infty}^{\infty}x^{2}\e^{-\beta x^{2}/(2\mu)}\dr x
\non\\
&=&c\mu\beta^{-1}\e^{-\mu^{-1}t},
\non
\eeqa
where $c=a_{0}^{0}\phi_{1}(1)$ 
is a constant involving the initial point for the contact Hamiltonian
system. The physical interpretation of $\ol{\mbbE}_{t}[x]$ is,
by recalling \fr{microscopic-hamiltonian-example}, 
the average of the momentum $x$. To obtain more accurate expressions,  
fast time-scales with $\lambda_{2},\lambda_{3},\ldots$ should be taken into 
account. 

Consider the case where dynamics involves $q\cdot B$ as discussed in
Section\,\ref{section:dynamics-depends-on-q}.
The most critical parameter is the $1$st eigenvalue of $\triangle_{\wt{\G}}$.
For the case that $n=1$, $B=x$, and 
$$
\wt{h}(x)=\frac{x^{2}}{2\mu}-\beta^{-1}qx,
$$
one starts with 
$$
\wt{\rho}_{\G}
=\wt{Z}_{\G}^{-1}\exp\left(-\beta h+qx\right)
=\wt{Z}_{\G}^{-1}\exp\left(-\frac{\beta}{2\mu}x^{2}+qx\right),
$$
where 
the partition function is calculated so that $\wt{\rho}_{\G}$ is normalized as
$$
\wt{Z}_{\G}
=\int_{-\infty}^{\infty}\exp\left(
-\frac{\beta }{2\mu}x^{2}+qx
\right)\dr x
=\exp\left(\frac{\mu}{2\beta}q^{2}\right)\sqrt{\frac{2\pi \mu}{\beta}}.
$$
The eigenvalue equation
$\dr_{\wt{\G}}^{\dagger}\dr\wt{\phi}_{s}=\wt{\lambda}_{s}\phi_{s}$,
$s=0,1,\ldots$, 
is written in coordinates as 
$$
-\wt{\phi}_{s}^{\prime\prime}+(\beta\mu^{-1}x-q)\wt{\phi}_{s}^{\prime}
=\wt{\lambda}_{s}\wt{\phi}_{s},\qquad s=0,1,\ldots.
$$
whose solutions for lower eigenvalue labels are 
\beqa
&&\wt{\lambda}_{0}=0,\qquad\qquad
\wt{\phi}_{0}(x)=1
\non\\
&&\wt{\lambda}_{1}
=\beta\mu^{-1},\qquad
\wt{\phi}_{1}(x)
=\wt{\phi}_{1}(1)\,(x-\beta^{-1}\mu q),\quad \cdots.
\non
\eeqa
with $\wt{\phi}_{1}(1)$ being a normalization constant. 
From Remark\,\ref{remark:Fokker-Planck-contact-q-dynamics}
with $\partial\wt{\lambda}_{1}/\partial q=0$, it follows that 
the dynamical system for $\wt{z}^{(1)}$ and $\wt{p}^{(1)}$ with $B_{1}(x)=x$ 
is a contact Hamiltonian system on $T^{*}\mbbR\times\mbbR$. 

\section{Discussions and Conclusion}
\label{section:conclusion}
In this paper, it has been shown
how the Fokker-Planck equation yields a class of 
contact Hamiltonian systems. 
A crucial point in this derivation has been to derive a diffusion equation 
and to employ several properties of a weighted Laplacian, where 
the diffusion equation with the weighted Laplacian
have been derived from the Fokker-Planck equation on a Riemannian manifold.
This Laplacian 
is 
symmetric   
with respect to an introduced global 
inner product for
forms, and the slowest time-scale eigenfunctions and eigenvalues
have been employed 
to construct closed dynamical equations for expectation values.
This set of dynamical
equations has been shown to be a contact Hamiltonian vector field.   

In the following contact geometric 
equilibrium statistical mechanics 
and contact geometric  nonequilibrium statistical mechanics are 
briefly summarized, so that the significance of the claims in 
this paper is illustrated. 
\begin{itemize}
\item
Equilibrium statistical mechanics provides  
links between phase space of 
dynamical systems and
equilibrium thermodynamic phase space. To show this schematically,
let $\cM$ be phase space of a 
dynamical system, $x$ be
a set of coordinates of $\cM$, 
$\mbbR^{n}$ be the space for externally applied field,
$\cN=T^{*}\mbbR^{n}\times\mbbR$ be thermodynamic phase space,
and $\cL\subset\cN$ be a Legendrian submanifold expressing
a set of equilibrium states. Since the space of Hamiltonians
is $\GamLamM{0}$, one of the roles 
of equilibrium statistical mechanics is summarized as the diagram, 
\beq
\GamLamM{0}\times\mbbR^{n}
\xdashrightarrow[\text{statistical mechanics}]{\text{equilibrium}}
\cL\subset \cN,
\label{diagram-equilibrium-statistical-mechanics}
\eeq
where $A\xdashrightarrow[]{\text{C}}B$
means that C provides $B$ from  $A$.   
To express \fr{diagram-equilibrium-statistical-mechanics} in coordinates, 
let $h$ be a Hamiltonian,  $q\in\mbbR^{n}$, 
and $\cL_{\psi_{0}}$ be the Legendrian submanifold
generated by some function $\psi_{0}$. Then
\fr{diagram-equilibrium-statistical-mechanics} is expressed in coordinates
as
\beq
(h(x),q)
\xdashrightarrow[\text{statistical mechanics}]{\text{equilibrium}}
\cL_{\psi_{0}}(q).
\label{diagram-equilibrium-statistical-mechanics-coordinates}
\eeq
The diagram\,\fr{diagram-equilibrium-statistical-mechanics-coordinates}
is a standard treatment of contact geometric thermodynamics\,
\cite{Mrugala1990}.

\item
Nonequilibrium statistical mechanics provides
links between 
dynamical systems and nonequilibrium thermodynamic
systems.
When restricting attention to derivations 
of 
thermodynamic dynamics from 
dynamical systems, by 
identifying dynamical systems with vector fields on manifolds,  
the role of nonequilibrium statistical mechanics is summarized as the 
diagram,  
\beq
\GTM\times\mbbR^{n}
\xdashrightarrow[\text{statistical mechanics}]{\text{nonequilibrium}}
\GT{\cN}.
\label{diagram-nonequilibrium-statistical-mechanics}
\eeq
For the case that $\cM$ is a symplectic manifold,
a Hamiltonian on $\cM$ determines a vector field uniquely, and
a contact Hamiltonian on $\cN$ determines a vector field uniquely.
Note that, by recalling the use of a Hamiltonian $h$ 
in equilibrium statistical mechanics,  
this $h$ as an element of $\GamLamM{0}$ 
need not induce a vector field.   
When a Hamiltonian and external fields are enough to specify 
dynamics,   
\fr{diagram-nonequilibrium-statistical-mechanics} can be expressed as
\beq
\GamLamM{0}\times\mbbR^{n}
\xdashrightarrow[\text{statistical mechanics}]{\text{nonequilibrium}}
\GamLam{\cN}{0}.
\label{diagram-nonequilibrium-statistical-mechanics-function}
\eeq
In coordinates,
when a pair $(h,q)$ is given,  
\fr{diagram-nonequilibrium-statistical-mechanics-function}
is expressed with $x$ for $\cM$ as 
\beq
(h(x),q)
\xdashrightarrow[\text{statistical mechanics}]{\text{nonequilibrium}}
\cH(\ol{p},q,\ol{z}),
\label{diagram-nonequilibrium-statistical-mechanics-function-coordinates}
\eeq
where $\cH\in\GamLam{\cN}{0}$ is a contact Hamiltonian
and $(\ol{p},q,\ol{z})$ is a point of $\cN$.
\end{itemize}
In this paper, one explicit
realization of \fr{diagram-nonequilibrium-statistical-mechanics-function-coordinates} has been proposed, as has been summarized as
Theorem\,\ref{fact:Fokker-Planck-contact-Hamiltonian} 
with 
Remark\,\ref{remark:Fokker-Planck-contact-Hamiltonian}. 
One recognizes that one open problem in 
contact geometric thermodynamics had been to systematically
determine a contact Hamiltonian $\cH$ from a given $h$. 
A solution to this problem has been given in this paper.  

There remain unsolved problems that have not been addressed in this paper.
They include the following.
\begin{itemize}
\item
  Obtain a contact Hamiltonian from a given dynamical spin system,
  where phase space of the spin system is discrete. For instance,
  consider the Glauber system that is a well-studied toy model
  in nonequilibrium statistical mechanics. Since 
  its dynamics is based on the master equation, and is defined on discrete
  phase space,  the reduction method presented in this paper is inapplicable. 
  An approach for this discrete system 
  is expected to be established by combining the study in
  Section\,\ref{section:diffusion-to-contact} of this paper and 
  techniques developed in \cite{Goto2020JMP}, where   
  in \cite{Goto2020JMP},
  a discrete diffusion equation was shown to be derived from a master equation. 
  
\item
  Establish a perturbation theory for
  Section\,\ref{section:diffusion-to-contact} of this paper, so that
  the faster time-scale modes of the weighted Laplacian
  are incorporated order by order.

\item
  Establish a method to obtain contact Hamiltonians from 
  nonlinear Fokker-Planck equations\,\cite{Frank2005,Wada2009},   
  since a natural extension of the linear Fokker-Planck equation
  is a nonlinear one. 

\item
  Extend the existing supersymmetric method for solving the Fokker-Planck
  equation by means of the present method\,\cite{Rosu1997},
  since the present method may include the so-called supersymmetric 
  method that can be applied to the Fokker-Planck equation. 

\item  
  Explore a relation between the present study and
  information geometry\,\cite{Ito2023}.
  This is because information geometry is compatible with nonequilibrium
  statistical mechanics\,\cite{Goto2020,Goto2016},
  clearer relations are expected 
  to be found with the present approach.

\item
  Show how symmetries associated with $\cM$ are reflected in 
  $\cN$.
  In particular for the case when $\cM$ is a symplectic manifold,  
  how symplectic reductions in $\cM$ are reflected in $\cN$ should be
  clarified\,\cite{Silva2008,Mcduff2016}.

\item
  Extend the present theory to describe  
  systems with phase transitions\,\cite{Goto2022,GLP},
  where singularities are developed on free-energy curves. 

\item   
  Explore how symplectic topology and contact topology\,\cite{Mcduff2016}
  can be applied to the present theory.
  
\end{itemize}
By addressing these together with the present study,
it is expected that a relevant and sophisticated geometric methodology
will be established for dealing with nonequilibrium phenomena.

\subsection*{Data availability statement}
No new data were created or analyzed in this study.

\subsection*{Acknowledgments}
The author would like to thank 
Leonid Polterovich for fruitful discussions, comments on this study,
and for useful references.
The author also would like to thank Minoru Koga for
indicating various mistakes found in the previous manuscript, for giving thoughtful comments, and for
fruitful discussions. Finally, the author would like to thank anonymous
referees for indicating mistakes and meaningful comments. 
These discussions and thoughtful comments helped significantly 
improve the paper.

\subsection*{ORCID ID}
Shin-itiro Goto\quad \verb| https://orcid.org/0000-0002-5249-1054|

\appendix
\section{Appendix}

\subsection{A simplified proof for Theorem\,\ref{fact:diffusion-from-Fokker-Planck}}
\label{section:appendix-simplified-proof-1}
The following is another proof for Theorem\,\ref{fact:diffusion-from-Fokker-Planck}.
\begin{Proof}
Using
\beqa
\dr \rho_{t}
&=&\Phi_{t}\dr\rho_{\G}+\rho_{\G}\dr \Phi_{t}
=\Phi_{t}(-\beta\rho_{\G}\dr h)+\rho_{\G}\dr \Phi_{t},
\non\\
\rho_{t}\dr h
&=&\rho_{\G}\Phi_{t}\dr h,
\non\\
\beta^{-1}\dr \rho_{t}+\rho_{t}\dr h
&=&\beta^{-1}\rho_{\G}\dr\Phi_{t},
\non\\
\dr^{\dagger}(\beta^{-1}\dr \rho_{t}+\rho_{t}\dr h)
&=&\beta^{-1}\dr^{\dagger}(\rho_{\G}\dr\Phi_{t}),
\non
\eeqa
and substituting $\rho_{t}=\rho_{\G}\Phi_{t}$ 
into the Fokker-Planck equation, \fr{Fokker-Planck-0},   
one obtains that 
$$
\rho_{\G}\frac{\partial}{\partial t}\Phi_{t}
=-\beta^{-1}\dr^{\dagger}(\rho_{\G}\dr\Phi_{t}).
$$
To reduce this equation further, the identity 
$(\dr^{\dagger}\rho_{\G})\alpha=(\rho_{\G}\dr_{\G}^{\dagger})\alpha$,
for all $\alpha\in\GamLamM{k}$, $k=0,1,\ldots,m$ is applied. This
holds due to  
$$
\dr^{\dagger}\rho_{\G}\alpha
=\rho_{\G}\rho_{\G}^{-1}\dr^{\dagger}\rho_{\G}\alpha
=\rho_{\G}\dr_{\G}^{\dagger}\alpha,
$$
where \fr{d-dagger-conversion} has been employed.  
Applying the identity with $\alpha=\dr\Phi_{t}$, one has that
$$
\frac{\partial}{\partial t}\Phi_{t}
=- \beta^{-1}\dr_{\G}^{\dagger}\dr\Phi_{t}.
$$
This is identical to the modified diffusion equation
\fr{diffusion-from-Fokker-Planck} 
with the
diffusion constant $\beta^{-1}$ due to $\triangle_{\G}=\dr_{\G}^{\dagger}\dr$. 
\qed
\end{Proof}

\subsection{Various weighted Laplacians }
\label{section:appendix-various-Laplacian}
The operator $\triangle_{\G}=\dr_{\G}^{\dagger}\dr$ in \fr{weighted-Laplacian},
with $\dr_{\G}^{\dagger}=\rho_{\G}^{-1}\dr^{\dagger}\rho_{\G}$ 
is closely related to   
the so-called Witten Laplacian,
and there are several variants of the Witten Laplacian.
In this subsection some of them are briefly discussed.

\begin{enumerate}
\item
Firstly,  define $\dr_{\W}$ as 
\beqa
\dr_{\W}:\GamLamM{k}&\to&\GamLamM{k+1},\quad k=0,1,\ldots,m
\non\\
\alpha&\mapsto&\dr_{\W}\alpha
\non
\eeqa
with 
\beq
\dr_{\W}
:=\e^{\beta h}\dr \e^{-\beta h}.
\label{Witten-exterior-2}
\eeq
For example, 
the action of \fr{Witten-exterior-2} to a function $\Phi$ is
$$
\dr_{\W}\Phi
=\e^{\beta h}\dr (\e^{-\beta h}\,\Phi)
=\dr\Phi-\beta\Phi\dr h,\quad
\forall\Phi\in\GamLamM{0},
$$
and $\dr_{\W}\dr_{\W}\Phi=0$ can be verified.
The corresponding adjoint operator of \fr{Witten-exterior-2}
with respect to $\inp{}{}_{\G}$ in \fr{inner-product-G} is
denoted by $\dr_{\W}^{\dagger}$.
Note that if another global inner product was chosen,
such as the standard one $\inp{}{}$ in \fr{inner-product-standard-k},
then the following argument would be significantly changed.  
The adjoint operator of $\dr_{\W}$ with respect to $\inp{}{}_{\G}$ 
is shown simply as
\beq
\dr_{\W}^{\dagger}
=\dr^{\dagger}.
\label{Witten-exterior-2-adjoint}
\eeq
To verify \fr{Witten-exterior-2-adjoint},
let $\alpha^{\I}\in\GamLamM{k-1}$ and
$\alpha^{\II}\in\GamLamM{k}$. Assume that the boundary term vanish, and write $\rho_{\G}=\exp(-\beta h+c)$ 
with $c=-\ln Z_{\G}$. Then
it is verified by 
\beqa
\inp{\dr_{\W}\alpha^{\I}}{\alpha^{\II}}_{\G}
&=&\int_{\cM}(\e^{\beta h}\dr\e^{-\beta h}\alpha^{\I})\wedge\star
\e^{-\beta h}\e^{c}\alpha^{\II}
=\e^{c}\int_{\cM}\dr(\e^{-\beta h}\alpha^{\I})\wedge\star\alpha^{\II}
\non\\
&=&\e^{c}\e^{-\beta h}\alpha^{\I}\wedge\star\alpha^{\II}|_{\partial \cM}
+(-1)^{k}\e^{c}\int_{\cM}\e^{-\beta h}\alpha^{\I}\wedge\star(\star^{-1}\dr\star\alpha^{\II})
\non\\
&=&
\int_{\cM}\alpha^{\I}\wedge\star\e^{-\beta h}\e^{c}\dr^{\dagger}\alpha^{\II}
=\inp{\alpha^{\I}}{\dr^{\dagger}\alpha^{\II}}_{\G}.
\non
\eeqa
By extending the standard procedure to define the Laplacian on a
Riemannian manifold, one defines 
  $\triangle_{\W}:\GamLamM{0}\to\GamLamM{0}$. 
This $\triangle_{\W}$ acting on a function is defined as
$\triangle_{\W}:=\dr_{\W}^{\dagger}\dr_{\W}$, which is 
  $$
  \triangle_{\W}
  =\dr^{\dagger}\dr_{\W}.
  $$
It can be shown that eigenvalues of 
this Witten Laplacian are equal or greater than $0$. 
Both
$\triangle_{\G}$ and $\triangle_{\W}$ can be extended for $k$-forms if needed. 

The explicit form of the Witten Laplacian $\triangle_{\W}$ acting on
a function $\rho$ is
$$
\triangle_{\W}\rho
=\dr^{\dagger}\dr_{\W} \rho
=\dr^{\dagger}\dr\rho-\beta\dr^{\dagger}(\rho\dr h).
$$
There is a sign difference between the equation 
$\dot{\rho}=-\beta^{-1}\triangle_{\W}\rho$   
and the Fokker-Planck equation \fr{Fokker-Planck-0}.

\item
Secondly, define 
\beqa
\Dr:\GamLamM{k}&\to&\GamLamM{k+1},\qquad k=0,1,\ldots,m
\non\\
\alpha&\mapsto&\Dr\alpha
\non
\eeqa
with
\beq
\Dr:=\e^{-\beta h}\dr\e^{\beta h}. 
\label{Witten-exterior-3}
\eeq
This is also employed in the literature. 
The adjoint of $\Dr$ with respect to the standard or unweighted
global inner product
$\inp{}{}$ in \fr{inner-product-standard-k} 
is obtained. Write this adjoint of \fr{Witten-exterior-3}
as $\Dr^{\dagger}$, and it is
\beq
\Dr^{\dagger}
=\e^{\beta h}\dr^{\dagger}\e^{-\beta h}.
\label{Witten-exterior-3-adjoint}
\eeq
When the boundary term vanishes, 
$\Dr^{\dagger}$ in \fr{Witten-exterior-3-adjoint} is verified
to be the adjoint of $\Dr$
for $\alpha^{\I}\in\GamLamM{k-1}$ and $\alpha^{\II}\in\GamLamM{k}$ 
as 
\beqa
\inp{\Dr\alpha^{\I}}{\alpha^{\II}}
&=&\int_{\cM}(\e^{-\beta h}\dr\e^{\beta h}\alpha^{\I})\wedge\star\alpha^{\II}
=\int_{\cM}\dr(\e^{\beta h}\alpha^{\I})\wedge \e^{-\beta h}\star\alpha^{\II}
\non\\
&=&\int_{\cM}\dr(\e^{\beta h}\alpha^{\I}\wedge\e^{-\beta h}\star\alpha^{\II})
+(-1)^{k}\int_{\cM}\e^{\beta h}\alpha^{\I}\wedge\star
(\star^{-1}\dr\star\e^{-\beta h}\alpha^{\II})
\non\\
&=&\alpha^{\I}\wedge\star\alpha^{\II}|_{\partial\cM}
+(-1)^{k}\int_{\cM}\alpha^{\I}\wedge\star
(\e^{\beta h}\star^{-1}\dr\star\e^{-\beta h}\alpha^{\II})
\non\\
&=&\inp{\alpha^{\I}}{\e^{\beta h}\dr^{\dagger}\e^{-\beta h}\alpha^{\II}}.
\non
\eeqa
There are several ways to express $\Dr$ and $\Dr^{\dagger}$, including
$$
\Dr=\dr+\beta\dr h\wedge,
\qquad
\Dr^{\dagger}
=\dr^{\dagger}+(-1)^{m+1}\beta\ii_{(\dr h)^{\sharp}}, 
$$
where the following isomorphism has been introduced: 
\beqa
{}^{\sharp}:\GamLamM{1}&\to&\GTM\non\\
\alpha&\mapsto&\alpha^{\sharp},
\non
\eeqa
with $\alpha^{\sharp}\in\GTM$ being such that 
$\alpha=g(\alpha^{\sharp},-)\in\GamLamM{1}$. 
The action of $\Dr^{\dagger}$ to a $k$-form $\alpha$ is calculated
from \fr{Witten-exterior-3-adjoint}  
as 
\beqa
\Dr^{\dagger}\alpha
&=&\e^{\beta h}(-1)^{k}\star^{-1}\dr(\e^{-\beta h}\star \alpha)
=(-1)^{k}\star^{-1}\e^{\beta h}
[\e^{-\beta h}(-\beta \dr h)\wedge\star\alpha
+\e^{-\beta h}\dr\star\alpha]
\non\\
&=&(-1)^{k}\star^{-1}
[(-\beta \dr h)\wedge\star\alpha
  +\dr\star\alpha]
=(\dr^{\dagger}\alpha)+
(-1)^{k+1}\beta\star^{-1}(\dr h\wedge\star\alpha). 
\non
\eeqa
The action of $\Dr^{\dagger}\Dr$ to a function is 
calculated from
$$
\Dr\Phi
=\dr\Phi+\beta \Phi\dr h,\quad \forall \Phi\in\GamLamM{0}
$$
as
\beqa
\Dr^{\dagger}\Dr\Phi
&=&\Dr^{\dagger}(\dr\Phi+\beta \Phi\dr h)
=\dr^{\dagger}(\dr\Phi+\beta \Phi\dr h)+
\beta\star^{-1}[\dr h\wedge\star
(\dr\Phi+\beta \Phi\dr h)]
\non\\
&=&
\dr^{\dagger}\dr\Phi+\beta \dr^{\dagger}(\Phi\dr h)
+\beta\star^{-1}(\dr h\wedge\star
\dr\Phi)+ \beta^{2}\Phi\star^{-1}(\dr h\wedge\star\dr h)
\non\\
&=&
\dr^{\dagger}\dr\Phi-\beta
\star^{-1}(\dr h\wedge \star\dr \Phi+\Phi\dr\star\dr h)
+\beta\star^{-1}(\dr h\wedge\star\dr\Phi)
+ \beta^{2}\Phi\star^{-1}(\dr h\wedge\star\dr h)
\non\\
&=&
\dr^{\dagger}\dr\Phi-\beta\Phi\star^{-1}\dr\star\dr h
+ \beta^{2}\Phi\star^{-1}(\dr h\wedge\star\dr h)
\non\\
&=&
\dr^{\dagger}\dr\Phi+\beta\Phi\dr^{\dagger}\dr h
+ \beta^{2}\Phi\,g((\dr h)^{\sharp},(\dr h)^{\sharp}), 
\non
\eeqa
where the following have been employed:
$$
\dr^{\dagger}\Phi\dr h
=-\star^{-1}\dr(\Phi \star\dr h)
=-\star^{-1}(\dr\Phi\wedge \star\dr h+\Phi\dr\star\dr h)
=-\star^{-1}(\dr h\wedge \star\dr \Phi+\Phi\dr\star\dr h),
$$
and
$$
\star^{-1}(\dr h\wedge\star \dr h)
=g((\dr h)^{\sharp},(\dr h)^{\sharp}).
$$
Note that
it is unclear how $\Dr^{\dagger}\Dr$ is exactly corresponding
to the Fokker-Planck equation. 
\end{enumerate}
%
%

\subsection{An expression of the Fokker-Planck equation}
\label{section:appendix-Fokker-Planck-d-and-d-dagger}
There are several expressions of \fr{diffusion-from-Fokker-Planck}. 
In this section one of them is shown. Specifically, 
a PDE for $\Phi_{t}=\rho_{\G}^{-1}\rho_{t}$ 
without use of $\triangle_{\G}$ is shown below.
\begin{Proposition}
The Fokker-Planck equation \fr{Fokker-Planck-0} 
can be written in terms of $\Phi_{t}$ and 
$\triangle:=\dr^{\dagger}\dr$ as
\beq
\frac{\partial}{\partial t}\Phi_{t}
=-\beta^{-1}\triangle\Phi_{t}
-g((\dr\Phi_{t})^{\sharp},(\dr h)^{\sharp}),
\label{Li-eq}
\eeq
with $\alpha^{\sharp}\in\GTM$ being such that
$\alpha=g(\alpha^{\sharp},-)\in\GamLamM{1}$. 
\end{Proposition}    
\begin{Proof}  
  The derivation of \fr{Li-eq}  
  is to substitute the assumed form
$\rho_{t}=\rho_{\G}\Phi_{t}$ into the Fokker-Planck
equation\,\fr{Fokker-Planck-0}. 
The details of calculations are as follows.

Since $\rho_{\G}=\exp(-\beta h+c)$ with $c=-\ln Z_{\G}$ and
$\rho_{t}=\rho_{\G}\Phi_{t}$ in  \fr{Gibbs-distribution} and \fr{rho-rho-G-Phi}, 
one has
$$
\dr\rho_{t}
=\e^{-\beta h+c}(\dr\Phi_{t}-\beta\Phi_{t}\dr h), 
$$
and then
\beqa
\dr^{\dagger}(\rho_{t}\dr h)
&=&-\star^{-1}\dr(\rho_{t}\star\dr h)
\non\\
&=&-\star^{-1}(\dr\rho_{t}\wedge\star\dr h+\rho_{t}\dr\star\dr h) 
\non\\
&=&-\star^{-1}[\e^{-\beta h+c}(\dr\Phi_{t}-\beta\Phi_{t}\dr h)\wedge\star\dr h
  +\e^{-\beta h+c}\Phi_{t}\dr\star\dr h ]
\non\\
&=&-\e^{-\beta h+c}\star^{-1}[(\dr\Phi_{t}-\beta\Phi_{t}\dr h)\wedge\star\dr h
+\Phi_{t}\dr\star\dr h],
\non
\eeqa
and
\beqa
\beta^{-1}\dr^{\dagger}\dr\rho_{t}
&=&-\beta^{-1}\star^{-1}\dr\star(\dr\rho_{t})
=-\beta^{-1}\star^{-1}\dr[\e^{-\beta h+c}\star (\dr\Phi_{t}-\beta\Phi_{t}\dr h)]
\non\\
&=&-\beta^{-1}\star^{-1}[
-\beta\e^{-\beta h+c}\dr h\wedge\star(\dr\Phi_{t}-\beta \Phi_{t}\dr h)
+\e^{-\beta h+c}\dr\star(\dr\Phi_{t}-\beta\Phi_{t}\dr h)
]
\non\\
&=&\e^{-\beta h+c}\star^{-1}[
\dr h\wedge\star(\dr\Phi_{t}-\beta \Phi_{t}\dr h)
-\beta^{-1}\dr\star(\dr\Phi_{t}-\beta\Phi_{t}\dr h)
].
\non
\eeqa
Summing the two terms above, one has that
\beqa
\beta^{-1}\dr^{\dagger}\dr\rho_{t}
+\dr^{\dagger}(\rho_{t}\dr h)
&=&-\beta^{-1}\e^{-\beta h+c}\star^{-1}\dr\star
\left(\dr\Phi_{t}-\beta\Phi_{t}\dr h\right)
-\e^{-\beta h+c}\Phi_{t}\star^{-1}\dr\star\dr h
\non\\
\e^{\beta h-c}\left(\beta^{-1}\dr^{\dagger}\dr\rho_{t}
+\dr^{\dagger}(\rho_{t}\dr h)\right)
&=&\beta^{-1}\dr^{\dagger}\left(\dr\Phi_{t}-\beta\Phi_{t}\dr h\right)
+\Phi_{t}\dr^{\dagger}\dr h
\non\\
&=&\beta^{-1}\dr^{\dagger}\dr\Phi_{t}-\dr^{\dagger}\Phi_{t}\dr h
+\Phi_{t}\dr^{\dagger}\dr h
\non\\
&=&\beta^{-1}\dr^{\dagger}\dr\Phi_{t}+\star^{-1}
\left(\dr\Phi_{t}\wedge\star\dr h\right),
\non
\eeqa
where in the last line 
$$
\dr^{\dagger}\left(\Phi_{t}\dr h\right)
=-\star^{-1}\dr \left(\Phi_{t}\star\dr h\right)
=-\star^{-1}\left(\dr\Phi_{t}\wedge\star\dr h+\Phi_{t}\dr\star\dr h\right)
=-\star^{-1}\left(\dr\Phi_{t}\wedge\star\dr h\right)+\Phi_{t}\dr^{\dagger}\dr h,
$$
has been used. 
Since
$$
\star^{-1}(\dr\Phi_{t}\wedge\star\dr h)
=g((\dr\Phi_{t})^{\sharp},(\dr h)^{\sharp}),
$$
the Fokker-Planck equation \fr{Fokker-Planck-0} 
can be written in terms of $\Phi_{t}$ and 
$\triangle:=\dr^{\dagger}\dr$ as \fr{Li-eq}. 
\qed
\end{Proof}
This derived PDE for $\Phi_{t}$, \fr{Li-eq}, is the PDE
discussed in Section 2 of \cite{Li2011}, up to some constant and the 
convention of the sign for the Laplacian.
  In \cite{Li2011}, the transform $\rho_{t}\mapsto\Phi_{t}=\rho_{\G}^{-1}\rho_{t}$
  is called the {\it ground state transform}. 
  In addition to \cite{Li2011}, several studies of \fr{Li-eq} 
  exist in the literature\,\cite{Villani2000}.  

  To see a property of the right hand side of \fr{Li-eq},
  define
\beqa
\Lr_{\beta,h}:\GamLamM{0}&\to&\GamLamM{0}
\non\\
\Phi&\mapsto&\Lr_{\beta,h}\Phi,
\non
\eeqa
with 
\beq
\Lr_{\beta,h}\Phi
:=\beta^{-1}\dr^{\dagger}\dr \Phi+\star^{-1}(\dr \Phi\wedge \star\dr h),
\quad\forall\Phi\in\GamLamM{0},
\label{F-beta-h}
\eeq
  so that \fr{Li-eq} is written as
  $$
  \frac{\partial}{\partial t}\Phi_{t}
  =-\Lr_{\beta,h}\Phi_{t}.
  $$
  The following is known in \cite{Villani2000}, and its proof is written 
  with the notation employed in this paper as follows.  
\begin{Proposition}  
\label{fact:F-self-adjoint}
  When the boundary term vanishes, 
  this $\Lr_{\beta,h}$ is symmetric 
  with respect to $\inp{}{}_{\G}$,
  $\Lr_{\beta,h}^{\dagger}=\Lr_{\beta,h}$: 
  $$
  \inp{\Lr_{\beta,h}\Phi^{\I}}{\Phi^{\II}}_{\G}
  =\inp{\Phi^{\I}}{\Lr_{\beta,h}\Phi^{\II}}_{\G},
  \qquad\forall\Phi^{\I},\Phi^{\II}\in\GamLamM{0},
  $$
  in addition, the following holds: 
  $$
  \inp{\Lr_{\beta,h}\Phi^{\I}}{\Phi^{\II}}_{\G}
  =\beta^{-1}\inp{\dr\Phi^{\I}}{\dr\Phi^{\II}}_{\G},
  \qquad\forall\Phi^{\I},\Phi^{\II}\in\GamLamM{0}.
  $$
  \end{Proposition}
\begin{Proof}
They are proved as follows.
  The main part of the proof is to show
  $\inp{\Lr_{\beta,h}\Phi^{\I}}{\Phi^{\II}}_{\G}
  =\beta^{-1}\inp{\dr\Phi^{\I}}{\dr\Phi^{\II}}_{\G}$. After showing this,
  by exchanging $\Phi^{\I}$ and $\Phi^{\II}$, the proof is completed. 

  First, one calculates 
  $\inp{\Lr_{\beta,h}\Phi^{\I}}{\Phi^{\II}}_{\G}$ straightforwardly.  
  From the identity 
  $\dr(\e^{-\beta h+c}\Phi^{\II})
  =\e^{-\beta h+c}(\dr\Phi^{\II}-\beta \Phi^{\II}\dr h)$,  
  it follows that 
\beqa
\beta^{-1}\inp{\dr^{\dagger}\dr\Phi^{\I}}{\Phi^{\II}}_{\G}  
&=&-\beta^{-1}\int_{\cM}(\e^{-\beta h}\Phi^{\II})\dr\star\dr\Phi^{\I}
\non\\
&=&\beta^{-1}\int_{\cM}\dr(\e^{-\beta h+c}\Phi^{\II})\wedge\star\dr\Phi^{\I}
-\beta^{-1}\e^{-\beta h+c}\Phi^{\II}\star\dr\Phi^{\I}|_{\partial\cM}
\non\\
&=&\beta^{-1}\int_{\cM}\dr(\e^{-\beta h+c}\Phi^{\II})\wedge\star\dr\Phi^{\I}
\non\\
&=&\beta^{-1}\inp{\dr\Phi^{\II}}{\dr\Phi^{\I}}_{\G}
-\inp{\Phi^{\II}\dr h}{\dr\Phi^{\I}}_{\G},
\non
\eeqa
and
\beqa
\inp{\star^{-1}(\dr\Phi^{\I}\wedge\star\dr h)}{\Phi^{\II}}_{\G}
&=&\int_{\cM}\e^{-\beta h+c}\Phi^{\II}(\dr\Phi^{\I}\wedge\star\dr h)
=\int_{\cM}\dr\Phi^{\I}\wedge\star\e^{-\beta h+c}\Phi^{\II}\dr h 
\non\\
&=&\inp{\Phi^{\II}\dr h}{\dr\Phi^{\I}}_{\G}.
\non
\eeqa
Summing these two obtained equations, one has that
$$
\inp{\Lr_{\beta,h}\Phi^{\I}}{\Phi^{\II}}
=\beta^{-1}\inp{\dr\Phi^{\II}}{\dr\Phi^{\I}}_{\G}.
$$
Exchanging $\Phi^{\I}$ and $\Phi^{\II}$ in the equation above, one has 
$\inp{\Lr_{\beta,h}\Phi^{\I}}{\Phi^{\II}}_{\G}=\inp{\Phi^{\I}}{\Lr_{\beta,h}\Phi^{\II}}_{\G}$.
\qed
\end{Proof}
From Proposition\,\ref{fact:F-self-adjoint},
the eigenvalues of $\Lr_{\beta,h}$ are greater or equal to zero,
which can be proved by a similar manner as 
in the proof of Lemma \ref{fact:eigenvalue-0-positive}. 

The relation between $\Lr_{\beta,h}$ and $\triangle_{\G}$ is given as follows.
\begin{Proposition}
\label{fact:F-weighted-Laplacian}
  $$
  \triangle_{\G}\Phi
  =\beta \Lr_{\beta,h}\Phi,\qquad
  \forall\Phi\in\GamLamM{0}.
  $$
\end{Proposition}
\begin{Proof}
  By straightforward calculations with \fr{weighted-Laplacian} and
  \fr{F-beta-h}, one has  
\beqa
\triangle_{\G}\Phi
&=&\dr_{\G}^{\dagger}\dr\Phi
=-\e^{\beta h-c}\star^{-1}\dr\left(\e^{-\beta h+c}\star\dr\Phi\right)
\non\\
&=&-\e^{\beta h-c}\star^{-1}\e^{-\beta h+c}\left(-\beta \dr h\wedge\star\dr\Phi
+\dr\star\dr\Phi\right)
\non\\
&=&\beta\star^{-1}(\dr\Phi\wedge\star\dr h)-\star^{-1}\dr\star\dr\Phi
\non\\
&=&\beta g\left((\dr\Phi)^{\sharp},(\dr h)^{\sharp}\right)+\dr^{\dagger}\dr\Phi
\non\\
&=&\beta \Lr_{\beta,h}\Phi,
\non
\eeqa
which is the desired equality. 
\qed
\end{Proof}
In the case of compact manifolds,  
the lower bound of the non-trivial first eigenvalue of $\Lr_{\beta,h}$ 
was estimated in \cite{Futaki2013}.  
Since the eigenvalues of  
$\triangle_{\G}$ coincide with the ones of $\beta\Lr_{\beta,h}$ 
due to Proposition\,\ref{fact:F-weighted-Laplacian}, 
this estimate and its  
related studies are expected to clarify a general property 
of relaxation processes that take place on compact manifolds.  

\subsection{Several choices of $z$}
\label{section:appendix-z=-F}
In the literature\,\cite{Entov2023,GLP}, the case with 
$z=-F$ is studied, where $F$ is given 
in \fr{F=S-H}.   
In this subsection, for the case of $z=-F$, it is shown that 
a construction of a closed dynamical system for $\ol{z}$ is 
impossible in general,  
where $\ol{z}$ obeys the slowest time-scale dynamics. 
In addition $z=\inp{\Phi_{t}}{h}_{\G}$ is also considered, 
where $h$ is the 
Hamiltonian.  
By recalling \fr{F=S-H}, 
this $z$ is interpreted as internal energy, $z=H$. 

Choose $z$ to be $-F$, that is expressed in terms of $\inp{}{}_{\G}$: 
$$
z:=-F
=\beta^{-1}\inp{\Phi_{t}}{\ln(\rho_{\G}\Phi_{t})}_{\G}+\inp{\Phi_{t}}{h}_{\G}.
$$
With the slowest time-scale dynamics developed in
Section\,\ref{section:derivation-contact-Hamiltonian}, introduce
$$
\ol{z}(t)
:=
\beta^{-1}\inp{\ol{\Phi}_{t}}{\ln(\rho_{\G}\ol{\Phi}_{t})}_{\G}
+\inp{\ol{\Phi}_{t}}{h}_{\G},
$$
where $\ol{\Phi}_{t}$ obeys \fr{slowest-mode}. 
Notice from Remark\,\ref{remark:averaged-values-asymptotic-same} that 
$$
\lim_{t\to\infty}z(t)
=\lim_{t\to\infty}\ol{z}(t)
  =\ol{z}(\infty),
$$
where 
$$
  \ol{z}(\infty)
  :=\beta^{-1}\inp{1}{\ln\rho_{\G}}_{\G}+\inp{1}{h}_{\G}.
$$
In addition the following holds:
$$
\ol{z}(t)-\ol{z}(\infty)
=\beta^{-1}\inp{\ol{\Phi}_{t}-1}{\ln\rho_{\G}}_{\G}
+\inp{\ol{\Phi}_{t}-1}{h}_{\G}+\beta^{-1}\inp{\ol{\Phi}_{t}}{\ln\ol{\Phi}_{t}}_{\G}.
$$
To derive an ODE for $\ol{z}$, 
differentiation of $\ol{z}$ with respect to $t$
is calculated as  
$$
\frac{\dr}{\dr t}\ol{z}
=
\beta^{-1}\left[
  \inp{\frac{\partial}{\partial t}\ol{\Phi}_{t}}{\ln(\rho_{\G}\ol{\Phi}_{t})}_{\G}
  +\inp{\frac{\partial}{\partial t}\ol{\Phi}_{t}}{1}_{\G}
  \right]
+\inp{\frac{\partial}{\partial t}\ol{\Phi}_{t}}{h}_{\G}.
$$
The second term in $[\cdots]$ above vanishes: 
$$
\inp{\frac{\partial}{\partial t}\ol{\Phi}}{1}_{\G}
=-\beta^{-1}\inp{\triangle_{\G}\ol{\Phi}}{1}_{\G}
=-\beta^{-1}\inp{\ol{\Phi}}{\triangle_{\G}1}_{\G}
=0,
$$
where Proposition\,\ref{fact:slowest-mode-diffusion-equation}
and the symmetric property of $\triangle_{\G}$ have been employed.  
From this vanishing term and \fr{slowest-mode-ODE}, 
the equation for $\dot{\ol{z}}$ reduces to    
\beqa
\frac{\dr}{\dr t}\ol{z}
&=&
\beta^{-1}\left[
  -\beta^{-1}\lambda_{1}\inp{\ol{\Phi}_{t}-1}{\ln(\rho_{\G}\ol{\Phi}_{t})}_{\G}
  \right]
-\beta^{-1}\lambda_{1}\inp{\ol{\Phi}_{t}-1}{h}_{\G}.
\non\\
&=&
-\beta^{-1}\lambda_{1}
\left[
  \beta^{-1}\inp{\ol{\Phi}_{t}-1}{\ln(\rho_{\G}\ol{\Phi}_{t})}_{\G}
+\inp{\ol{\Phi}_{t}-1}{h}_{\G}
  \right]
\non\\
&=&
-\beta^{-1}\lambda_{1}\left[
  \beta^{-1}\inp{\ol{\Phi}_{t}-1}{\ln\rho_{\G}}_{\G}
  +\beta^{-1}\inp{\ol{\Phi}_{t}-1}{\ln\ol{\Phi}_{t}}_{\G}
+\inp{\ol{\Phi}_{t}-1}{h}_{\G}
\right]
\non\\
&=&
-\beta^{-1}\lambda_{1}\left(\ol{z}-\ol{z}(\infty)
-\beta^{-1}\inp{1}{\ln\ol{\Phi}_{t}}_{\G}
\right).
\non
\eeqa
If the term being proportional to $\beta^{-2}$ is ignored above, then the 
right hand side in the obtained equation 
is written in terms of $\ol{z}$ only, and in this sense
this equation is a closed system.
  Meanwhile in general, the equation for this choice of $z$, $z=-F$,
  is not closed.
As another example, choose
$z:=\inp{\Phi_{t}}{h}_{\G}$, and let
$$
\ol{z}
:=\inp{\ol{\Phi}_{t}}{h}_{\G}
=\ol{\mbbE}_{t}[h]. 
$$
In this case, one has the closed dynamical system for $\ol{z}$: 
$$
\frac{\dr}{\dr t}\ol{z}
=-\beta^{-1}\lambda_{1}\left(\ol{z}-\inp{1}{h}_{\G}\right).
$$
The asymptotic limit of $\ol{z}$ is obtained as $\ol{z}\to \inp{1}{h}_{\G}$, 
$t\to\infty$, which is
$$
\lim_{t\to\infty}\ol{E}_{t}[h]
=\int_{\cM}h\rho_{\G}\star1.
$$
This can be compared with full time-scale dynamics as follows
(see also Remark\,\ref{remark:averaged-values-asymptotic-same}).  
Without introducing a contact  manifold, the asymptotic limit of
full time-scale dynamics is written for
$z=\inp{\Phi_{t}}{h}_{\G}$, 
as $z\to \inp{1}{h}_{\G}$, $t\to\infty$ from Corollary\,\ref{fact:limit-Phi}.
Thus the asymptotic limit of $z$ is the same as that of $\ol{z}$.



\begin{thebibliography}{99}
\bibitem{Abraham1991}
  Abraham R, Marsden JE, and Ratiu T 1991 
  {\it Manifolds, Tensor Analysis, and Applications 2nd Edition}, 
  Springer 
  
\bibitem{Bravetti2019}
  Bravetti A 2018
  Contact geometry and thermodynamics,
  {\it Int. J. Geom. Meth. Mod. Phys.} 1940003 (51pp)

\bibitem{Callen}
  Callen HB 1985 
  {\it Thermodynamics and an introduction to thermostatistics 2nd Eds}  
  John Willy and Sons 

\bibitem{Chavel1984}
 Chavel I 1984
  {\it Eigenvalues in Riemannian geometry}
  Academic Press  
  
\bibitem{Silva2008}
  Cannas da Silva A 2008 
  {\it Lectures on Symplectic Geometry} Springer
  
\bibitem{Choquet-Bruhat1997}
  Choquet-Bruhat Y, DeWitt-Mortte C, and Dillard-Bleick M 1997 
  {\it Analysis, Manifolds and Physics\ Part I:Basics}  
  Elsevier
  
\bibitem{Entov2023}
  Entov M and Polterovich L 2023
  Contact topology and non-equilibrium thermodynamics
  {\it Nonlinearity} {\bf 36} 3349-3375  

\bibitem{Entov2023arxiv}  
  Entov M, Polterovich L, and  Ryzhik L 2023
  Geometric aspects of a spin chain 
  arXiv:2311.14205

\bibitem{Ezra2002}
  Ezra GS 2002 
  Geometric approach to response theory in non-Hamiltonian systems 
  {\it J. Chem. Math.} {\bf 32} 339-360 
  
\bibitem{Frank2005}
  Frank TD 2005
  {\it Nonlinear Fokker-Planck equations: fundamentals and applications} 
  Springer  

\bibitem{Frankel2012}  
 Frankel T 2012
  {\it The geometry of physics An introduction  } Third edition 
  Cambridge University Press

\bibitem{Futaki2013}
  Futaki A, Li Haizhong, and Li Xiang-Dong 2013 
On the first eigenvalue of the Witten–Laplacian and the diameter of compact shrinking solitons 
{\it  Ann. Glob. Anal. Geom.} {\bf 44} 105–114
  
\bibitem{Gay2018}
  Gay-Balmaz F and Yoshimura H 2018
  Dirac structures in nonequilibrium thermodynamics,
  {\it J. Math. Phys.} {\bf 59} 012701 (29pp) 
  

\bibitem{Glauber1963}
  Glauber RJ 1963 Time-dependent statistics of the Ising model
  {\it J. Math. Phys.} {\bf 4}, 294–307

\bibitem{Goto2015}
  Goto S 2015
  Legendre submanifolds in contact manifolds as attractors and geometric nonequilibrium thermodynamics,
  {\it J. Math. Phys.} {\bf 56} 073301 (30pp)

\bibitem{Goto2016}
  Goto S 2016
  Contact geometric descriptions of vector fields on dually flat spaces and their applications in electric circuit models and nonequilibrium statistical mechanics, 
  {\it J. Math. Phys.} {\bf 57} 102702 (40pp)
  
  
\bibitem{Goto2020}
  Goto S and Hino H 2020
  Information and contact geometric description of expectation variables exactly derived from master equations,
{\it  Phys. Scr.} {\bf 95}  015207 (14pp)
  
\bibitem{Goto2020JMP}
  Goto S and Hino H 2020
   Diffusion equations from master equations -- A discrete geometric approach
  {\it J. Math. Phys.} {\bf 61} 113301 (27pp)

\bibitem{Goto2022}  
  Goto S 2022   
  Nonequilibrium thermodynamic process with hysteresis and metastable states—
  A contact Hamiltonian with unstable and stable segments 
  of a Legendre submanifold
  {\it J. Math. Phys.} {\bf 63} 053302 (25pp) 

\bibitem{GLP}
  Goto S, Lerer S, and Polterovich L 2023 
  Contact geometric approach to Glauber dynamics near a cusp and its limitation,
  {\it J. Phys. A} {\bf 56} 125001 (16pp) 


\bibitem{Grmela2014}
  Grmela M 2014 
  Contact geometry of mesoscopic thermodynamics and dynamics
  {\it Entropy} {\bf 16} 1652–1686 

\bibitem{Gromov2015}
  Gromov D and  Caines PE 2015 
  Stability of composite thermodynamic systems with interconnection
  constraints {\it IET Control Theory \& Applications} {\bf 9} 1629-1636
  
\bibitem{Haslach1997}
  Haslach HW Jr  1997 
  Geometric structure of the non-equilibrium thermodynamics of homogeneous 
  systems {\it Rep. Math. Phys.} {\bf 39} 147–62
  
\bibitem{Helffer2005}
  Helffer B and Nier F 2005 
  {\it Hypoelliptic Estimates and Spectral Theory for Fokker-Planck Operators
    and Witten Laplacians} 
  Springer

\bibitem{Hsu2002}
  Hsu EP 2002 
  {\it Stochastic Analysis on Manifolds},
  American Mathematical Society 

\bibitem{Ito2023}
  Ito S 2023 
  Geometric thermodynamics for the Fokker–Planck equation:
  stochastic thermodynamic links between information geometry and
  optimal transport  
  {\it Information geometry} {\bf 7} S441–S483 

\bibitem{Jost2017}
Jost J 2017
  {\it Riemannian Geometry and Geometric Analysis Seventh Edition}
  Springer

  
\bibitem{Otto1997}
  Jordan R, Kinderlehrer D, and   Otto F 1997 
  Free energy and the Fokker-Planck equation  
  {\it Physica D} {\bf 107} 265-271
 
\bibitem{Kubo1991}
  Kubo R, Toda M, and Hashitsume N 1991 
  {\it Statistical Physics II}  Springer

\bibitem{Kubo1962}
  Kubo R 1962 
  Generalized Cumulant Expansion Method  
  {\it J.Phys.Soc.Jpn} {\bf 17} 1100-1120

\bibitem{Li2011}
  Li XD 2011 
  Perelman's $W$-entropy for the Fokker-Planck equation over
  complete Riemannian manifold, 
 {\it  Bull. Sci. Math.} {\bf 135} 871-882 

\bibitem{Lotto2008}
  Lotto J 2008
  Some Geometric Calculations on Wasserstein Space, 
  {\it Commun. Math. Phys.} {\bf 277} 423-437
  
\bibitem{Mcduff2016}  
  McDuff D and Salamon D 2016
  {\it Introduction to Symplectic topology 3rd ed}, Oxford University Press 

\bibitem{Mrugala1990}
  Mrugala R,  Nulton JD, Sch\"on JC, and Salamon P 1990   
  Statistical approach to the geometric structure of thermodynamics,
  {\it Phys. Rev. A} {\bf 41} 3156-3160

\bibitem{Mrugala2000chapter}
  Mrugala R 2000 
  Geometrical Methods in Thermodynamics. In: Sieniutycz, S., De Vos, A. (eds)
  {\it Thermodynamics of Energy Conversion and Transport} Springer

\bibitem{Nishimori2001}
  Nishimori H 2001 
  {\it Statistical Physics of Spin Glasses and Information Processes} 
  Oxford University Press
    
\bibitem{Olver1986}
  Olver PJ 1986 
  {\it  Applications of Lie Groups to Differential Equations}
  Springer 

\bibitem{Pavliotis2014}
  Pavliotis GA 2014 
  {\it Stochastic Processes and Applications}
  Springer

\bibitem{Risken1989}
  Risken H 1989
  {\it The Fokker-Planck Equation Second Edition}  
  Springer    

\bibitem{Rosu1997}
  Rosu HC 1997  
  Supersymmetric Fokker-Planck strict isospectrality 
  {\it Phys. Rev. E} {\bf 56} 2269-2271 

\bibitem{Ruelle1999}
  Ruelle D 1999
  Smooth Dynamics and New Theoretical Ideas in Nonequilibrium Statistical
  Mechanics 
  {\it J. Stat. Phys.} {\bf 95}, 393-468 

\bibitem{Salazar2023}
  Salazar DSP 2023
  Bound for the moment generating function from the detailed fluctuation theorem
{\it Phys. Rev. E} {\bf 107} L062103
  
\bibitem{Schaft2018}
  van der Schaft A and Maschke B 2018 
  Geometry of Thermodynamic Processes  
  {\it Entropy} {\bf 20}  925 (23pp)

\bibitem{Villani2000}  
  Markowich PA and Villani C 2000
  On the trend to equilibrium for the Fokker-Planck equation:
  an interplay between physics and functional analysis
  {\it Mat. Contemp} {\bf 19} 1-29 

\bibitem{Villani2008}  
  Villani C 2008  
  {\it Optimal transport, old and new}  Springer

\bibitem{Vu2023PRX}
  Vu TV and Saito K 2023
  Thermodynamic Unification of Optimal Transport: Thermodynamic Uncertainty Relation, Minimum Dissipation, and Thermodynamic Speed Limits 
  {\it Phys. Rev. X} {\bf 13} 011013 (45pp)

\bibitem{Wada2009}
  Wada T and Scarforne AM 2009
Asymptotic solutions of a nonlinear diffusive equation in the framework of $\kappa$-generalized statistical mechanics  {\it Eur. Phys. J. B} {\bf 70} 65-71   
  
\bibitem{Zwanzig2001}
  Zwanzig R 2001 
  {\it Nonequilibrium statistical mechanics} 
  Oxford University Press 

\bibitem{Zetti2005}
 Zetti A 2005
  {\it Sturm-Liouville Theory} 
  American Mathematical Society   

\end{thebibliography}
\end{document}